\documentclass[fleqn]{annalen}
\usepackage{graphics}
\usepackage{epsffrag}
\usepackage{psfrag}
\usepackage{rotate}
\usepackage{graphicx}
\usepackage{amssymb}
\newcommand{\bm}[1]{\mbox{\boldmath$#1$}}
\def\part#1#2{ {\partial {#1} \over \partial {#2}} }
\def\pard#1#2{ {d {#1} \over d {#2}} }
\begin{document}
\newcommand{\volume}{9}              %sets current volume,
\newcommand{\xyear}{2000}            %sets year in header
\newcommand{\issue}{1}               %sets current issue,
\newcommand{\recdate}{24 February 2000}    %sets received date,
\newcommand{\revdate}{dd.mm.yyyy}    %sets revised date,     
\newcommand{\revnum}{0}              %number of revisions,
\newcommand{\accdate}{17 April 2000}    %sets accepted date,
\newcommand{\coeditor}{U. Eckern}           %sets (co)editor,
\newcommand{\firstpage}{1}           %first page number,  
\newcommand{\lastpage}{5}            %last page number,
\setcounter{page}{\firstpage}        %sets page counter to first page number 
\newcommand{\keywords}{Tunnelling, Superfluids, Vortex} 
\newcommand{\PACS}{66.35.+a, 67.40.Vs, 67.40.Hf}
\newcommand{\shorttitle}{Tunnelling of 
%topological %line 
defects 
in 
%strongly coupled 
superfluids} 
%\newcommand{\shorttitle}{Tunnelling of %line %topological line 
%defects in superfluids} 
%\newcommand{\shorttitle}{Tunnelling of topological line 
%defects in superfluids} 
%% sets the header on oddpage
\title{Tunnelling of topological line defects\\
 in strongly coupled superfluids}
\author{Uwe R. Fischer} 
\newcommand{\address}{Institut f\"ur Theoretische Physik der  
Eberhard-Karls-Universit\"at T\"ubingen \\
Auf der Morgenstelle 14, D-72076 T\"ubingen, Germany}
%%%%%%%%%%%%%%%%%%%%%%%%%%%%%%%%%%%%%%%%%%%%%%%%%%%%%%%%%%%%%%%%%%%%%%%%%%%%%%
\newcommand{\email}{\tt uwe.fischer@uni-tuebingen.de} 
\maketitle
\begin{abstract}
%The present work investigates the theoretical framework governing the 
%quantum process of tunnelling for a vortex moving in a superfluid liquid. 
The geometric theory of vortex tunnelling in superfluid liquids  
is developed.   
%in the semiclassical limit
Geometry rules the tunnelling process 
%of %the topological line defect quantized vortices 
in the approximation 
of an incompressible superfluid, which yields the 
identity of phase and configuration space in the vortex 
collective co-ordinate.
% of these topological line defects. 
%moving in an incompressible superfluid.  
%Assuming a domination of the Magnus force contribution in
%the Euclidean action, in three dimensions a volume traced out by the vortex 
%on its tunnelling path  
To exemplify the implications of %the geometric 
this approach to tunnelling, 
we solve explicitly for the two-dimensional motion of 
a point vortex in the presence of an ellipse, showing that the hydrodynamic 
collective co-ordinate description limits the constant energy paths 
allowed for the vortex in configuration space. 
%This restriction imprints 
%itself on the possible values of the semiclassical tunnelling exponent, which 
%has a lower limit. 
We outline the experimental procedure used in helium II 
to observe tunnelling events,  
and compare the conclusions we draw to the experimental 
results obtained so far. 
%The vortex dynamical equations governing t
Tunnelling in Fermi superfluids is discussed, where it is assumed that 
%solely 
the low energy quasiparticle excitations localised in the vortex core govern 
the vortex dynamical equations. 
%dynamical equations.   
%The peculiarities of these superfluids, as regards a number of energy 
%scales, which occur for 
%Different underlying order parameter symmetries   
%introduce different parameter ranges relevant for the nature of the 
%tunnelling process. 
%This process %of tunnelling 
The tunnelling process can be dominated by Hall or dissipative terms,  
respectively be under the influence of both, with a possible %experimental 
realization of this last intermediate case in unconventional,  
high-temperature  superconductors.  
\end{abstract}
%%%%%%%%%%%%%%%%%%%%%%%%%%%%%%%%%%%%%%%%%%%%%%%%%%%%%%%%%%%%%%%%%%%%%%%%%%%%%
\section{Introductory considerations}
%\addcontentsline{toc}{chapter}{\protect Introductory considerations}
%\markboth{\protect Introductory considerations}{\protect 
%Introductory considerations}
%\addcontentsline{toc}{chapter}{\protect Preface {\it cum} Introduction}
%\markboth{\protect Preface {\it cum} Introduction}{\protect Preface {\it cum} Introduction}
%\section{The dense superfluid}
%\subsection{General setting}
\addcontentsline{toc}{section}{\protect General setting}
The primary notion we have of a superfluid is that it shows no
dissipation of the flow, under certain, well-defined conditions. 
Friction can be caused by the creation of elementary quasiparticle
excitations in the superfluid, that is, irreversible
energy transfer from the coherently 
moving superfluid %vacuum 
state to incoherent degrees of freedom.
%representing the heat bath. 
This can occur if the superflow 
(relative to some reference frame) exceeds the Landau critical velocity,
creating the excitation, and thereby reducing %the superfluid density in 
the superfluid current. 
A different kind of dissipation  
%, which is not accompanied by entropy 
%production in the bulk of the superfluid, 
 %`friction' 
can be caused by a topological excitation,
%the transfer of part of the superflow energy to 
the quantized vortex, representing a travelling defect structure 
in the order parameter of the superfluid. 
The %energy conversion 
dissipation mechanism is then represented 
by the vortex crossing the streamlines of the flow, %thereby also 
diminishing %reducing the superflow velocity in 
the superfluid current by reducing the superfluid 
phase difference between points 
on a line perpendicular to the vortex motion. % in the superfluid.
To cause current reduction, %dissipation, 
a vortex first has to be generated.  
%The defect we will be dealing with in this thesis is the quantized vortex, 
%in three spatial dimensions a linear defect structure.
The task of this paper is to develop a formalism describing a 
quantized vortex entering a superfluid at the absolute zero of temperature 
by the quantum mechanism of tunnelling. 

The nucleation theory of quantized vortices in the Bose superfluid 
helium II has been an elusive subject ever since the existence of quantized
vortices was conjectured by Lars Onsager in %a short note dating from 
1949 \cite{onsager}. 
The difficulty to grasp their coming into existence  
in a quantitative manner 
from first principles has one fundamental reason: 
There is no microscopic theory of this dense superfluid. 
We do not know how to describe the motion of a vortex on atomic
scales,  %of the order of the coherence length
where this motion is governed 
by the full quantum many-body structure of the superfluid.
% and the interaction of the vortex with the microscopic excitations. 
One can even go further, and state that we even do not 
know precisely what a vortex should {\em be} on these small scales. 
%As we will argue in section \ref{sectiontopological} and 
Its very definability as a stable topological object essentially depends
on the usage of a large scale approach.  

Before we describe the problems inherent in vortex nucleation theory, 
to clarify terminology we use here and further in this
work, it is advisable to fix some notions.  
%\subsubsection{Definition of terms}
The treatment of a fluid will be called {\em hydrodynamic} or {\em large
scale}, with the frequency of excitations of the fluid approaching zero
for large wavelengths,  
if the underlying atomic structure of the fluid %, of extension $a$,
is not relevant  % and carrying any weight 
for the phenomena under investigation. The fluid is reasonably well,
on these large scales, 
approximated to consist of structureless, pointlike particles.
In particular, a flow {\em field} can be defined, with the prescription
that a volume element moving with a certain flow velocity contains
enough particles for the hydrodynamic formulation to make sense. 

The definition of {\em semiclassicality} is related to the existence
of quantum `fluctuations' or, better, the fact that a quantum
mechanical variable is indeterminate in its value with respect to 
the outcome of different measurements. If quantum fluctuations are
small against the expectation value of a quantum operator,
we speak of the semiclassical limit. 
The notions of hydrodynamic and semiclassical 
in the dense superfluid helium II 
are mutually corresponding to each other,  
and a hydrodynamic treatment on macroscopic scales has semiclassical accuracy.
%\footnote{The so-called macroscopic quantum phenomena like quantization of 
%circulation are thus also understood here as being semiclassical,
%whereas in mesoscopic physics matters are more intricate.}
%, given that proper conditions are imposed).
A {\em mean field} %or semiclassical 
theory asserts that it is reasonable to replace the 
field operator %(respectively a product of its spinor components) 
by its expectation value, in particular 
within expressions of order higher than the second
in the operator and its conjugate (an example is to replace 
two of the field operators in the quartic interaction 
term of the second quantized Hamiltonian by their expectation value). 
It therefore makes
the assumption that the system can be treated on any scale by making use of 
%semiclassically with respect to 
such a replacement.
%As we already expounded above, s
This procedure certainly cannot be used in the case of helium II.  
Though, on large scales, this dense superfluid %helium II %this superfluid 
is, to a good approximation, described by an
order parameter function 
%resulting from the expectation value of the 
%quantum field operator, 
$\phi$, % = <\!\hat \phi \!>$, where
%$<\!\cdots\!>$ denotes an average with many-particle wavefuctions 
%over a suitably large volume \cite{anderson}. 
on the atomic scales
of order the coherence length $\xi$, %of the order parameter, %in contrast, 
it is necessary to solve the full second 
quantized problem. There the indeterminacy of, {\it e.g.}, 
the operator particle density is as large as its expectation value. 
%$\hat\phi$ is as large as its expectation value $\phi$. 
%On the other hand, if $\xi $ is large
%against interaction lengths ({\it e.g.}, the $s$-wave scattering length
%\cite{ll}, to be discussed below), in dilute superfluids,
%or weakly coupled pair-correlated Fermi superfluids, 
%the mean field theory works on scales $O(\xi)$.  

A measure of the applicability of mean field theory can be
obtained if we consider the effective strength $g$ 
of interaction. It can, in the limit of long wavelengths, be defined to be the 
spatial integral  of the two-body interaction of $^4\!$He atoms, 
%where $\sigma_{\rm LJ}=2.556$ \AA \, 
%is  the Lennard-Jones length of the $^4\!$He atomic interaction.  
and is related to the $s$-wave scattering length $a$ \cite{ll}, 
the speed of sound $c_s$, and the bulk density $\rho_0$
by $g=4\pi \hbar^2 a /m=m{c}^2_s/\rho_0$ (repulsive interaction,
$a>0$, \cite{fetterwal1}).
A mean field treatment is useful if $\rho_0 a^3\ll 1$ \cite{BECreview}, 
which implies $a=\pi (m c_s/h)^2/\rho_0 \ll d$, 
%$\pi^3 (mc_s/h)^6/\rho_0^2(\sqrt \pi c_s /(\kappa d))^6\ll1$, 
with $d\equiv \rho_0^{-1/3}$ the interparticle spacing.
%and $\kappa = h/m = 10^{-7}$ m$^2$/s the quantum of circulation. 
%(cf. section \ref{sectiontopological}). 
Using the data  of \cite{donnelly2}, we have in helium II 
(at $p\simeq 1$ bar), $d\simeq 3.58$ \AA, $c_s\simeq 238$ m/s and thus 
$a \simeq  8 $ \AA, increasing further to approximately
double this value at solidification pressure. 
The scattering length, that is, the effective 
range of the scattering potential of the atoms for 
long wavelength excitations, is larger than  the interparticle
spacing, and the condition for the applicability of mean field theory is 
%grossly 
violated. 
%This provides us with the definition of a {\em dense}
%superfluid, namely the condition that $a\sim O(d)$. 
This fact entails that one has to consider helium II as a strongly 
coupled system, which has caused quite 
formidable computational efforts trying to understand 
its behaviour on sub-$a$ scales \cite{apaja}--\cite{dalfovo2}.
%, there has been compiled a far from exhaustive list of such efforts.  
Based on the above considerations, we will understand  a {\em dense}, strongly 
coupled superfluid as one fulfilling the condition $a\gtrsim d$ and,
correspondingly, a {\em dilute}, weakly coupled 
superfluid as one having $a\ll d$, the 
Bose-Einstein condensed atomic vapours \cite{BECreview} 
being examples of this kind of superfluid.
The condition  $a\gtrsim d$ implies %that $\xi\sim O(d)$, so 
that, in strongly coupled superfluids, the coherence length is of order
the interparticle spacing. 
%\begin{center}
%\begin{figure}[hbt]\label{scales}
%\begin{center}
%\rotate[r]
%{\includegraphics[width=0.7\textwidth]{regimesfullnew.eps}}\vspace*{-1em}
%-167.0 580.0 translate
%\end{center}
%\caption{Length scales of vortex motion and interaction with elementary 
%excitations in the dense  superfluid helium II. The radius of a domain
%indicates the length scale of vortex curvature radius respectively
%distance between vortices. It is only in the regions I and II where 
%a semiclassical %(string) 
%picture of vortex motion is applicable. The vortex in the central
%region of extension the atomic scale %$O(d)$ 
%is called virtual because it is no
%well-defined topological object in this region.}\vspace*{1em}
%\end{center}
%\end{figure}
%\end{center}  

We have already mentioned at the beginning that there are two classes
of excitations related to dissipation, the first being Landau quasiparticles, 
%excitations above the superfluid ground state, 
the second one defects in the order parameter. 
There arises the question if these different branches 
of excitations merge in a certain manner, approximately for a 
wavevector $2\pi/d$ corresponding to the interparticle spacing. 
%In particular, it would be of interest if they have the same critical
%current densities.  
%The measurable excitation spectrum of helium II is
%qualitatively given as follows. It begins with the linear phonon
%branch (because of  the Landau criterion \cite{landau1}, necessarily),
%has a maximum (the {\em maxon} excitation), then a minimum at k=  
The roton, {\it i.e.} the excitation at the local minimum of the 
excitation spectrum of superfluid helium II, has a wavevector $k_r$ which is 
very close to this value, $k_r\simeq 2\pi /d$. The idea that the roton
could correspond to a vortex ring of the smallest possible size (namely,
such that just one atom can pass through it), goes back to the seminal
papers of Feynman \cite{feyn1}--\cite{feyn3}. 
Evidence for such an identification would indeed provide justification
to speak of some kind of, % unification of, 
at least, similarity on the 
scale of the interparticle distance, of the two types
of excitations. 
%A proof that such a similarity may be asserted, however, 
%necessitates to show that there is both a backflow current structure
%resembling that of a vortex ring, and that (some kind of)
%quantized circulation may be shown to exist. 
There is still a debate going on about this possibility (recent
contributions are found, {\it e.g.}, in
\cite{apaja},\cite{galli}) %,\cite{Gov}). 
In this work, we will, however, not enter into such a discussion,  
for it is evidently still a long way to a complete understanding of 
the microscopic {dynamical} behaviour of vortices, 
as even their stationary microscopic character (provided that a vortex ring 
of atomic size is properly definable at all), is not yet
completely clear. We will take the (classic) point of view that a vortex is a
topological object, well-defined as a defect structure in the order
parameter. Then, a vortex ring of atomic size is termed virtual, because
it does not constitute a topological object on this scale. 
%The different scales one can distinguish for 
%the vortex dynamical behaviour in a qualitative way 
%are shown in 
%cf. Fig. \ref{scales}.

The lack of a microscopic idea of vortex motion makes it necessary to
resort to a hydrodynamic theory of the motion of a 
vortex, be this motion in real or imaginary time, the latter of interest 
for the tunnelling processes we intend to investigate. 
In a hydrodynamic treatment, 
the existence of the vortex as a semiclassical object 
has to be assumed {\it ab initio}. 
No details of the underlying microscopic dynamics, {\it i.e.} of the actual 
nucleation event, are to be described in such a theory, 
but only the laws which rule vortex motion on curvature scales well beyond
the atomic one. 
%two central regions in Fig. \ref{scales}. 
The microscopic dynamical behaviour of the vortex is, in such a
description, bound to appear only in cutoff parameters determining 
the borderline to the microscopic realm. 
We will consequently consider the vortex motion as it
results from the Lagrangian in terms of a collective co-ordinate %$\vec X$ 
for the vortex, which is useful as 
long as the curvature radii of the line described by this co-ordinate 
are much larger than $\xi$.
Such an approach in terms of a collective co-ordinate 
makes sense and is physically meaningful, if we additionally assume that the 
potential barriers, through which the vortex has 
to tunnel, have themselves effective curvature radii well beyond a 
scale $O(\xi)$.
%microscopic scale. 
%The length scales, which can be relevant for
If tunnelling events can be actually observable, then, 
depends for a given superfluid at a given temperature $T$
on the ratios $\xi/d$, $T^*/T$, where
$k_B T^*\equiv \hbar^2/(m d^2)$ is a characteristic quantum 
energy of the quantum fluid constituted by particles of mass $m$  
and $T^*$ an associated temperature;  
%equalling the Fermi energy in the case of Fermi superfluids. 
for helium II, $T^*\approx 1$ K. 
%(at normal pressures). 
%, and a slightly larger value for $^3\!$He.
This dependence can qualitatively be understood as follows \cite{parts}. 
Consider a vortex ring of radius $R_0$, which has in the bulk an energy
$E(R_0)=(m\rho_0\kappa^2/2) R_0 \ln (R_0/R_c)$ (see section
\ref{introhalfring}), where $R_c = O(\xi)$ 
is the vortex core radius.  \label{energyconsider}
The relevant quantity to compare this energy with, is the thermal
energy $k_B T$. If $E(R_0)\gg k_B T$, quantum tunnelling is exponentially
suppressed. Writing the energy in terms of the above quantities, 
$E(R_0)/k_B T = (T^*/T)(R_0/d)\ln (R_0/R_c)$ (barring a factor of $1/4$ for
pair-correlated superfluids). 
The ratio $\xi/d$ then effectively enters, % the ratio $E(R_0)/k_B T$, 
because the smallest possible value of the radius of the vortex
appearing in the fluid is $R_0\gtrsim  R_c = O(\xi)$. 
Considering the fact that thus simultaneously $T\sim O(T^*)$ 
and $R_0\sim O(d)$ are to be fulfilled, 
helium II is the most promising candidate for quantum tunnelling to
happen. The conventional superconductors and $^3\!$He, with their 
large $T^*/T$ as well as $\xi/d$ are, already
on this ground, ruled out. Hence the only possible candidates remaining for 
an observation of quantum tunnelling of vortices are, save for helium II,
high-$T_c$ superconductors. \label{mentiontunnelhigh}
\subsection*{Overview}
\addcontentsline{toc}{subsection}{\protect Overview}
To give the reader a concise impression 
%of the structural and physical contents 
of what follows, we provide here an overview of the 
principal directions to be pursued, and ideas to be developed in the three
sections of this work to follow. 
The theory of  quantum tunnelling is developed 
in the next section. 
It is shown that in the limit of long wavelengths, which we 
are required to be using in a dense superfluid,  
the probability of quantum tunnelling is predominantly determined 
by external geometrical quantities connected to the geometry of the flow. 
The tunnelling exponent is separated into a dominant volume contribution  
solely associated with the tunnelling path of the vortex, which is a 
contribution independent from the fact that the fluid is compressible
or incompressible, and a subdominant 
area contribution associated with that same path {\em and} the fact that the
fluid is compressible. 
The dependence of the dominant volume term in the tunnelling exponent 
on geometrical quantities is exemplified 
by the analytically solvable problem of a point vortex in the presence of an
ellipse, where the long wavelength, collective co-ordinate treatment
is shown to impose strict constraints on the motions allowed for the vortex. 
In the third section the experimental procedure to observe the temperature
independent quantum mechanical triggering of vortex generation, 
which we wish to explain, 
is demonstrated. We discuss the data obtained in these experiments, with 
particular emphasis on the applicability of our predictions 
in this work.
The fourth section describes some aspects of pair-correlated, charged 
Fermi superfluids. We discuss, in particular, the role which might be
played by the existence of bound quasi-particle states for observable
vortex tunnelling phenomena in superconductors. The high-$T_c$ 
superconductors, on account of their small coherence lengths, play 
a prominent role in these considerations, as emphasised above.  
%on page  \pageref{mentiontunnelhigh}. 
It is explained that, even for very low 
temperatures, in the case of unconventional ($d$-wave) high-$T_c$ 
superconductors of practically feasible purity, 
the tunnelling phenomenon is  
not adequately described by the theory 
of sections one and two, as this were the case for conventional ($s$-wave), 
extreme type II superconductors, on a length scale well below that of 
the magnetic penetration depth.  
%with the intention to describe the pecularities of
%these superfluids with respect to the motion of vortices in them. 
%%The possibility to observe quantum tunnelling under certain circumstances
%%is discussed. 

%\newpage
%\thispagestyle{empty} 

\section{Quantum tunnelling of vortices}
\subsection{General introduction}
%\subsection{Tunnelling}
The quantum mechanical phenomenon of tunnelling
%%\footnote{Orthographical note: 
%%Throughout this work, the british spelling of 
%%`tunnelling', to be contrasted with the american `tunneling', is used.}
has attracted attention from researchers in theoretical 
and experimental condensed matter physics, field theory and other areas.  
% (cf. the corresponding bibliography section). 
%, Refs. \cite{coleman1}-\cite{freire}). 
It belongs to the most remarkable properties solely pertaining to
systems obeying the laws of quantum mechanics.   
Essentially, quantum mechanical tunnelling in a condensed matter system 
is described by the motion of a few % singled-out 
degrees of freedom (subsumed in what follows into the term `particle') 
under a potential barrier in % some (multidimensional) 
configuration space. 
The particle is assumed to have less energy than represented by the height of 
the barrier (setting the bottom of the potential equal to the zero of
energy).  Because of wave-particle duality, a wave function can be
associated with the particle. 
This wave function is damped under the barrier or, in other words, the
particle travels with imaginary momentum there.  
To some extent and with some probability, the particle is thus located
under the barrier.  It can then even completely penetrate it, if the damping
is small enough,  getting with some nonzero probability  
from one side of the barrier to the other, 
whereas classically this is a completely forbidden process. 
%This motion of constant energy under the barrier 
%is called quantum tunnelling. 

There are different means to describe the tunnelling motion. The canonical way
is to calculate wave functions for the problem,  solving, {\it e.g.}, 
the Schr\"odinger equation for the potential of interest with
appropriate boundary conditions imposed \cite{schmid,eckern}. 
The most popular formal means to investigate tunnelling, however, is 
provided by the calculation of the Euclidean action of the system
along the tunnelling trajectory \cite{coleman,coleman1,coleman2}. 
The Euclidean action is obtained
from the Minkowski action by rotating to purely imaginary times
$S_e\equiv -i S[t \rightarrow -it_e]$. The time on the imaginary axis 
of the complex $t$-plane will be denoted $t_e$. 
If the action is dominated by the
classical path in Euclidean time, the tunnelling probability can be 
calculated in the semiclassical approximation. The 
corresponding solution of the second order 
Euclidean classical (field) equations of motion with finite action 
is called instanton. The name stems from the fact that the instanton
is a particle-like object localised in Euclidean time. It exists, so
to speak, just long enough for the actual particle to tunnel. 

In the semiclassical limit, the tunnelling 
probability for a given energy $E$ is taking the form 
\begin{equation}\label{PASe}
P(E)=A(E)\exp\left[-\frac{S_e(E)}\hbar\right]\,.
\end{equation}  
In this relation, $S_e(E)$ 
is the Legendre transform of the Euclidean action $S_e(T_e)\gg \hbar$ 
as a function of $T_e$, the Euclidean period of motion 
%(see, in particular, the textbook 
\cite{ll}:
\begin{eqnarray}\label{SE}
S_e(E)= S_e(T_e)-\frac{\partial S_e}{\partial T_e} T_e= S_e(T_e)- ET_e\,.
\end{eqnarray}
The quantity $A(E)$, frequently called {\em prefactor}, 
 represents the influence of fluctuations around
the classical path. It is essentially given by the inverse 
determinant of the co-efficients of the second order deviations
from the classical path in the action \cite{coleman,kleinert}. 
%We will concentrate on the pure quantum tunnelling case, that is, we
%will consider the case of zero environmental temperature. 
%Fluctuations
\subsection{The collective co-ordinate action of the vortex}
It is a quite commonly accepted wisdom that
any complex condensed matter problem
%in principle requiring to solve (in three spatial dimensions) $3N$ equations
%(where N=O$(10^{20}\ldots 10^{23}) is the number of particles),  
remains intractable if we do not single out certain central, collective degrees 
of freedom, termed in general `{collective co-ordinates}'. 
This is possible because there are conservation laws and 
symmetries governing the behaviour of the system as a whole: 
We can actually describe essential features while not 
referring explicitly to the $10^{20}\cdots 10^{23}$
particles and their interaction.  

The obvious choice for the vortex collective co-ordinate is its center
$X^i(t,\sigma)$, which also indicates the center of topological stability
and thus conserved topological charge. 
% (cf. subsections \ref{sectiontopological}, \ref{subsectionMagnus} 
%and, for visualization, the central dots in Figures 
%\ref{winding} and \ref{vortexisolated}).  
That this co-ordinate represents the vortex sufficiently accurate
%a singular, well-defined line
requires that we consider scales much larger than
the vortex core size of order the coherence length $\xi$. Furthermore,
we assume that there is a canonical collective vortex momentum related to
this central co-ordinate.
% \cite{annalspaper}. 
%as expounded in subsection \ref{subsectionLagrangian}. 
The action (\ref{SE}) then is 
\begin{eqnarray}\label{SEKP}
S_e(E) & = &  -i\int_0^{T_e}\!\!\oint\! dt_e \, d\sigma \,\dot {\bm X}\cdot
{\bm P} 
%\nonumber\\&=& 
=-i \oint\! d\sigma\!\oint d{\bm X}\cdot {\bm P}%(\sigma)
\nonumber\\&=& 
\oint\! d\sigma\!\oint d{\bm K}\cdot {\bm P}%(\sigma)
\,,\label{SeEKP}
\end{eqnarray}
where we defined the imaginary differential co-ordinate vector
$d{\bm K}=-i d{\bm X}$ of the
vortex.\footnote{$\bm K$ is not to be confused with a wave vector. We could
have chosen as well to incorporate the $-i$ into the (imaginary) 
momentum ${\bm P}$. Crucial is only that $S_e$ is a real quantity.}
The parameter $\sigma$ labels points on the vortex string, and ${\bm P}$
is the canonical momentum per $\sigma$-length.

%\footnote{

Note that the co-ordinate 
{\em differential} vectors in (\ref{SeEKP})
are no function of $\sigma$, as the co-ordinate 
{\em position} vectors themselves of course are. 
The closed time integral indicates that we take the integral
over a full period of the motion.
That such a periodic motion exists,  
is of course a highly nontrivial assumption for arbitrary dimension 
of the phase space. Only in an effectively one-dimensional problem 
(one spatial dimension), respectively for multidimensional
systems separable into such one-dimensional problems (cf. \cite{ll} 
\S 48), closed phase space trajectories necessarily exist. 
\subsubsection{Contributions in the tunnelling action}\label{contributions}
In two dimensions and in a conventional, {\em electrically 
uncharged} superfluid, 
vortices and charged particles have identical dynamical 
equations in the hydrodynamic limit.  
According to the three-dimensional 
extension of this duality, {\it i.e.} as a vortices--charged strings analogy, 
the vortex Hamiltonian takes the form \cite{annalspaper} 
%\begin{eqnarray}
$$
H_V %[{\bm Q}, {\bm P}]
=   \oint\! d\sigma %\left\{
\sqrt\gamma\, \left[M_0 c_s^2  
+ \frac 1{2\gamma M_0} \left( {\bm P}
-q{\bm a}\right)^2 
%\right.\nonumber \\
%-{\bm P}^{\rm inc}\right)^2\right.\nonumber \\
%& & \quad \,\quad \qquad \left. 
+\frac{M_0 c_s^2}{2\gamma}\,
{\bm Q}'{}^2\right] 
+ q\int\! d\sigma \left(\frac12 a^0_C 
 +a^0_{u}\right)\,.\nonumber \\
\hspace*{-3em}
%- m \rho_0 \Gamma_s\int\! d\sigma \left(\frac12 {\bm \psi}_C 
% +{\bm \psi}_{u}\right)\cdot  {\bm X}'\,. 
\vspace*{-1em}$$
\begin{eqnarray}
\label{HV}
%\right\}
\end{eqnarray}
The first integral represents the self energy of the vortex, which has static, 
kinetic, and elastic contributions, respectively. 
The arc length of the line is written as $\sqrt\gamma d\sigma$,  
and the vector ${\bm Q}$ is a perturbation of some equilibrium string 
configuration perpendicular to the tangent vector ${\bm X}'$ of the line. 
The rest frame mass $M_0$ is given by $M_0=E_0/c_s^2$ \cite{duan}, 
where $E_0$ is the logarithmically divergent static self energy of the vortex 
(per unit length). In helium II this energy reads 
$E_0 = [{N_v^2 \kappa^2m\rho_0}/{(4\pi )}]\ln \left({8R_c}/{\xi e^C}
\right)\,$, 
with $\rho_0$ the bulk number density, $m$ the helium mass, 
$\kappa = h/m$ the velocity circulation quantum,  
$R_c$ the infrared cutoff (the local curvature radius of the line), 
$\xi$ the ultraviolet cutoff (the core size), and $C$ a constant 
characterising the core structure.
The second integral represents the `electrostatic' energy, in which 
we split the scalar potential $a^0$ into a Coulomb contribution from the 
interaction with other vortices, $a^0_C$ and the interaction 
with a nonvortical background flow, $a^0_u$. The vectorial 
generalization %\cite{annalspaper} 
of the stream function $\psi$ 
of classical hydrodynamics \cite{milne-t}, is related to $a_0=-a^0$ by 
$a_0=\rho_0 {\bm \psi}\cdot  {\bm X}'$.       
The factor 1/2 in front of $a^0_C$ stems from the Coulomb gauge for 
the vector potential, div\,${\bm a}=0$,  and makes sure that the energy of 
each vortex is counted only once.  
The `charge' $q=N_v h $ is given by 
the %(superfluid phase) 
winding number $N_v$ multiplying Planck's quantum of action.\footnote{  
In pair-correlated Fermi superfluids, $q=N_v h/2$, where the number density
in (\ref{rotaBX'})  and (\ref{Efield}) is understood to refer to  
the ``elementary'' 
particles constituting the superfluid, and not to the Cooper pairs.} 

The gauge potentials are derived from the external flow field at the position 
of the line element at $\sigma$ by a gauge invariant 
duality relation \cite{annalspaper}. 
In its nonrelativistic form needed here, this reads
\begin{eqnarray}
{\rm rot}\, {\bm a} & = & - \rho  {\bm X}'\,,\label{rotaBX'}\\
\partial_t {\bm a} +\nabla a^0 & = & {\bm X}'\times {\bm j}\label{Efield}\,,  
\end{eqnarray} 
where ${\bm j}=\rho{\bm v}$ 
is the conserved particle current,  
%and $\rho$ the densityof the fluid,  
$\partial_t\rho + {\rm div}{\bm j} = 0$. 
%\begin{equation}
%\partial_\mu a_\nu -\partial_\nu a_\mu = 
%\epsilon_{\mu\nu\alpha\beta} j^\alpha X'^\beta\,. 
%\end{equation}
The quantity of crucial importance in the Euclidean action 
of constant energy (\ref{SeEKP}), determining 
the tunnelling exponent, is the canonical momentum 
\begin{equation}\label{Pcanon}
\bm P = \bm P^{\rm inc} + \bm P^{\rm kin}
%\vec{e}_A \frac{\delta L}{\delta \dot Q_A(t,\sigma)}
=q \bm a + M_0 \sqrt\gamma \dot {\bm X}
\, .
\end{equation}
It consists of a contribution $\bm P^{\rm inc}$,  
which is related to the Magnus force acting on the vortex,  
%in an incompressible ideal fluid 
and a second contribution  $ \bm P^{\rm kin}$ related to the existence
of a nonzero vortex mass, that is, to a finite compressibility and thus 
finite $c_s$. It follows by integration of (\ref{rotaBX'}) 
that the ratio of the momentum contributions  
${\bm P}^{\rm kin}$, ${\bm P}^{\rm inc}$ 
is in order of magnitude 
$\approx (N_v \kappa)/(c_s|\bm X|)(|\dot {\bm X}|/c_s)$ 
(neglecting the vortex energy logarithm).   
%\cite{annalspaper}   
%\begin{equation}\label{ratioP}
%\left|\frac{{\bm P}^{\rm kin}}
%\right|/\left|
%{{\bm P}^{\rm inc}}\right|
%\approx \frac{N_v \kappa}{c_s|\bm X|}\frac{|\dot {\bm X}|}{c_s}\,.
%\end{equation}
Hence for large scales (large curvature radii),  
and small velocities, that is in the hydrodynamic limit, 
$ \bm P^{\rm kin}$ is dominated by $\bm P^{\rm inc}$.
    
%The self-action is dominated by the self-energy of the line. 
%Likewise for t
The Euclidean action is given by 
%$(\dot {\bm X}= \dot {\bm Q})$: 
\begin{eqnarray}
S_e(T_e) & = &  
\int_0^{T_e}\!\oint dt_e d\sigma\sqrt{\gamma(t_e,\sigma)}M_0 c_s^2\left\{ 
1+\frac1{2c_s^2} \dot{\bm Q}{}^2 +\frac1{2\gamma}\, 
{\bm Q}'{}^2\right\}\nonumber\\
& & \quad + q\int_0^{T_e}\!\oint dt_e d\sigma\left[ a^0 -i \bm a\cdot
\dot {\bm X}\right]\,.\label{SEQ}
\end{eqnarray} 
In case that a tunnelling process of constant energy is under
consideration, % which is usually the case of interest, 
the quantity of interest is the action as a function 
of constant energy  (\ref{SEKP}).
%, as discussed above. 
According to the relation (\ref{Pcanon}), this action consists of 
a part related to the vector potential and another part related to 
the vortex effective mass $(\dot {\bm X}= \dot {\bm Q})$: 
\begin{eqnarray}
S_e(E) & = & S^{\rm inc}_e(E)+S^{\rm kin}_e(E)=
\int_0^{T_e}\!\oint dt_e d\sigma \dot{\bm X}\cdot \left[\bm P^{\rm inc}
+ \bm P^{\rm kin}\right]\nonumber\\ 
& = & \int_0^{T_e}\!\oint dt_e d\sigma \dot{\bm X}\cdot \left[
-iq \bm a + M_0 \sqrt\gamma \dot {\bm X}\right]\,.\label{SEtotal}
\end{eqnarray}
The Euclidean action splits into a part $S_e^{\rm inc}$, 
due to the interaction of the vortex with an (approximately) 
incompressible background superfluid, and a part $S_e^{\rm kin}$ 
which can be ascribed to the kinetic (`vortex matter') 
term in  the vortex momentum.    
%According to (\ref{ratioP}), 
%$S^{\rm inc}_e(E)\gg S^{\rm kin}_e(E)$. 
We will now show that 
$S^{\rm inc}_e(E)$ is given by a volume associated with the path 
the vortex line traces out in configuration space, whereas $S_e^{\rm
kin}$ is connected with an area associated with that path.

To demonstrate this, we integrate relation (\ref{rotaBX'}), multiplied with
$q$, to obtain ($\rho=\rho_0$) 
\begin{equation}
-\oint\!\!\oint P^{\rm inc}_{\it a} {d}X^{\it a} {d}\sigma 
= {N_v}h\rho_0\int\!\!\int\!\!\int\!
 \sqrt g \, {d}X^{\it 1}{d}X^{\it 2}{d}\sigma\,,
\end{equation}
where $g$ is the determinant of the coordinate basis on the line
(unity for a triad).  
%which is equivalent to the statements made in equations (\ref{bintVol})
%and (\ref{p1p2}).
The closed surface with surface elements of magnitude 
${d}X^{\it a} {d}\sigma$ ({\it a= 1,2} is the index of the two 
${\bm Q}$-directions), 
encloses the total volume with local element 
$\sqrt g \, {d}X^{\it 1}{d}X^{\it 2}{d}\sigma$
traced out by the line on its path.     
Further, using the gauge freedom for the momentum,  
we can express the action $S^{\rm inc}_e(E)$ by the volume integral  
\begin{equation}
\label{volume}
\frac{S^{\rm inc}_e (E)}\hbar =2\pi{N_v}\, 
%\rho_0\int {\bm d} V  =
\rho_0\,\int\!\!\!\int\!\!\!\int\! \sqrt g\,
%\, n_{ijk}\, {\bm d}Z^i\wedge {\bm d} Z^j {\bm d} Z^k
dZ^1 dZ^2 \, d\sigma\, \,, 
\end{equation}
wherein the co-ordinate differentials  %in the surface integral 
are defined to be 
\begin{eqnarray}
dZ^1 & = & \cos\alpha \, dK^{\it 1} + \sin\alpha\, dK^{\it2}
= -i \left(\cos\alpha \, dX^{\it 1} + \sin\alpha\, dX^{\it
2}\right)\,,
\nonumber\\
dZ^2 & = & -\sin\alpha\, dX^{\it 1} + \cos\alpha\, dX^{\it 2}\,.
\end{eqnarray}
The angle $\alpha (\sigma)$ parameterizes in these differentials 
the local (co-ordinate) gauge freedom for the momentum, of rotations 
about the  line tangent ${\bm X}'$. %, contained  in (\ref{p1p2}). 
It expresses the 
fact that one degree of freedom is still available, namely that for
the direction of the 
local gauge dependent momentum ${\bm P}^{\rm inc}(\sigma)$, even after a 
local basis on the string has been chosen. 
For the components of $\bm P^{\rm inc} $ in the two  
$\bm Q$-directions ${\bm e}_{\it 1}$ and ${\bm e}_{\it 2}$ the relation
\begin{equation}\label{p1p2}
\partial_{\it 2} P^{\rm inc}_{\it 1} -\partial_{\it 1} P^{\rm inc}_{\it 2}
% = -\sqrt g\, ({\rm rot} {\bm P})\cdot {\bm e}_\sigma
%= \frac{N_v}{N_s} 
= N_v h \rho_0\sqrt g\,
\end{equation}  
obtains \cite{geo}, cf. relation (\ref{rotaBX'}). 
The simplest example is 
a rectilinear line in $z$-direction, for which the local momentum can rotate
in the $x$-$y$ plane. 
The gauge invariant quantity is the integral
$\oint\!\!\oint P^{\rm inc}_{\it a} {d}X^{\it a} d\sigma$, which is 
left unchanged by the rotation freedom. 
The relation (\ref{p1p2}) implies that 
in the hydrodynamic limit of $|{\bm P}^{\rm kin}|/|{\bm P}^{\rm inc}|
\rightarrow 0$, phase space and configuration space 
become indistinguishable, since the momentum 
components then become functions of the co-ordinates alone, and are 
no longer independent variables \cite{onsagerstathydro}.  

It is illuminating to go back to our `electrodynamic' quantities 
and rewrite 
\begin{equation}
S_e^{\rm inc}(E)= iq\, \int\!\!\!\int\!\!\!\int\! B_\sigma \,\sqrt g\, 
dX^{\it 1} dX^{\it 2} \, d\sigma \,,
\end{equation} 
where $B_\sigma = - \rho_0$  is the (nonrelativistic)
`magnetic' field,  
pointing antiparallel to the direction of the line tangent. 
%, and $q=N_v h $ the `charge'. 
The part $S_e^{\rm inc}(E)$
is thus the Aharonov-Bohm type Berry phase \cite{berry} 
in the Euclidean wave function 
of the adiabatically moving quantum object vortex. 
 
The part of the action $S_e^{\rm kin}$ explicitly involves the vortex
dynamics, %{\it i.e.} velocity of vortex,  
and thus can not be calculated by knowledge of the vortex 
path {\em geometry} alone, as this was possible for the part $S_e^{\rm inc}$. 
Treating the influence of the mass as a small perturbation on the
vortex path, the ratio of the actions is of the same order. 
More exactly, 
\begin{eqnarray}
S_e^{\rm kin}(E) & 
= & \oint\!\!\oint P^{\rm kin}_{\it a} {d}X^{\it a} {d}\sigma
=\oint\!\!\oint M_0\sqrt \gamma\, {\dot Q}_{\it a} {d}X^{\it a} {d}\sigma
\nonumber\\
& = & \frac{N_v h\rho_0}{4\pi}\,\frac{\Gamma_s}{c_s}
\oint\!\!\oint\ln (\cdots)\,({\dot Q}_{\it a}/{c_s})
%\frac{{\dot Q}_{\it a}}{c_s}
\, \sqrt\gamma\, dX^{\it a}d\sigma\nonumber\\
& \simeq & h\rho_0\frac{\ln(\cdots)}{2\pi}\,\xi \oint\!\!\oint 
({\dot Q}_{\it a}/{c_s})\sqrt\gamma\, dX^{\it a}d\sigma\;
\qquad (\mbox{$\textstyle N_v=1$, He II}).
\label{SEkin}
\end{eqnarray}
The last line is valid for a unit circulation vortex in helium II, in
which the approximate equality 
\begin{equation}\label{kappacs}
\kappa\simeq  c_s 2\xi \qquad (\mbox{in helium II})
\end{equation}
holds.\footnote{This relation 
assumes that $\sigma_{\rm LJ}< \xi < a$, with 
$\xi$ nearer to the lower bound ($a$ is the interparticle spacing, 
$\sigma_{\rm LJ}=2.556 \,$\AA\, the Lennard-Jones parameter of the atomic
helium interaction), 
which is consistent with quantum many-body and density-functional
calculations \cite{ortiz,vitiello,sadd,dalfovo2}.}
The logarithm is assumed to vary only slightly during the
tunnelling process, so that it is written in front of the integral. 
In the length scale domain of
interest, $\ln (\cdots)/2\pi = {O}(1)$. The dots, $\cdots$, 
indicate an average over the argument of the vortex energy logarithm.

Essentially, (\ref{SEkin}) tells us that the kinetic part of the
Euclidean action in units of $\hbar$ depends on an area (element)
multiplied by the (local) velocity of the vortex in units of the speed of
sound. The coherence length $\xi$ times this area gives 
a volume, which multiplied by the bulk number density finally yields a
dimensionless action.
The full Euclidean action (\ref{SEtotal}) thus takes the schematic form 
(for simplicity again displayed in the unit circulation case of helium II) 
\begin{equation}\label{SEgeo}
\frac{S_e(E)}\hbar % = S^{\rm inc}_e(E)+S^{\rm kin}_e(E)
=\rho_0\left( 2\pi \Omega^{(d)}+ \ln (\cdots)\, \xi\,
\Sigma^{(d)}\left[\partial {\bm X}/(c_s \partial t) \right]
\right)\,. 
\end{equation}
The volume $\Omega^{(d)}=-i\int\!\!\int\!\!\int\! \sqrt g \, 
{d}X^{\it 1}{d}X^{\it 2}{d}\sigma$. The effective surface 
\begin{equation}
\Sigma^{(d)}=\oint\!\!\oint 
({\dot Q}_{\it a}/{c_s})\, \sqrt\gamma\, dX^{\it a}d\sigma
\end{equation}
obtained by integrating over the surface enclosing  $\Omega^{(d)}$
is a functional of the vortex velocity scaled by the speed of sound
($d$ indicates the spatial dimension). 

To summarize, in a dense, strongly coupled superfluid, 
the first term dominates the second term in (\ref{SEgeo}) 
for the following reasons: 
\begin{itemize}
\item[a.] The large scale, collective co-ordinate 
limit requires that the scales to be
considered are much larger than $\xi$. The corresponding volumina
and areas have to be very much larger than $\xi^3$ and $\xi^2$,
respectively, and hence $\Omega^{(d)}\gg\xi\, \Sigma^{(d)}$ .    
\item[b.] The area contribution of $S_e^{\rm kin}$ is additionally
suppressed by the vortex velocity divided by the speed of sound, {\it
i.e.} $\Sigma^{(d)}\ll \partial\Omega^{(d)}$, where $\partial\Omega^{(d)}$
is the proper surface area enclosing $\Omega^{(d)}$.
\end{itemize}
This dominance of $S_e^{\rm inc}$ over $S_e^{\rm kin}$ is in 
contrast to the case of relativistic (string) objects moving with speeds 
of order $c_s$ %on scales of order $\xi$ 
\cite{davis3,kaolee}. 
Under this circumstance,
$S_e^{\rm kin}$ %, where it is understood that for strings 
%this now takes the Nambu form (\ref{Nambu}),  
is of the same order as $S_e^{\rm inc}$ and the action
is of order $S_e/\hbar \approx \rho_0\ln(\cdots)\,\xi\, \partial\Omega^{(d)}$. 
%, where $\partial\Omega^{(d)}$ is the proper surface area enclosing
%$\Omega^{(d)}$. 
If one  neglects the dependence of the
logarithm on the co-ordinates, %as we have done, 
this is essentially the Nambu action (\cite{nambustrings}), 
in units of $\hbar$, 
up to a factor of order unity.
% (in the ultra-relativistic case, 
% this factor can be even less than unity \cite{teitelboim}). 
 
We assumed that the vortex path in phase space is closed. As a
consequence, the number of particles in the effective volume on the 
right-hand side of (\ref{SEgeo}) is quantized according to 
\begin{equation}\label{N}
S_e(E) = (N^{(d)}+\alpha)h\,\quad
\Leftrightarrow \quad S_e(E)/\hbar = 2\pi (N^{(d)}+\alpha).
\end{equation}
The number of particles in the effective volume (including
the small kinetic contribution on the right-hand side of (\ref{SEgeo})),
plays the part of the Bohr-Sommerfeld quantum number in semiclassical 
quantization. 
The number $\alpha$ is of the order one and signifies 
the onset of the microscopic quantum regime.
In the semiclassical approximation, $N^{(d)}\gg \alpha $ must hold, so
that $N^{(d)}\gg 1$ gives, as usual, a direct measure of semiclassicality.

\subsection{Geometry of Quantum Tunnelling}
%\subsection{Solutions of the geometric vortex tunnel\-ling problem}
\subsection{Galilean invariance violation}
At the absolute zero of temperature, a homogeneous nonrelativistic 
superfluid has Galilean invariance, that is, physical contents
are invariant under co-ordinate 
transformations to any reference frame moving at constant
velocity. If we approach absolute zero, which is  
what is actually realized in experiment, we expect the tunnelling rate  
to make no abrupt change as the temperature is lowered. 
Thus the result for the rate we obtain at $T=0$ 
should also be valid for temperatures slightly above zero 
(we will estimate the temperatures, for which this is no longer the
case, later on). 

Because we can always transform to the
rest frame comoving with the superfluid, 
the tunnelling probability at $T=0$ equals zero if Galilean invariance 
remains unbroken: 
In the rest frame there is a 
tunnelling barrier of infinite height, the logarithmically diverging
vortex self energy. Hence it is necessary to explicitly include the
violation of Galilean invariance by a flow obstacle into 
any calculation of tunnelling rates for Galilei invariant superfluids
at absolute zero. 
%This is, in addition, also particularly appropriate because 
%the very nature of the problem is a geometric one in the semiclassical
%limit.     
The necessity of %Galilei or Lorentz 
invariance violation for tunnelling to be energetically allowed
thus stems in the real superfluid from the fact %requirement 
that it is possible to invariantly transform to the rest frame of the
superfluid, for any allowed velocity of flow.
%without changing any physical procedure.

%\subsection{Solutions of the geometric vortex tunnel\-ling problem}
Having thus shown that it is strictly required that Galilean invariance
be violated, we are in demand of constructing an explicit solution 
of the following problem. Given a vortex in the presence of an
invariance breaking flow obstacle, we have to calculate the Hamilonian 
energy $H_V$ of the vortex in the Hamiltonian (\ref{HV}) 
as a function of the co-ordinates, by solving one of  
the equations of motion corresponding to this Hamiltonian 
respectively the Euclidean action (\ref{SEQ}). 
%(\ref{dPdtHamilton}), (\ref{Lagrangeeqmotion}) or (\ref{originaleqmotion}).  
We then set $H_V$ equal to a constant $E$, to calculate the vortex
trajectory of constant energy, which finally yields the action
(\ref{SEKP}) respectively (\ref{SEQ}), in the form of (\ref{SEgeo}).
It is obvious that without a very high degree of symmetry, this is 
a task necessitating a quite formidable computational effort.
 
We have seen that the problem of determining the dominant contribution
$S_e^{\rm inc}$ is a geometrical problem, because the phase space co-ordinates 
are functionally dependent on the configuration space co-ordinates.  
%(cf. subsection \ref{Gaugedepend}).
We will therefore restrict the discussion in what follows to cases 
which elucidate in particular the geometric nature of quantum 
tunnelling, that is, concentrate on the behaviour of $S_e^{\rm inc}$,
which is a functional  of geometrical quantities. 

\subsection{The vortex half-ring case}\label{introhalfring}
To flesh out the discussion which has been so far quite abstract, we 
now give some quantities relevant in the tunnelling problem for 
a vortex. Imagine, for concreteness, a singly quantized 
vortex half-ring of radius $R$
with its circulation axis ${\bm e}_\phi$  
standing perpendicular on the plane $y=0$. Over the plane is flowing 
liquid with a velocity $u$ at infinity from right to left (in the
negative $x$-direction). Consider, first, the case that
the invariance-breaking asperities on the surface are small-scale 
(it will become clear in a moment what `small-scale' means). 
Neglecting the elastic and kinetic %and potential
terms in the Hamiltonian, 
as well as perturbations of the flow generated by the asperities, 
the conserved vortex energy is simply 
\begin{equation}
H_V =  h\rho_0\int_0^\pi d\phi \left[
\frac{\kappa}{4\pi}\, R \, \ln \left(\frac{8R}{\xi e^C}\right)
- \frac12\, u R^2 \right]\,.
\end{equation}
If we normalize the energy by 
\begin{equation}\label{normE}
\tilde H_V  \equiv \frac{4\pi m}{h^2\rho_0}\, H_V\,,
\end{equation} 
and solve for the path of constant total energy $\tilde E_0$, we get 
the relation
\begin{equation}
R = \frac{\kappa}{2\pi u} \, \ln \left[ \frac{8R/\xi}{\exp( C+\tilde
E_0 /(\pi R))}\right]\,.
\end{equation}
This equation has two solutions for the radius $R$. One of them is, 
in the case of $u\ll \kappa/2\pi \xi$ and small $\tilde E_0/\pi R \ll C$,
located far away from the surface $x=0$, at 
\begin{equation}\label{defR0}
R_0\simeq \frac\kappa{2\pi u}\,\ln \left( \frac{8\kappa /(2 \pi u \xi)}
{\exp(C)}\right)\,.
\end{equation}
The other solution is a half-ring of radius
in the order of the coherence length, $R_\xi= {O}(\xi )$ (which
will not be of further interest here; it signifies the path of the (virtual)
vortex trapped at the flat boundary).

The closed path in phase space, needed to evaluate (\ref{SEKP}),  
can be obtained as follows. Assume that
the small asperities are approximately of the shape of oblate 
rotation ellipsoids 
with half-axes $b$ perpendicular and $a$ parallel to the flow and define 
the {\em sharpness} $\beta = b/a > 1$. 
Then, the half-ring trajectory over such a small ellipsoid, collinear
with the ring axis, were given by 
\begin{equation}\label{RbK}
R^2 = b^2 + \beta^2 K^2\,,  
\end{equation}
using the  complex co-ordinate $K=-iZ$, and provided that the 
vortex exactly follows the surface (up to a constant $O(\xi$)). 
%That this trajectory 
%is correct only in lowest order of $\beta -1$ will be shown 
%below. In particular, for $\beta =1$, the sphere, the result is exact.   
This trajectory hits the one far away from the surface, 
given by $R_0=$const., at $K_0 %= \sqrt{R_0^2-a^2}
\simeq  R_0/\beta $ (provided that $b \ll R_0$). The closed phase
space trajectory thus begins at the ellipsoid top ($K=0,R \simeq b$), 
then propagates along the line given by (\ref{RbK}), 
meets the constant $R_0$ at 
$K_0$, then follows this constant to $-K_0$, and finally follows the
branch of (\ref{RbK}) for negative $K$ back to the ellipsoid top. 
 
The tunnelling exponent, using the gauge $P_X= (1/2) h\rho_0 R^2$
for the momentum, %and neglecting the kinetic term, 
is thereby
\begin{eqnarray}
S_e^{\rm inc}(\tilde E_0)& = & 2h\rho_0 \int_0^\pi\! d\phi \int_0^{K_0}\!\! dK
 \,\frac12\left(R_0^2-\beta^2 K^2\right)\nonumber\\
& = & h\rho_0\frac{2\pi}3 R_0^3 /\beta
= h\rho_0\Omega^{(3)}
\,.   \label{SEusmall}
\end{eqnarray}
The tunnelling volume and thus the exponent may consequently be reduced
and the tunnelling probability enhanced if a surface with sufficiently 
sharp peaks perpendicular to the flow is present. In particular, if we
were able to reduce the small half-axis
to $a\gtrsim \xi$, and still have $b\gg a $, we could reduce the
tunnelling volume to $\Omega^{(3)}\approx (2\pi /3)R_0^2\,\xi (R_0/b)$. 
It is worthwhile to point out that this reduction is not due to an
enhancement of the flow velocity at the ellipsoid top: For any value
of $\beta$, because of cylindrical symmetry, the velocity at the top 
is exactly $2u$, just as for the half-sphere.  

%However, 
%we will show in the analytical calculation to follow for the 
%two-dimensional case that the sharpness $\beta$ can not be reduced to the
%lower limit just described, but is limited by a lower bound connected
%with the validity of the semiclassical approximation. 
The trajectory (\ref{RbK}) is correct only in lowest order of $\beta -1$. 
In particular, for $\beta =1$ ($a=b$), the half-sphere, the result 
for the tunnelling exponent (\ref{SEusmall}) is exact in the low
velocity limit.  
This can be shown by solving the  
problem for the sphere exactly, which has been done %by Volovik 
in \cite{volovik}. The result for the stream function part of the potential
in the Hamiltonian (\ref{HV}), $a^0=- \rho_0 \psi$,  
corresponding to Stokes' stream function $\psi$ \cite{milne-t}, 
has been checked by the author, and is in Coulomb gauge 
\begin{equation}
\frac12\psi_C +\psi_u = R\, \frac{\kappa}{4\pi} Q_{1/2}(w)
- \frac12 u R^2 \left(1- \left(\frac
a{\sqrt{R^2+Z^2}}\right)^{3}\right) \,,\label{3dspherical}
\end{equation}
with $w\equiv 1+ (1/2)[(Z^2+R^2-a^2)/(aR)]^2$. 
The function $Q_{1/2}$ is a Legendre function of the second kind 
(\cite{GR}, No. 8.821), a solution of the Poisson  
equation $\Delta \psi = \kappa \delta^{(2)}(\bm x -\bm X)$
for the combined spherical-cylindrical symmetry of the problem at
hand.

The problem of a undeformed, massless 
half-ring situated at a half-sphere, with 
their axes coinciding, is actually the only nontrivial problem in
three spatial dimensions solvable with reasonable effort
analytically. It has just the maximal symmetry in three dimensions. 
It does not, however, contain the
crucial ingredient for a full geometrical analysis of quantum tunnelling:
The curvature of the flow obstacle is constant.  
% The solution of a problem of varying 
The result that, for the sphere, the vortex half-ring follows exactly the
surface (up to a {\em constant} distance of order $\xi$), has led
in \cite{volovik} to the assumption that it is admissible to
continue this result to the ellipsoid case, in the manner of
(\ref{RbK}). We will show in the subsection to follow that this is
incorrect: The analytic continuation of the sphere result to the
ellipsoid case of arbitrary $\beta$ is incompatible with the geometric
requirements imposed by a %semiclassical 
hydrodynamic, collective co-ordinate treatment. 
\subsection{Analytical solution in two dimensions}
A treatment of our (boundary) problem in two spatial dimensions \cite{geo} 
is advantageous for the following reasons:
\begin{itemize}
\item[{\it i}.] The additional spacelike co-ordinate, namely the arc length
parameter, complicates the analysis enormously because of the
locality of any variable in $\sigma$ and the existence of elastic energy. 
\item[{\it ii}.] Even if we treat an undeformed ring or half-ring, 
{\it i.e.} neglect elastic energy and locality altogether,  
any  3d problem which has not the maximal symmetry described above, requires 
a multitude of image vortices and gets quite intractable.
\item[{\it iii}.] In two dimensions, we have available the tools of 
conformal transformation, allowing for comparatively simple
calculations of %a number of interesting 
analytical solutions.   
\item[{\it iv}.] The geometrical features, which are dominant in the
hydrodynamic, collective co-ordinate limit, 
as expounded at length above, are most clearly
seen in two dimensions. 
%The extension to three dimensions is then
%merely a question of the (tedious) calculation of order
%unity prefactors in the action.  
\end{itemize} 
%The treatment to follow is an extended account of work by the
%author in Refs. \cite{geo,lammi}.
\subsubsection{The solution for the circle}
\begin{figure}[hbt]
%\psfrag{a}{\small $\alpha(\sigma)$}
\psfrag{sigma}{$\sigma$}
\psfrag{+k}{+$\kappa$}
\psfrag{-k}{-$\kappa$}
\psfrag{Z1}{$Z_1$}
\psfrag{Z1bar}{$\bar Z_1$}
%\psfrag{ab2}{$1/|Z_1|$}
\psfrag{ab2}{$d^2/|Z_1|$}
\psfrag{X}{$X$}
\psfrag{Y}{$Y$}
\psfrag{R}{$d$}
\psfrag{Zplane}{$Z$-plane}
\begin{center}
%\parbox{
%\rotate[r]{
\includegraphics[width=0.65\textwidth]{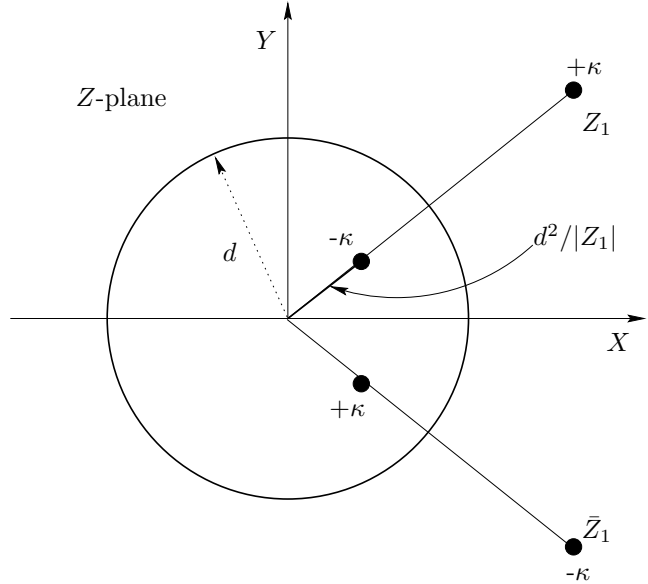}
%}
%}
\end{center}
\caption{\label{circle} The most simple nontrivial boundary problem solvable 
by the image technique: A unit circulation point vortex in the half space
$Y>0$, which is filled with liquid, moving  
near a (half-)circle. 
The boundary conditions are satisfied by 
an image vortex at the plane and two image
vortices of opposite strength inside the circle.}
\end{figure}%\vspace*{2em}    
The basic solution from which we start is that for a vortex 
in the presence of a half-circle of radius $d$ 
at an otherwise flat boundary (cf. Fig. \ref{circle}\,).
%, the radius of the circle is scaled to unity). 
The complex plane of this original solution is called the $Z$-plane
(the uppercase letter does here not imply that a vortex position is meant). 
The imaginary part of the complex potential \cite{milne-t} 
gives the stream function $\psi = \Im [w]$,  
whereas the real part is the usual velocity potential. 
%($=(\hbar/m)\theta$). 
It follows that a single vortex at $Z_1$ has 
complex potential $ w(Z)=-i(\kappa/2 \pi)\ln [Z-\bar Z_1]$.  

The boundary condition to be fulfilled 
is obviously that there be no flow into the 
surface consisting of the line $Y=0$ and the half-circle. This
amounts to the requirement that the line $Y=0$ (for $|Y|>d$), $Y=\sqrt
{d^2-X^2}$ (for $|Y|\le d$) is a streamline of constant $\psi\equiv
0$. Such a requirement can be met by using the technique of image
charges quite familiar from electrostatics: Our vortex problem is completely
equivalent to that for a `charge' situated near a perfectly conducting
surface, with no tangential `electric' field. 

The complex potential generated by the image vortices and
acting on the vortex at $Z_1$ is then given by 
\begin{equation}
w_i(Z_1)=-i\,\frac\kappa{2\pi}\ln \left[ 
\frac{\left(Z_1-\bar Z_1\right)
%\left(1/Z_1-\bar Z_1\right)}{{1}/Z_1-Z_1}
\left(d^2/Z_1-\bar Z_1\right)}{{d^2}/Z_1-Z_1}
\right]\,.
\end{equation}
The first factor in the numerator stems from the
image vortex at the plane $Y=0$ with complex potential $w(Z)=-i(\kappa/2 \pi)
\ln [Z-\bar Z_1]$ (which has to be present even without the circle), 
the second one is obtained by the circle theorem 
\cite{milne-t} as the image of the original vortex at the circle.
Finally, the potential of the remaining $+\kappa$-circulation vortex
inside the circle, contributing in the denominator of the logarithm, 
completes the image vortex system, again by the 
circle theorem. 
The first term in the denominator of the logarithm is incorporated into the 
static self energy of the vortex,
$E_{\rm self} = (m\rho_0\kappa^2/4\pi)\ln \left(|Z_1-\bar Z_1|/\xi\right)$\,,
which is cut off by $\xi$ and equal to half the energy of a vortex
pair separated by $|Z_1-\bar Z_1|$. The expression for the   
potential of  (\ref{HV}) is thus 
\begin{equation}
\psi_C 
%-(\kappa /2\pi) \Im [i\ln [((d^2/Z_1-\bar Z_1)/({d^2}/Z_1-Z_1)]]
=-(\kappa /2\pi) \ln\left( \left|(d^2/Z_1-\bar Z_1)/({d^2}/Z_1-Z_1)
\right|\right)\,,
\end{equation} 
% The imaginary part of this solution 
which is just the counterpart of $\psi_C$ in 
(\ref{3dspherical}) in two dimensions. The complex 
counterpart of $\psi_u$, in turn, is
\begin{equation}
w_u(Z)=u \left(Z+%\frac{1}Z\right)\,.
\frac{d^2}Z\right)\,,
\end{equation} 
giving the complex potential of flow from right to left (in the
negative $X$-direction). 
  
The full energy of the point vortex, neglecting compressibility
effects, and using the normalisation (\ref{normE}), 
is thence given by the expression
\begin{eqnarray}\label{tildeEZ}
\tilde E(Z_1)= \ln\left| \frac{(Z_1-\bar Z_1)
\left(d^2/Z_1-\bar Z_1\right)}{\xi\left({d^2}/Z_1-Z_1\right)}\right|
%\nonumber\\
-\frac{4\pi u}\kappa \Im\left(Z_1 + \frac{d^2}{Z_1}\right)\,.
\end{eqnarray} 
\subsubsection{Conformal transformation}
A conformal transformation is a co-ordinate transformation leaving angles 
invariant, that is, the metric is multiplied by a conformal factor which 
is % the square root of 
the transformation's Jacobian determinant.  
What we want to do is to map by a conformal transformation 
the region outside a boundary 
surface with varying curvature radius, lying in the (target) $z$-plane, 
to the domain outside the circle, which is in the (original) $Z$-plane. 
Any such transformation can be written as a holomorphic 
function of $Z$ (save for singular points, such as vortex centers), 
%, cf. \cite{milne-t}), % and \cite{CKP}), 
in the form
\begin{equation}\label{conformaltrans} 
z= a_0 Z +\sum_{n=0}^\infty b_n Z^{-n}\,, 
\end{equation} 
where $a_0, b_n$ are some coefficients and $Z=d\exp(i\chi)$ is on the
circle (we omitted a constant, indicating 
a change of $z$-plane origin). 
\subsubsection{The ellipse solution}
We would like to invert relation (\ref{conformaltrans}) 
to obtain the solution for the boundary surface directly from that for the
circle, which we have already obtained above. Easiest to perform
is this inversion if we let $b_0=0$, $b_n=0$ for $n>1$. 
Furthermore, if $a_0\equiv 1$ by
proper normalizing choice of scale, 
we are led to the celebrated Joukowski transformation
\begin{eqnarray}    \label{Joukowski} 
z=Z-l^2/4Z\,,
\end{eqnarray}
which maps the outside of an ellipse with half-axes $a,b$ (where $a<b$) 
to the outside of a circle of radius $d=(a+b)/2$. The parameter $l$ is 
defined by $l^2=b^2-a^2$. The inversion of this transformation is cast
into the form
\begin{equation}
Z=\frac12\left( z + \sqrt{z^2+l^2}\right)\equiv  \frac l2
\left(\sinh\zeta + \cosh\zeta\right)=\frac l2 \exp[\zeta]\,.
\end{equation}
with the aid of elliptic co-ordinates, defined by ($\zeta=\chi+i\eta$)  
\begin{equation}
z=l\sinh\zeta=l\left( \sinh\chi \cos \eta +i \cosh\chi\sin \eta\right)
=x+iy\,.
%\qquad (\zeta=\chi+i\eta)  
\end{equation}
The lines of constant $\eta$ and $\chi$ are confocal ellipses and
hyperbolae, as follows from 
\begin{equation}
\frac{y^2}{l^2\sin^2\eta}-\frac{x^2}{l^2\cos^2\eta}=1\,,
\qquad\frac{x^2}{l^2\sinh^2\chi}+\frac{y^2}{l^2\cos^2\chi}=1
\end{equation}
The co-ordinate basis ${\bm e}_\chi \equiv \partial/\partial\chi$, 
 ${\bm e}_\eta \equiv \partial/\partial\eta$ 
is an ortho-basis with the conformal metric 
$g_{ij}=l^2(\cosh^2\chi\cos^2\eta +
 \sinh^2\chi\sin^2\eta)\delta_{ij}$.

%In what follows, we consider the solution (\ref{tildeEZ}) in the
%half-plane $y>0$. 
The normalized energy (\ref{tildeEZ})
as a function of the elliptic vortex 
co-ordinates $\chi_1,\eta_1$ takes the form
%, from (\ref{tildeEZ}),  
\begin{eqnarray}%\vspace*{3em}
\label{tildeE}
\tilde{E} (\chi_1,\eta_1) 
 = %\equiv \frac{4\pi \, E}{m\rho_0 {\kappa}^2}=
\ln\left[\frac{a+b}{\xi}\,\frac{\exp(\chi_1-\chi_0)|\sin\eta_1| \sinh (\chi_1-\chi_0)}{(\sinh^2(\chi_1-\chi_0)+\sin^2\eta_1)^{1/2}}\right]
\nonumber\\
%\hspace*{-4em}
-\frac{4\pi u(a+b)}\kappa\sinh
(\chi_1-\chi_0)|\sin\eta_1|\,,%\nonumber\\
\end{eqnarray}
where $\chi_0 ={\rm artanh} (a/b)$ is the co-ordinate specifying the 
(half-)ellipse surface at the boundary, {\it i.e.} the half-axes
of the ellipse are given by $a=l\sinh \chi_0$, $b=l\cosh\chi_0$.

%\subsubsection{The low velocity limit}
For direct comparison with the half-ring case treated in 
subsection \ref{introhalfring}, which had a quite simple geometrical meaning, 
we will investigate the paths of constant energy $\tilde E (\chi_1,\eta_1) 
\equiv \tilde E_0$ mainly 
in the case of small velocity $u$. The notion of
`small' requires some more care in two dimensions as compared to the 
cylindrically symmetric half-ellipsoid case in three dimensions. Whereas in the
latter the velocity enhancement at the top is always $2u$, in the 2d
case it is $2(b/a)u$. Hence it is required that we restrict the
velocity at infinity to 
\begin{equation}\label{uabvl}
u \ll \frac  a {2b}\, v_L\,,
\end{equation}
such that the velocity at the top is well below the critical velocity 
\begin{equation}\label{defvL}
v_L\equiv \frac\kappa{2\pi\xi}\,, 
\end{equation}
which gives a measure of the onset of many-body quantum physics in the
atomic superfluid helium II. For Fermi superfluids, the corresponding 
critical velocity is the pair-breaking velocity of Cooper pairs.     
We will henceforth refer to $v_L$ as `Landau
velocity' and generally scale velocities with it. Numerically, 
the actual Landau critical velocity of roton creation $\simeq$ 59 m/s 
at $p\simeq$ 1 bar equals $v_L$\, if \,$\xi\simeq 2.7 $ \AA\, is
taken.\footnote{It should, however, be pointed out that the very notion of
(classical) velocity becomes questionable on  
$\xi$-scales. On these scales, it is more appropriate to refer
to (mean) current densities.}

%The velocity around the ellipse is highest at the top $(\eta=\pi/2)$.
The approach of the vortex to the ellipse surface 
is closest at the top $(\eta=\pi/2)$. 
Let us calculate the distance at the top as a function of the constant
energy $\tilde E_0$. In general, a small distance interval 
$\delta s $ is given by 
$\delta s = \sqrt{g_{\chi\chi}(\chi_0)} \,\delta\chi=
(a^2\sin^2\eta + b^2 \cos^2\eta)^{1/2}\delta \chi$.
%, provided that $\delta\chi\equiv \chi-\chi_0 \ll 1$. 
We define a quantity $s$ by 
\begin{equation}\label{sdef}
s =a \delta\chi_c \equiv  a\delta\chi(\eta = \pi/2)\,.
\end{equation}
At the ellipse top, 
$s = \delta s (\eta=\pi/2)$, provided that 
$\delta\chi_c\equiv \chi_1(\eta=\pi/2)-\chi_0 \ll \chi_0$. 
The actual distance of the vortex to the top is somewhat different
if this inequality does not hold.  
We do not further dwell on this difference here for the sake of
simple argument and take (\ref{sdef}) as a definition of the 
quantity of distance $s$.
%, requiring only that always $\delta \chi \ll 1$.   
 
Evaluating (\ref{tildeE})  
at $\eta=\pi/2$ for small velocities, 
and assuming that $\tilde E_0$ is small enough for 
$\delta\chi_c\ll 1 $ to hold, we get $\delta\chi_c \simeq
\xi\exp[\tilde E_0]/(a+b)$ and therefore
\begin{equation}\label{deltasab}
s \simeq \frac a{a+b} \,\exp[\tilde E_0]\,\xi\simeq 
\frac ab \,\exp[\tilde E_0]\,\xi\qquad (\eta=\pi/2,\,\mbox{$u$ small})\,.
\end{equation}
On the other hand, far away from the half-ellipse, at the flat
boundary, the distance is, in the low velocity limit,
given by $\delta s \simeq (1/2) \xi \exp[\tilde E_0]$. We observe
that the vortex comes closer to the ellipse by a ratio $2a/b$ (for
$b\gg a$) than it was far away from the ellipse. If $a=b$, that is, in
the case of the circle, the distance remains the same. 
\subsubsection{Geometric restrictions}  
Now, if we are bound to remain in the realm of the 
hydrodynamic description of a collective co-ordinate vortex, 
we have to impose that the total distance $\Delta s$ of the vortex 
to the ellipse always exceeds a quantity  ${O}(\xi)$:
\begin{equation}\label{scond}
\Delta s = \int^{\chi_1}_{\chi_0}\delta s\ge {O}(\xi)\,.
\end{equation}
The quantity $O(\xi)$ means `a value in the order of $\xi$ by definition', 
as there can obviously be 
no sharp distinction between the `inside' and `outside' 
of a quantum vortex.     
Additionally, the choice for the lower limit value of the 
total distance $\int\delta s$ in units of $\xi$ depends on the
value of the core constant $C_0=O$(1), parameterizing the many-body 
core structure in the vortex energy logarithm 
$\ln[R/(\xi\exp [C_0])]$. For the 
point vortex considered, which has $R = 2Y$,$\exp C_0 =1$; 
for a ring vortex we chose the parameterization 
$8\exp [C_0]=\exp[C]$, % in (\ref{Energy}), 
cf. \cite{robgrant}. 

It is apparent from (\ref{deltasab}) that if the energy of the vortex
is sufficiently small (in particular, if it is zero) and the ratio 
$b/a$ is sufficiently large,
the condition (\ref{scond}) will be violated and a hydrodynamic 
collective co-ordinate description invalid. 
Thus, there is a minimum vortex energy required 
for the whole  formalism we employ here to retain its validity. 
Expressing this energy
as a function of the parameters involved, we have, for general
velocities  ($\delta\chi_c\equiv \delta\chi(\eta=\pi/2)$):
\begin{eqnarray}
\tilde E_0 & = &
\ln\left[\frac{a+b}{\xi}\,\exp\delta\chi_c\, \tanh \delta\chi_c\right]
-\frac{4\pi u(a+b)}\kappa\sinh
\delta\chi_c\,\label{E0}\\
& \simeq & %=\tilde E_0 (a,b,s,\xi, u) 
\ln\left[\left(1+\frac{b}{a}\right)\frac {s} \xi \right]
-\frac{2u}{v_L}\left(1+\frac ba\right)\frac {s}\xi\,.
%-2\frac u{v_L}\frac{a+b}{a}\frac s \xi\,, \nonumber
\end{eqnarray}
the last line valid 
provided that we can approximate, with reasonable accuracy, {\it e.g},
$\tanh\delta\chi_c\simeq \delta\chi_c$.  
The energy $\tilde E_0$ is the energy needed 
by the vortex to remain completely describable in %semiclassical 
terms of a collective co-ordinate 
on its way along the ellipse. The quantity $s$ is defined in 
(\ref{sdef}), and has a lower bound related to 
the prescription (\ref{scond}).

\begin{figure}[hbt]
%\begin{center}
%\begin{figure}[hbt]
\psfrag{chi}{\normalsize$\chi_1$}
\psfrag{eta}{\normalsize$\eta_1$}
\psfrag{E-E0}{\normalsize$\tilde E-\tilde E_0$}
{\includegraphics[width=\textwidth]{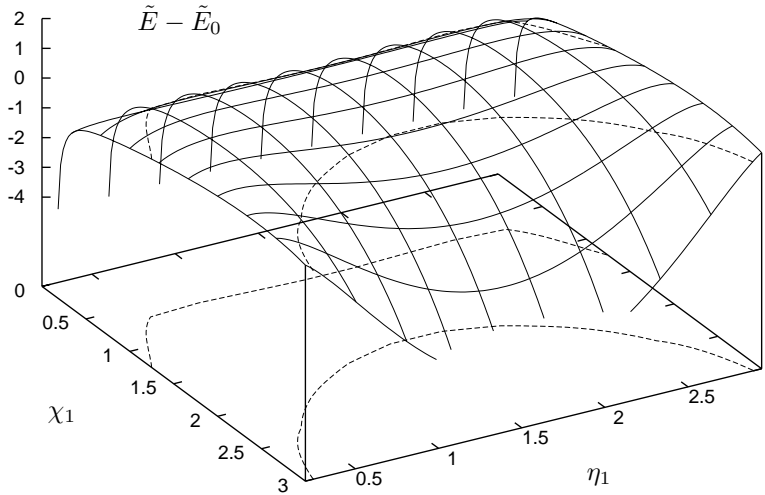}}
%{\includegraphics[width=\textwidth]{Escaled004|025.eps}}
\vspace*{-3em}\\
\caption{\label{Escaled004050} 
Shape of the potential barrier (\ref{tildeE})
with the choice $\chi_0=0.175$ for the velocity $u=0.04$ in units of 
$v_L=\kappa/2\pi\xi$. 
The corresponding real space ellipse with 
$b/a\simeq 5.7$ is shown in Fig. \ref{ellipse}. 
The ratios $a/\xi=2$, $s/\xi=1$ 
and thus $s/a=1/2$.
The zero of this normalised potential energy is shifted by 
$\tilde E_0$, defined in (\ref{E0}). 
The contour lines of the vortex paths of constant energy shown are 
those for $\tilde E=\tilde E_0$. 
%For simplicity, we neglect
%corrections in $E_0$ due to the fact that $s/a = 0.5 \lesssim 1$, 
%{\it i.e.} use the last approximation in (\ref{E0}).
}
%\vspace*{-3em}
\end{figure}

\begin{figure}[hbt]
\psfrag{chi}{\normalsize$\chi_1$}
\psfrag{eta}{\normalsize$\eta_1$}
\psfrag{E-E0}{\normalsize$\tilde E-\tilde E_0$}
{\includegraphics[width=\textwidth]{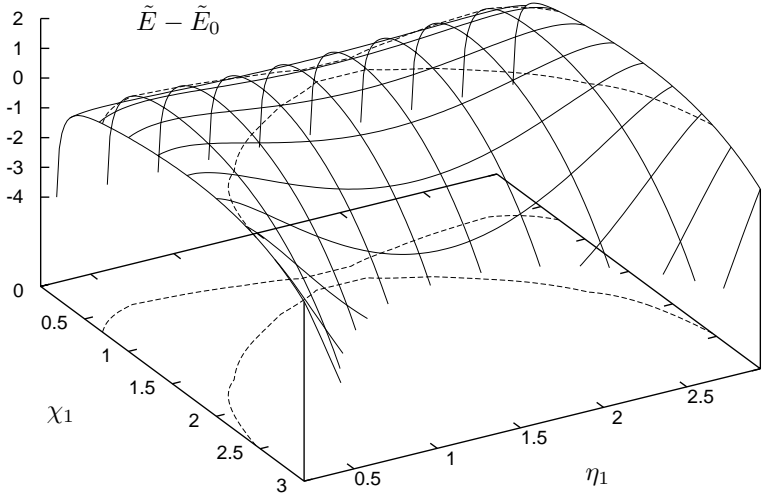}}\vspace*{-3em}\\
\caption{\label{Escaled008050} Same potential as in Figure
\ref{Escaled004050}, but for velocity $u=0.08$.}
\end{figure}

\begin{center}
\begin{figure}[hbt]
%\begin{center}
%\begin{figure}[hbt]
\psfrag{chi}{$\scriptstyle\chi_1$}
%\psfrag{eta}{$\eta_1$}
\psfrag{E-E0}{$\scriptstyle\tilde E-\tilde E_0$}
\hspace*{0.3em}
{\includegraphics[width=0.48\textwidth]{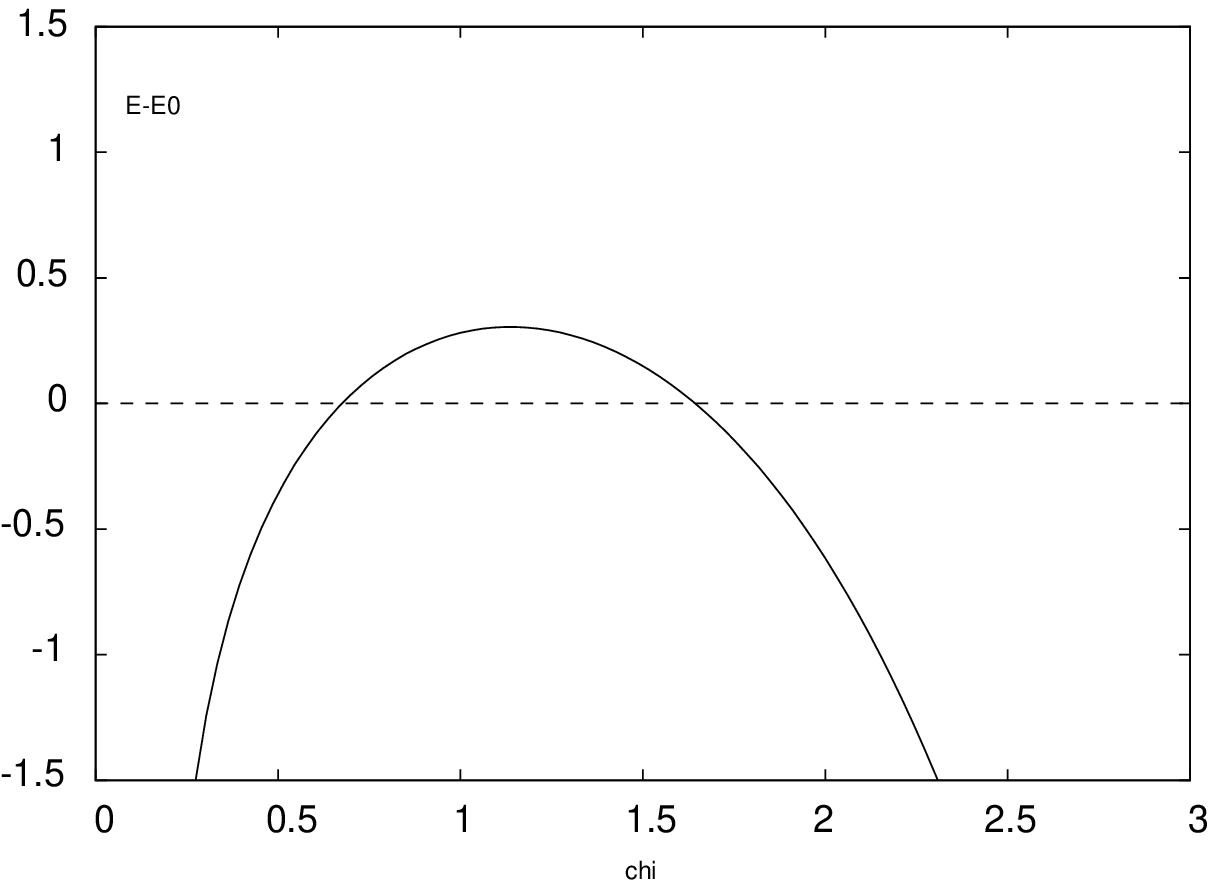}}
{\includegraphics[width=0.48\textwidth]{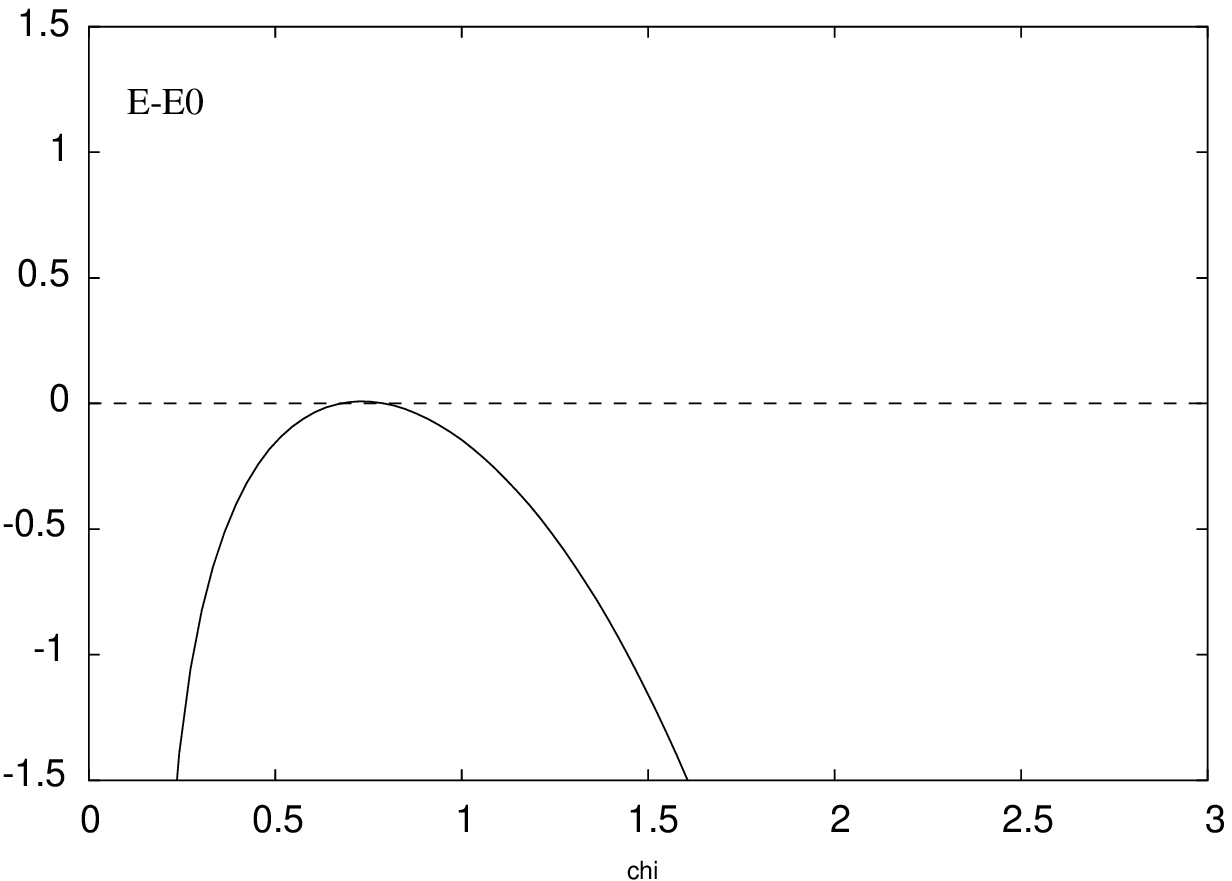}}
\caption{\label{Escaledpi2|050} The potential barriers of 
Figs. \ref{Escaled004050} and \ref{Escaled008050} in the predominant
direction of tunnelling along the ${\bm e}_\chi$-direction at
$\eta=\pi/2$ (the $y$-axis). Left: $u=0.04$, right: $u=0.08$.}
\end{figure}    
\end{center}

\begin{center}
\begin{figure}[hbt]
\begin{center}
{\includegraphics[width=0.65\textwidth]{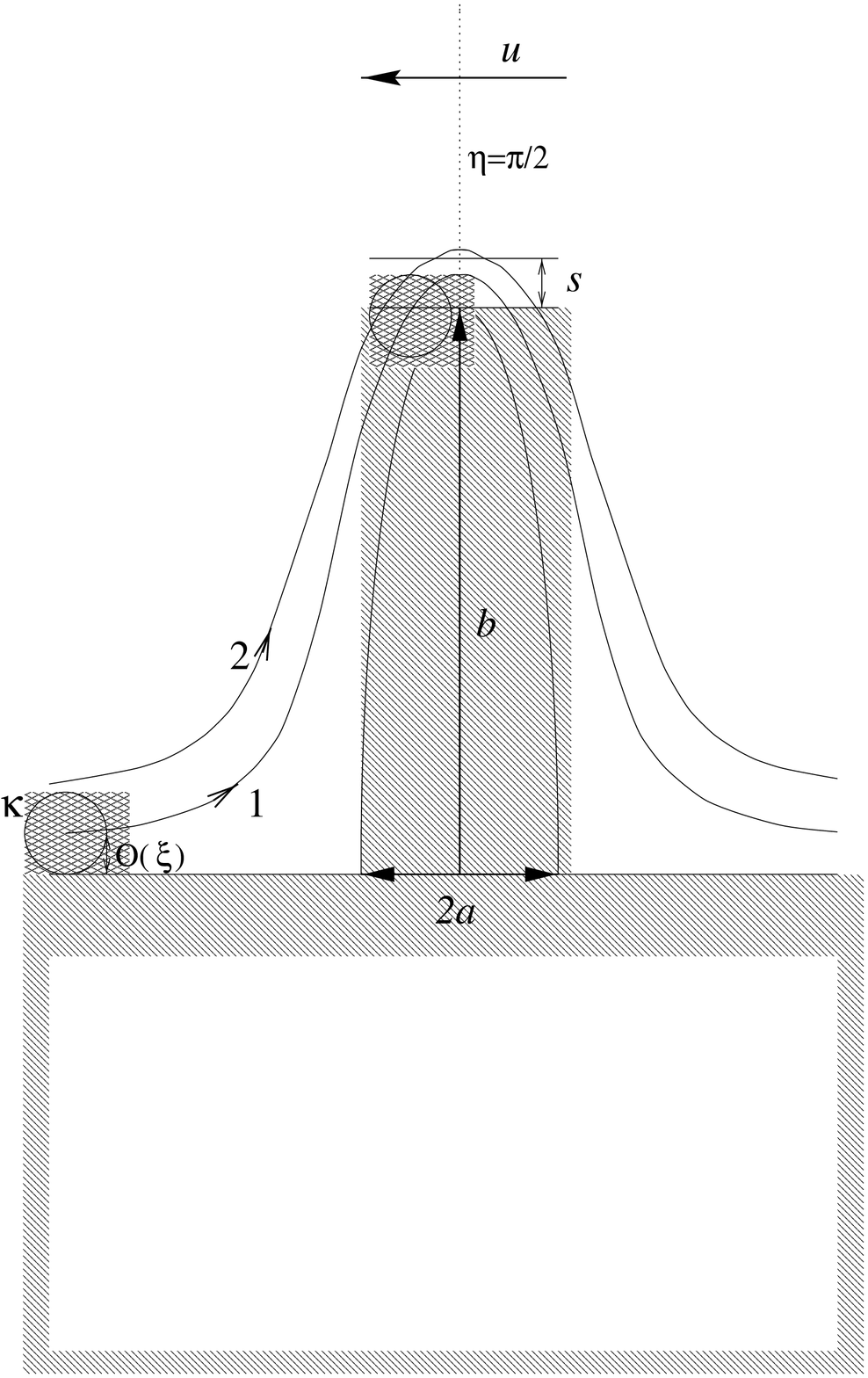}}
\end{center}
% -120 824.0 translate
\vspace*{-11em}
\caption{\label{ellipse} Two vortex paths of constant energy 
near the ellipse. Whereas
the vortex on path 1 with approximately zero energy, $\tilde E \simeq 0$, 
does not manage to pass by
without coming closer than $O(\xi)$, the second one, having  energy 
$\tilde E= \tilde E_0$, defined in (\ref{E0}), is able to do so.  
The velocity $u$ is to be understood that of the flow `at infinity'.}
\end{figure} %\vspace*{2em}
\end{center}

We have depicted the normalised potentials corresponding 
to different velocities in units of $v_L$ and ratios $s/a$
in Figs. \ref{Escaled004050} and \ref{Escaled008050}.
The potentials in these figures correspond to the real space ellipse shown in 
\ref{ellipse} which has $b/a\simeq 5.7$ and where $s/a \approx 2$. 
For clarity, we have additionally displayed the  shape of the barriers
in the direction of the $y$-axis ({\it i.e.}\,the ${\bm e}_\chi$-axis at
$\eta=\pi/2$) in Fig. \ref{Escaledpi2|050}. % and \ref{Escaledpi2|025}. 
%The area between the potential
%curves and the constant $\tilde E-\tilde E_0=0$
%gives a measure of the applicability of the semiclassical
%approximation. 
Whereas for $u=0.04$ the semiclassical approximation 
is still applicable, for $u=0.08$ this is no longer the case. 
It is approximately for this velocity that the barrier tends to zero
and the tunnelling distance approaches $\xi$.    
One can see here explicitly 
that $v_L$ indeed represents %to (\ref{uabvl}), the 
a sensible measure for a critical velocity, because the local velocity 
at the top is $\simeq v_L$, if the velocity 
at infinity takes the value $u\simeq 0.088$. \vspace*{-1.5em}%\indent 
%\setlength{\floatsep}{0em}
%\setlength{\textfloatsep}{0em}
%\setlength{\intextsep}{0em}
%\renewcommand{\floatpagefraction}{0.99}
%\begin{center}
%\end{center}
%\begin{center}
%\begin{figure}[hbt]
%\psfrag{chi}{\normalsize$\chi_1$}
%\psfrag{eta}{\normalsize$\eta_1$}
%\psfrag{E-E0}{\normalsize$\tilde E-\tilde E_0$}
%{\includegraphics[width=\textwidth]{Escaled004|025.eps}}\vspace*{-3em}\\
%\caption{\label{Escaled004|025} Same potential as in Figure
%\ref{Escaled004050}, but for $s/a=1/4$.}
%\end{figure} 
%\begin{figure}[hbt]
%\psfrag{chi}{\normalsize$\chi_1$}
%\psfrag{eta}{\normalsize$\eta_1$}
%\psfrag{E-E0}{\normalsize$\tilde E-\tilde E_0$}
%{\includegraphics[width=\textwidth]{Escaled008|025.eps}}\vspace*{-3em}\\
%\caption{\label{Escaled008|025} Same potential as in Figure
%\ref{Escaled008050}, but for $s/a=1/4$.}
%\end{figure} 
%\end{center} 
%\begin{center}
%\begin{figure}[hbt]
%\begin{center}
%\begin{figure}[hbt]
%\psfrag{chi}{$\scriptstyle\chi_1$}
%\psfrag{eta}{$\eta_1$}
%\psfrag{E-E0}{$\scriptstyle\tilde E-\tilde E_0$}
%{\includegraphics[width=0.48\textwidth]{Escaledpi2004|025.eps}}
%{\includegraphics[width=0.48\textwidth]{Escaledpi2008|025.eps}}
%\caption{\label{Escaledpi2|025} The potential barriers of 
%Figs. \ref{Escaled004|025} and \ref{Escaled008|025} in the predominant
%direction of tunnelling along the ${\bm e}\chi$-direction at
%$\eta=\pi/2$ (the $y$-axis). Left: $u=0.04$, right: $u=0.08$.}
%\end{figure}    
%\end{center}
%\hspace*{1.2em} 
The situation we encounter in real space is shown in Figure
\ref{ellipse}. We visualize the quantum core size as the shaded area 
around the vortex center, having the $O(\xi)$-radius used in
(\ref{scond}). 
The vortex on path 1 is not able to pass the ellipse without some part
of this shaded area covering the ellipse, but the vortex on path 2
with energy $\tilde E_0$ avoids the ellipse surface completely.
That the validity of the hydrodynamic approach enforces that we
introduce another geometrical quantity, $s$, the vortex distance of
closest approach, which implies a lower bound of vortex energy, 
$\tilde E_0(s)$, is an observation of general character.
It is of relevance for any attempt 
to describe tunnelling in a realistic, non-spherical 
geometry, {\it i.e.} when the boundary and thus the path of the
tunnelling object near it is not of $S^n$ symmetry. 
A hydrodynamic collective co-ordinate description 
is valid only if the quantum core structure of the tunnelling object 
is not touched upon during its motion along the boundary. 
A pinning potential for the vortex 
moving in the superfluid stems 
in general from some flow obstacle, in our case the ellipse. 
Any phenomenological {ansatz} for a pinning potential usually employed in 
% a collective co-ordinate approach to 
tunnelling calculations, which has curvature 
perpendicular to the applied flow larger than parallel to the flow 
will have to take into account that the object can approach the
surface within its core size. %, invalidating the collective co-ordinate
%description. 

We now come back to the determination of the two paths of constant
energy we need to construct the closed path in Euclidean phase space. 
Assuming $\delta \chi \gg 1$, we get the path far away from the
ellipse to which the vortex has to tunnel. 
It is, with decreasing magnitude  of $u$, given by the 
relations (as a reminder, $b\gg a$):
\begin{eqnarray}  
l\exp[\chi_1]|\sin \eta_1| & = &
\frac\kappa{2\pi u }\ln \left[\frac{l}{\xi\exp[\tilde E_0]}
\exp[\chi_1]|\sin \eta_1|\right]\\
2Y_1/\xi & = & \frac{v_L}{u}\ln\left[\frac{2Y_1}{\xi\exp[\tilde E_0]}\right] \\
2Y_0/\xi & = & \frac{v_L}{u}\ln\left[\frac{v_L}{u\exp[\tilde E_0]}\right]
\,,\label{Y0}
\end{eqnarray}
The relevant path is the solution  
which has $Y_1\gg b$ (the second solution of the equations above
is just the vortex trapped at the boundary, far from the ellipse). 
The last relation is exactly analogous to (\ref{defR0}), which is
valid in the low velocity limit of three dimensions. 

The path at the ellipse with energy $\tilde E_0$ does have
approximately constant $\delta\chi\ll 1$ if $s\ll a$. The shape of this 
path can be obtained (to lowest order in $\delta \chi$)
by solving the equation 
\begin{equation}
(a+b)|\sin\eta_1|=\frac\kappa{4\pi u\,\delta\chi}
\ln\left[\frac{(a+b)\delta \chi}{\xi\exp[\tilde E_0]}\right]
\end{equation} 
Around the ellipse top, within a large range of $\eta$-values, the 
solution is approximately given by $\delta\chi_c \simeq s/a$.  
This can also be inferred from  Figure \ref{Escaled004050}.
%,\ref{Escaled008}.
% $a/b$\ll 1, $u\ll (a/2b) v_L, $\delta\chi\ll 1$ 
\subsubsection{The tunnelling area}
To describe the motion of the vortex in momentum space, we choose a 
gauge for the momentum 
which is most appropriate to the symmetry of our problem. 
In Cartesian co-ordinates, this is $P_X=h\rho_0 Y$, {\it i.e.} the gauge
momentum equals the physical momentum ({\it i.e.} Kelvin momentum) 
for an unconstrained vortex in the bulk superfluid. 
We will evaluate the action in the limit of small velocities. In this
limit the path to which the vortex has to tunnel 
(named in what follows $Y_N$), is given by a
constant, $Y_0$ in (\ref{Y0}). We have also seen at the end of the
preceding section that the vortex remains approximately on the same
ellipse, having the elliptic co-ordinate
$\chi_1=\chi_0+\delta\chi_c(\pi/2)\simeq \chi_0 + s/a$,  
while it is moving around the ellipse (path 2 in Figure \ref{ellipse}).  
Calling this path $Y_E$, we can write  
\begin{eqnarray}
Y_{\rm E}^2 & = & l^2\cosh^2(\chi_0+\delta\chi) +
\tanh^{-2}(\chi_0+\delta\chi)K^2\nonumber\\ 
& \simeq & l^2\cosh^2(\chi_0+\delta\chi_c) +
\tanh^{-2}(\chi_0+\delta\chi_c)K^2\,,
\label{YEK} 
\end{eqnarray}
using the imaginary co-ordinate $K=-iX$. Then, the integral 
in (\ref{SEKP}) takes the form
\begin{eqnarray}\label{intYNYE}
\frac{S_e^{\rm inc}}{h}&=&2\rho_0 \int_0^{K_m}\!\!
(Y_{\rm N}-Y_{\rm E})dK\,,
\end{eqnarray}
where $K_m$ is the point at which the trajectories $Y_N$ and $Y_E$ 
meet in complex phase space. 
\begin{figure}[t]
\begin{center}
\includegraphics[width=0.65\textwidth]{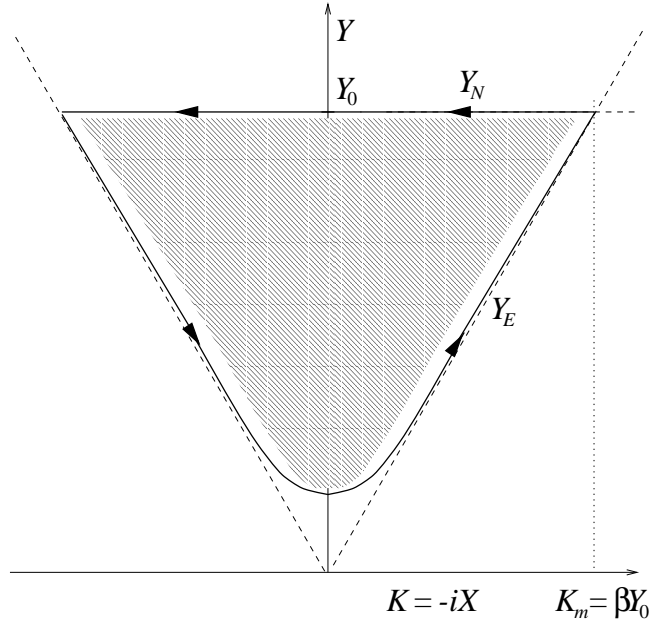}
\end{center}
\caption{\label{area} 
The closed vortex path giving the action (\ref{areabchiachi}) in the low
velocity limit. The first part $Y_E$ corresponds to the analytically continued
path 2 of Fig. \ref{ellipse} along the ellipse surface. 
The second part of the closed path, $Y_N\simeq Y_0 $, 
represents the border line to a free vortex. 
For higher velocities, this simple triangle shape is deformed. 
%The two parts $Y_E$ and $Y_N$ are shown in {\em real} $\chi_1,
%\eta_1$-space in Fig. \ref{Escaled}.
%The area $\simeq \beta Y_0^2=V^{(2)}_N$ 
%enclosed by the vortex path in complex space gives the action (\ref{tunnel}) 
%in two space dimensions. This area has a lower semiclassical limit  
%$(a/b+ \xi/a)Y_0^2$.
}  
\end{figure}
This point, determined in the low velocity limit by the solution 
of $Y_E=Y_0$, can be shown to be 
\begin{eqnarray}
K_m & \simeq & \tanh(\chi_0+\delta\chi_c)\, Y_0\nonumber\\
& \simeq  &\left(\frac ab + \frac sa\right)
\frac{v_L}{u}\,\ln\left[\frac{v_L}{u\exp[\tilde E_0]}\right]\,,
\end{eqnarray}
neglecting the first term in (\ref{YEK}), which is possible if
$Y_0\gg l\cosh(\chi_0+\delta\chi)\equiv b_\chi$. 
The last line is valid provided that both $\chi_0, \delta\chi_c\ll 1$.  

Now, the integral (\ref{intYNYE}), with $a_\chi=a(\chi_0+\delta\chi)
=l\sinh(\chi_0+\delta\chi)$, takes the form 
\begin{eqnarray}
\frac{S_e^{\rm inc}}{h}
%&=&
%2\rho_0\left(K_m Y_0 % \right.\nonumber\\ 
%% & & \left. 
%-\frac{b_{\chi_c}}2 \left[
%K_m\sqrt{1+a^{-2}_{\chi_c} K_m^2}
%\right.\right.\nonumber\\& & \qquad \qquad\qquad \qquad
%\left. \left. \,\,
%+\, a_{\chi_c}\ln\left|K_m a^{-1}_{\chi_c}
%+\sqrt{1+a^{-2}_{\chi_c} K_m^2}\,\right|\right]
%\right)\nonumber\\
& \simeq & 2\rho_0\left( K_m Y_0 -
\frac{b_{\chi_c}}{2 a_{\chi_c}}\,K_m^2\right)
%K_m^2/(2\tanh(\chi_0+\delta\chi_c))
%\right)\nonumber\\& = & 
=\rho_0\tanh (\chi_0+\delta\chi_c)\, Y^2_0\,.\label{areabchiachi}
\end{eqnarray}
The above integral corresponds to the shaded area in Figure
\ref{area}. 
Neglecting the small cusp at the bottom reducing slightly this area, 
as we did in the third line of (\ref{areabchiachi}), 
leads to a {\em volume law}\footnote{We refer to 
the contribution of $S_e^{\rm inc}$ as a volume contribution in
general, according to (\ref{SEgeo}).  
In two dimensions, of course, a volume is an area and an area is 
a length in conventional terms.} 
of the form 
\begin{equation}\label{SEOmega}
\frac{S_e^{\rm inc} (E)}h=\rho_0\Omega^{(d)}= \rho_0\beta V_N^{(d)}\,.
\end{equation}
Here, the effective 
sharpness $\beta = \tanh(\chi_0+\delta\chi_c)$ is a measure of 
the maximal eccentricity $\epsilon =\sqrt{1-\beta^2}\simeq 1-(1/2)\beta^2$ 
the vortex path is allowed to have, under the condition 
that the vortex should remain completely within the hydrodynamic, collective 
co-ordinate domain.  

The area of Fig. \ref{area} thus does have a lower limit. 
In general, $\beta$ characterizes the effective dimension of the vortex
escape path, {\it i.e.}\,\,the relative degree to which this path is
confined to $n$ dimensions by the presence of an 
asperity which is effectively $n$-dimensional. 
%The quantity $\beta$ is 
%bounded from below by the
%requirement of semiclassicality for the vortex path at the boundary,
%as expounded above for the two-dimensional case.   
The tunneling volume  $V^{(d)}_N$ is that for a vortex escape
path of O$(d-1)$ symmetry, which is the highest 
possible symmetry if one preferred direction, namely that of the external 
current, is given.
 % and small velocities.
In the $d=2$ case treated here at length, we have $n\gtrsim 1$, 
%$\beta \simeq a/b +s/a$ and 
$V^{(2)}_N = Y_0^2$ for a single vortex
and correspondingly $V^{(2)}_N = 2Y_0^2$ for a vortex pair.
In three dimensions ($d=3$) 
we have $V^{(3)}_N = (2\pi/3) R_0^3$ for a half-ring 
with radius $R_0$ and double this value for a full ring.
The effective sharpness will be reduced 
({\it i.e.} the value of $\beta$ larger) for the 3d
case (cf. the equation (\ref{SEusmall})), in analogy  
to our analytical findings for the vortex in the plane.  
In order of magnitude, the sharpness will in any dimension be
given by the product of the ratios of the curvature radii
parallel and perpendicular to the flow of the allowed vortex path.

We conclude with an estimation of the kinetic contribution in the
action, using the last line in (\ref{SEkin}). We will also use that 
in lowest order of perturbation theory the velocity of the vortex 
equals the local flow velocity ${\bm v}_s$ in the incompressible superfluid
(the Kelvin-Helmholtz theorem). 
The velocity maximum (at the ellipse top) 
we set  $v_{\rm max}\, =\, {\rm max}\,(|{\vec
v}_s|)$. Then,
\begin{eqnarray}
S_e^{\rm kin}(E) & = & h\rho_0  \frac{\ln(\cdots)}{2\pi}\,\xi \oint
({\dot Q}_{\it a}/{c_s})dX^{\it a}\nonumber\\
& \ll & 2h\rho_0  \frac{\ln(\cdots)}{2\pi}\,\xi\,\frac {v_{\rm max}}{c_s}\, K_m
\nonumber\\
& = & \frac{\ln(\cdots)}{\pi}\,\frac\xi{Y_0}\,\frac{v_{\rm max}}{c_s}\,
S^{\rm inc}_e(E)\,.\label{SEkinOmega}
\end{eqnarray}   
The contribution of the vortex mass is thus at least suppressed by two
small factors: $\xi/Y_0$ and $v_{\rm max}/c_s$, contributing both 
in equal measure to the fact that $S_e^{\rm kin}$ is negligible as
compared to $S_e^{\rm inc}$. 
%It is to be observed that the suppression factor $\xi/Y_0$ contains the  
\subsubsection{Summary}
We now summarize this extended investigation of the
analytically soluble 2d problem, emphasising in particular its crucial
outcome. 

Starting from the `electrostatic' problem of a single point vortex 
situated near a half-circle at a boundary, we derived the vortex energy 
by conformally transforming {\it via}
 the (inverse) Joukowski transformation to the 
half-ellipse solution. We concluded from the general expression for
the vortex energy (\ref{tildeE}), that it is necessary to introduce the
geometric constraint (\ref{scond}), expressing the limits of the 
hydrodynamic collective co-ordinate formalism under consideration. 
Calculating the tunnelling volume (area), we have seen that it assumes the
general form (\ref{SEOmega}). The tunnelling volume (area) $\Omega^{(2)}$
cannot be reduced to a lower bound given by the sharpness of the
ellipse, $\tanh \chi_0$, but has to be larger, 
$\beta=\tanh (\chi_0+\delta\chi_c)$, if we are bound to remain
within the domain of the approach which has been employed.

It is to be noted that we expressed the energy of the vortex 
in units of the energy $h^2\rho_0/(4\pi m)=
m\rho_0\kappa^2/4\pi\simeq 0.82 $ K/\AA\, (at $p\simeq
1$ bar and in three dimensions). 
For realistic values of the parameters in (\ref{E0}), 
the values of $\tilde E_0$
%, the energy needed by the vortex to remain describable in
%semiclassical terms, 
cover the same range as the phonon-maxon-roton spectrum. 
Energywise, the trapped small scale vortex thus cannot be
distinguished from an elementary excitation of the superfluid. 
It could have been excited thermally and remained  trapped at a pinning center
during the cool-down of the superfluid to very low temperatures.  

From the above analysis it thus follows that it is semantically in general 
not quite appropriate to employ the widely used term `vortex nucleation' 
for vortex tunnelling investigated within a hydrodynamic, collective 
co-ordinate theory. 
If we define `nucleation' to mean 
%`creation from nothing pre-existent', that is, 
creation from the zero of energy, we have seen that nucleation 
is not amenable to such a description under general circumstances. 
Experimentally, it will be impossible to distinguish the tunnelling 
of a %pre-existing 
small energy vortex at a rough boundary 
from the true nucleation event of a nascent vortex there,  
if no direct means to control the microscopic dynamics can be provided. 
It remains, of course, to be explained how a vortex of 
such small energy and size ({\it i.e.} one with small distance 
of $O(\xi)$ to the boundary), can be defined
and described quantitatively. The
present approach can only fix the tunnelling rates of a
vortex if one presupposes that the vortex can be
described by a collective co-ordinate along its whole path. This, 
however, will be true if the size of the vortex is sufficiently
larger than $O(\xi)$.  
%, at least that a nonzero (average) energy can be ascribed to it. 
%On the other hand, 
Then, the dissipation-free motion of the vortex 
with constant energy and the purely geometric nature of the problem 
lead to strict bounds for the possible values of the 
semiclassical tunnelling exponent, directly related to geometrical
quantities.
     
\subsection{The prefactor}
Any discussion of quantum tunnelling is incomplete without at least 
an estimation of the prefactor $A(E)$ in (\ref{PASe}). If the quantum 
fluctuations, or better, the indeterminacies 
of the vortex position and momentum
vanish, the tunnelling probability does the same, because the very
process of quantum tunnelling stems from this quantum  uncertainty 
of the vortex in phase space. The crucial advantage of our WKB-like
investigation lies in the fact that the behaviour of the exponent, the
Euclidean action of the instanton, dominates any dependencies of the
prefactor on observable quantities as long as $S_e(E)/\hbar\gg 1$, 
{\it i.e.} as long as we are in the semiclassical limit.
%coherence length scale of $\xi$. 
We will write (\ref{PASe}) in the form 
\begin{equation}\label{PSelnA}
P(E)= \exp\left[- S_e(E)/\hbar+\ln A(E)\right]\,,
\end{equation}
to compare $-S_e(E)/\hbar$ with the dimensionless
quantity $\ln A(E)$  
(the prefactor $A(E)$ is originally in units of Hz).
\subsubsection{Estimations}
Apart from the considerable difficulties in evaluating prefactors
in general, an accurate calculation of $A$ in a dense superfluid like 
helium II is in principle not possible at present, due to the lack of a 
microscopic theory. 
It is, however, feasible to get an idea about the value 
of this prefactor within about two orders of magnitude.

Let us begin with a physical picture for the origin of the 
prefactor in the semiclassical limit. 
The simplest possible idea about the prefactor is gained by
considering the % ground state 
frequency $\omega_a$  of a particle
oscillating in a metastable well. Then,
%, neglecting the influence of dissipation on the vortex motion, 
within about one order of magnitude,  
$A\sim \omega_a$ \cite{schmid}.
In the thermal activation limit, {\it i.e.}, 
in the Arrhenius law case we have more exactly 
$P=(\omega_a/2\pi)\exp [-U/k_B T]$, where
$\omega_a$ is the frequency of oscillations at the metastable well bottom
against a barrier of height $U$.      
The frequency $\nu_a=\omega_a/2\pi$ can generally be understood
as a measure of the number of times per second the vortex bounces
against the potential barrier, trying to get free. 

We have no possibility to describe the vortex state (at the boundary) 
quantitatively, but we are able to conclude on the order of magnitude
of $\omega_a$, if we take into account that there exists a surface layer 
of vorticity, of width $\xi$: Because the superfluid density goes to
zero at the boundary and heals back within $\xi$, the energy needed
for the activation of vortices vanishes within this distance
\cite{sonin,sonin3}. The frequency of motion of these % virtual 
vortices
%, one per surface area of order $\xi^2$, 
should then be of order 
\begin{equation}\label{omega0}
%\omega_c 
\omega_0
= \frac\kappa{\pi \xi^2}= 4.87 \cdot 10^{11} {\rm sec}^{-1}
\xi^{-2}[\sigma^{-2}_{\rm LJ}]\qquad(\mbox{helium II}),
\end{equation}
which is the cyclotron frequency of vortex motion \cite{donnelly1}.  
We scaled $\xi$ with the Lennard-Jones parameter $\sigma_{\rm LJ} 
= 2.556$ \AA\,\,
of the $^4\!$He atomic interaction.
The frequency $\omega_0$ is the natural vortex frequency 
associated with the scale $\xi$ alone. 

If we wish to connect $\omega_a$ with the quantity 
relating the strength of interatomic forces and compressibility,
namely the speed of sound $c_s$, which is involved in the vortex kinetic term,
the following phenomenological treatment is of use. 
The expression for the vortex self energy is
logarithmically divergent. If we consider $\xi$-scales, 
%this is no longer the case, because we explicitly have to take into account 
this may be cured in a heuristic manner by regularizing the logarithm 
under the condition that the vortex energy be zero at the boundary (that is, at
$Y=0$) \cite{DissVIAA}. 
This leads to the following total energy of a point vortex in the laboratory
frame, using the normalisation (\ref{normE}),   
\begin{equation}\label{EYdotX}
\tilde E (Y,\dot{\bm X})=\frac12\left((\dot{\bm X}/c_s)^2
+\ln \left(1+(Y/\xi)^2\right)-8\pi u Y/\kappa\right)\,.  
\end{equation}
The minima of the potential, existing for low enough velocity, are situated at 
\begin{equation}
Y_{\rm min}= \frac\kappa{8\pi u}\left(1- \sqrt{1-16\,(u/v_L)^2}\,\right)
\simeq \frac\kappa{\pi u} \left(\frac u{v_L}\right)^2
= 2\xi\, \frac u{v_L}\,,
\end{equation}
where in the final result we used an approximation for $u\ll v_L$. 
The curvature of the potential is
$\xi^{-2}(1-(Y/\xi)^2)/(1+(Y/\xi)^2)$ and hence at the minimum, for 
$u\ll v_L$, approximately $1/\xi^2$.  
We have thus found that the spring constant of an `elastic object'
vortex should scale with a quantity of order $1/\xi^2$. 
%The full energy may be
%written $(1/2c_s^2)(\partial{\bm X}/\partial t)^2
%+(1/2\xi^2)(Y-Y_{\rm min})^2$ 
Because the mass is $1/c_s^2$ in the same units, 
the frequency of oscillation is therefore 
%the frequency we already encountered in (\ref{omegasfirst}):
\begin{equation}\label{omegas}
\omega_s=
\frac{c_s}{\xi}= 9.4\cdot 10^{11} {\rm Hz}\, \frac{c_s[ 240\, {\rm m/s}]}
{\xi[\sigma_{LJ}]}
%\omega_s = \frac{c_s}\xi\simeq 9.23\cdot 10^{11}\mbox{sec}^{-1}
%\frac\sigma\xi\qquad (\mbox{$p$=1 bar, helium II}).
\end{equation} 
The ratio of the two estimates (\ref{omega0}), (\ref{omegas}) 
is given by $\omega_0/\omega_s=(\kappa/c_s)/(\pi\xi)$. 
They thus coincide in order of magnitude because the relation 
(\ref{kappacs}), valid in helium II, holds. It is quite obvious 
that both estimates can only give a rough idea about the actual attempt 
frequency in the {\em dense} superfluid, whereas at least the estimate
(\ref{omegas}) should be quite accurate in dilute
superfluids (for `nonrelativistic' vortex velocities \cite{arovas}). 
However, by argument of continuity, we do not expect the analytical 
relation for the frequency $\omega_a$ as a function of the
parameters $\xi,c_s$ to make abrupt changes, if we increase the 
density of the superfluid. We suspect that the density increase does not 
cause a change of more than, say, one order of magnitude in
$\omega_a$, than that predicted by the above 
estimates. 

Now, relying on the estimate (\ref{omegas}) and scaling the frequency 
with $1.13\cdot 10^{12}$\,Hz (the roton frequency at $p\simeq 1$\,
bar), so that $P(E)\equiv P(E)[{\rm Hz}]$ and 
$(c_s/\xi)_n \equiv({c_s}/\xi)\,[1.13\cdot 10^{12}\,{\rm Hz}]$, 
by use of (\ref{SEgeo}), we arrive at a tunnelling probability 
having the appearance of (\ref{PSelnA}):
\begin{eqnarray}
P(E)%[{\rm Hz}]
& = & %\frac{c_s}\xi
%\left[10^{11}\,{\rm Hz}\right]
\,\exp\left\{-\rho_0
\left( 2\pi \Omega^{(d)}+ \ln (\cdots)\, \xi\,\Sigma^{(d)}
%\left[\partial {\bm X}/(c_s \partial t) \right]
\right)
+27.75+\ln[({c_s}/\xi)_n]\right\} \nonumber\\
& \simeq  & %\frac{c_s}\xi
%\left[10^{11}\,{\rm Hz}\right]
\,\exp\left\{-\rho_0 \,\beta Y_0^2\left(2\pi+
2{\ln(\cdots)}\,\frac\xi{Y_0}\,\frac {v_{\rm max}}{c_s}\right)
+27.75+\ln[({c_s}/\xi)_n]\right\} \nonumber\\
& \simeq  & %\frac{c_s}\xi\,
\exp\left\{-2\pi N^{d)}\left(1+
\frac{\ln(\cdots)}\pi\,\frac\xi{Y_0}\,\frac {v_{\rm
max}}{c_s}\right)+27.75+\ln[({c_s}/\xi)_n]\right\}, \nonumber\\
\label{P(E)scaled}
\end{eqnarray}
where in the second line we inserted the results of the 2d
half-ellipse problem in equations (\ref{SEOmega}) and (\ref{SEkinOmega}),  
and the third line employs the Bohr-Sommerfeld quantization 
of (\ref{N}). 
%Assuming that the prefactor can vary % in its order of magnitude % at most 
%between $A\approx 10^{10}\cdots 10^{12}$ Hz, 
%its logarithm  ln$\,A \simeq  23\cdots 28$.
%It is of particular interest to compare $\ln A$ with the area 
%contribution of the vortex mass in (\ref{SEgeo}). Whereas $\ln A$ is
%present in any case because of intrinsic quantum fluctuations, 
%the contribution $S^{\rm kin}_e(E)$ does occur because we admitted
%small density and large phase fluctuations of the order parameter 
%to propagate.  
\section{Comparison to experimental results}
\begin{figure}[hbt]
\psfrag{IT}{$I_t$}
\psfrag{IW}{$I_w$}
\psfrag{IQ}{$I_q$}
\psfrag{IK}{$I_k$}
\psfrag{F}{$m_d,\eta_d,k,F$}
\psfrag{W}{$S_w,l_w$}
\psfrag{K}{$S_k,l_k$}
\psfrag{B}{\small He II}
\begin{center}
\includegraphics[width=0.72\textwidth]{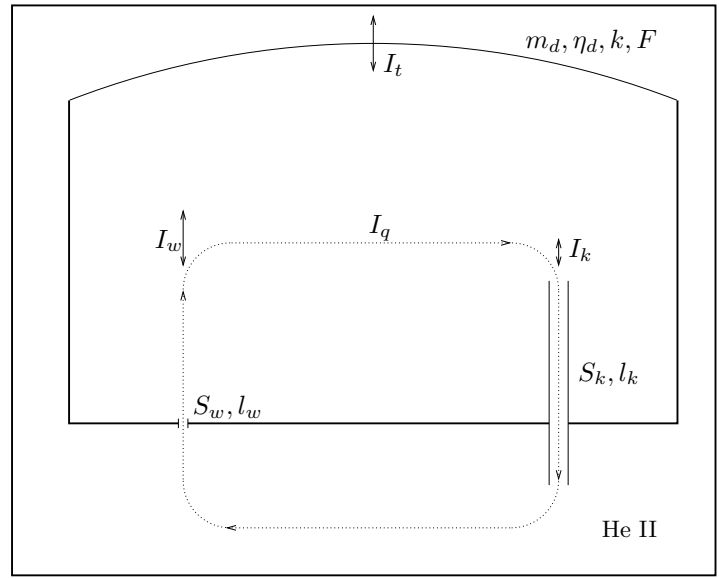}
\end{center}
\caption{\label{zavosz} The schematic arrangement of the 
ZAV (Zimmermann-Avenel-Varoquaux) oscillator 
\cite{AIV}.
A membrane, generating a current $I_t$ by an (electrical) driving force $F$, 
has mechanical 
parameters mass $m_d$, stiffness $k$ and damping constant
$\eta_d$. This
current divides into a current $I_w=I_t(L_k/(L_w+L_k))$ through a
micrometer-sized small hole and a current  $I_k=I_t(L_w/(L_w+L_k))$
through a long channel. Here, $L_{w,k}=l_{w,k}/(m \rho_0 S_{w,k})$ are
the hydrodynamic kinetic inductances, with $l_{w,k}$ the lengths and 
$S_{w,k}$ the cross-section areas of the micro-orifice and the long 
channel, respectively. The circulation threading the hole and 
channel, $I_q$, is quantized. The whole cavity is filled with He II.} 
\end{figure}

Having determined what we can expect for the magnitudes of
tunnelling probabilities we now come to discuss 
how experimental arrangements to measure such tunnelling events are 
constructed, and what the findings of such experiments are. 
In most of the measurements (cf. \cite{BBAV}--\cite{amar}), the
apparatus to measure phase-slip events is structured like that shown in 
Figure \ref{zavosz}. We note in passing that such an apparatus with an
effective torus geometry can be used to measure external rotation
({\it e.g.}, that of the Earth) quite sensitively (see 
Refs. \cite{AVearth}--\cite{AVrot}).   
%for a detailed discussion). %\vspace*{1em}

Essentially, it is observed in these experiments 
that at a well-defined value of the amplitude of the
diaphragm, driving the current through the micro-orifice (the small
hole on the left-hand side of Figure \ref{zavosz}), 
there is an instantaneous (on the scale of the
driving frequency) breakdown of the diaphragm amplitude, which is quantized. 
This quantized dissipation event is associated with 
a vortex generated at the orifice walls, subsequently 
crossing all the streamlines of the flow, thereby causing a {\em phase 
slip} event, which draws a quantized amount of energy from 
the flow.
We will first discuss these {phase slips} 
\cite{anderson}, which give the crucial physical argument for
the interpretation of the experimental results.
\subsection{Phase slips}
\begin{figure}[hbt]
\input psfwbdiss2
\begin{center}
\includegraphics[width=0.6\textwidth]{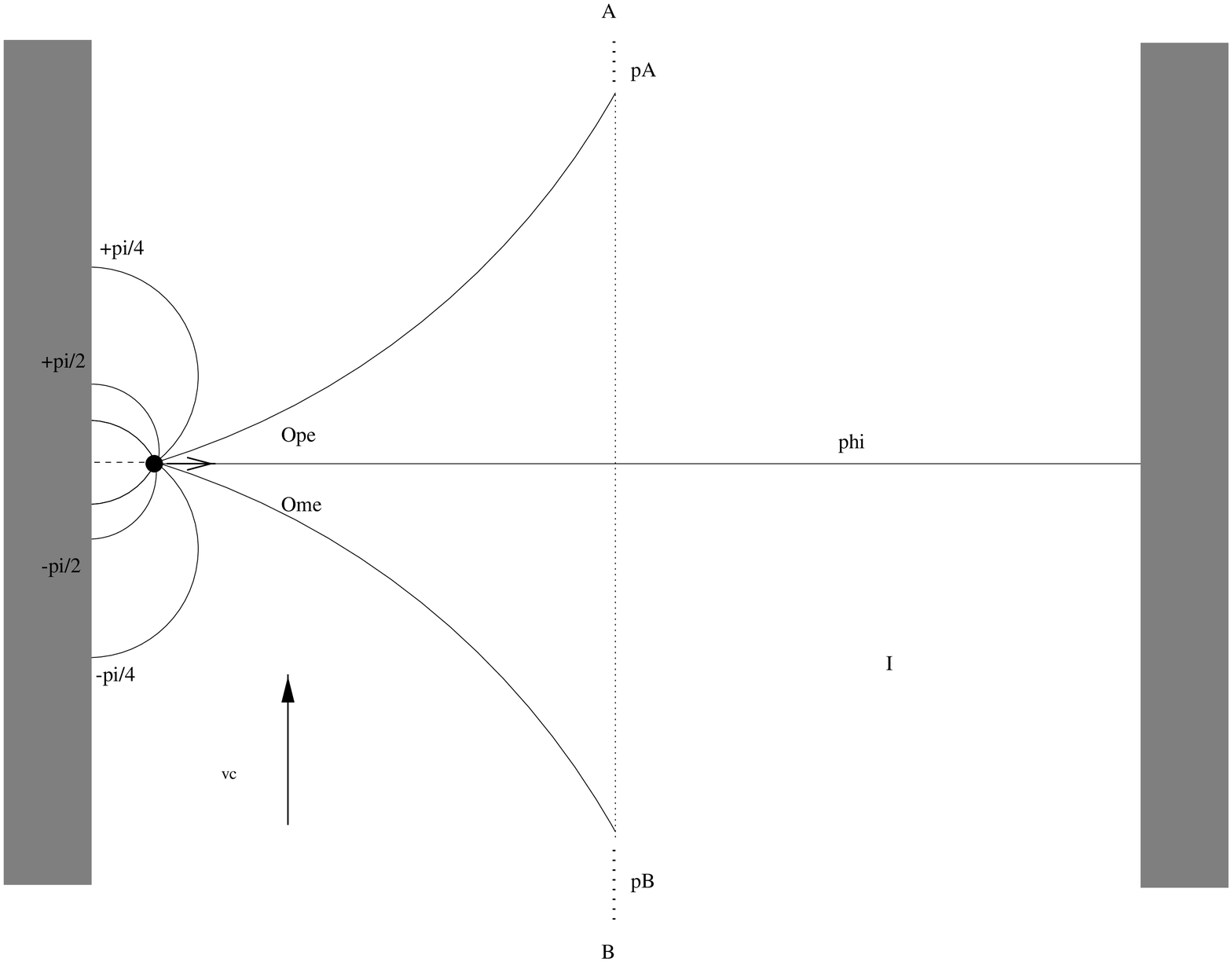}\vspace*{1.5em}
%\hspace*{0.2\textwidth}
%\includegraphics[width=0.8\textwidth]{wb2.eps}
\includegraphics[width=0.6\textwidth]{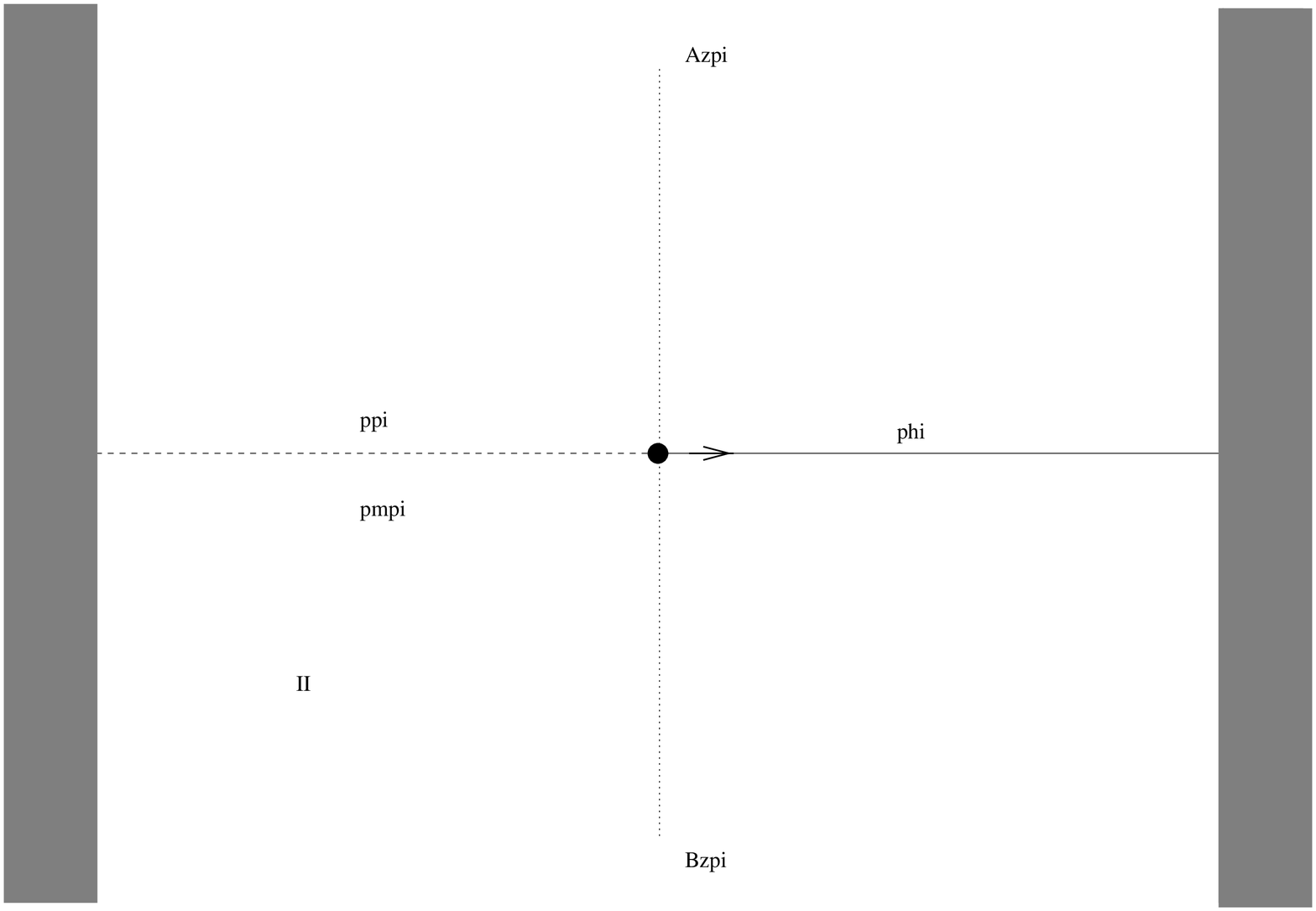}
\end{center}
\caption{\label{wirbelbewI} The early and intermediate 
stage of a phase slip process in the 
orifice \cite{anderson}. 
%A detailed explanation is contained below the caption of Figure 
%\ref{wirbelbewII}.
}
\end{figure}

\begin{figure}[hbt]
\input psfwbdiss2
\begin{center}
\includegraphics[width=0.6\textwidth]{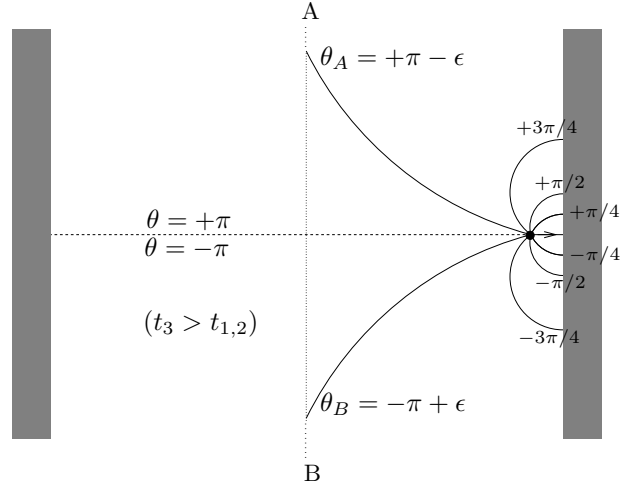}
\end{center}
\caption{\label{wirbelbewII} Final stage of the phase slip process. }
\end{figure}

The picture of phase slips is based on the fact that the phase and 
particle number are canonically conjugate,\footnote{This is true in the
hydrodynamic limit, after averaging over a cell much larger
than the atomic size. In the microscopic domain, we encounter
consistency problems related to the general problem of the existence 
of quantum mechanical phase operators (see, in particular, 
\cite{froh}; also \cite{pegg}).} that is, 
\begin{equation} 
[N,\theta]=i\,.
\end{equation}   
The quantum mechanical equation of motion of the phase is hence given by 
\begin{eqnarray}
%\hat{\!\dot N}&=&\frac1{i\hbar}[\hat H,\hat N]
%=\frac1\hbar\part{\hat H}{\hat\theta}\quad ,\label{dNdt}\\
{\!\dot \theta}=\frac1{i\hbar}[\theta,H]
=-\frac1\hbar\part{H}{N}\,. \label{dphidt}
\end{eqnarray}
Taking the {\em thermodynamic} average of this equation, we see that 
the time rate of change of the phase equals the (negative)
local chemical potential (defined by
$\tilde \mu =\mu + (1/2) m v_s^2 $, where $\mu$ 
is the chemical potential in the superfluid rest frame), 
divided by $\hbar$, a relation which became popular under the 
name `Josephson-Anderson equation' \cite{josephsonfluxflow,anderson}. 
We now consider the {\em time} average of this thermodynamic average 
over a long time span $\tau$ and take the
difference of the results for two points A,B in the superfluid:
\begin{eqnarray}
-\left\langle\tilde\mu({\bm r}_A)-\tilde\mu({\bm r}_B)\right\rangle_t & =&
\left\langle \hbar \left(\pard {(\theta_A-\theta_B)}t\right)
\right\rangle_t \nonumber\\
%\hspace{6.7em}\\
%& & \equiv\lim_{\tau\rightarrow\infty}\left[
%\frac1\tau\int_0^\tau\! dt \,\hbar \left(\pard {(\theta_A-\theta_B)}t\right)
%\right] \nonumber\\
& = &\lim_{\tau\rightarrow\infty}\left[\frac\hbar\tau 
\int_0^\tau\! dt \,\pard {}t\left(
\int_{{\cal C}_{\rm AB}}\nabla\theta\cdot d{\bm s}\right)
\right]\nonumber\\
& \equiv & h \left\langle \pard{n_w}t \right\rangle_t\,.
\end{eqnarray}
The last relation tells us that the the negative %superfluid 
chemical potential 
difference between the
points A and B, divided by Planck's quantum of action, is equal to the 
number of vortices crossing the 
line ${\cal C}_{\rm AB}$, joining A and B, per unit time, $dn_w/dt$. 
A single phase slip process, caused by one migrating vortex,
%mathematically expressed in this fashion,
may be visualized as shown  
in Figs. \ref{wirbelbewI} and \ref{wirbelbewII}. 
The Figs. \ref{wirbelbewI}, \ref{wirbelbewII} 
represent a pronounced simplification of the actual vortex
motion process. We visualize in these pictures a point vortex moving on a 
straight line across the superflow through the orifice, 
which gives a highly symmetric view of 
the process. The real process of vortex half-ring motion is 
investigated in Refs. \cite{schwarz2,BBAV}.  
The principal global (topological)
features of importance however remain untouched
and independent of the actual vortex motion trajectory: 
The phase difference 
between two stationary states (times $t\ll
t_1$ and $t\gg t_3$ in Figs. \ref{wirbelbewI} and \ref{wirbelbewII})
is exactly $2\pi$ and the process always sucks the same amount of
quantized energy from the flow, given by $\Delta E = m\rho_0 \kappa S_w
v_c = \kappa J_c $ \cite{huggins},  
with $v_c$ being the (mean) {\em critical velocity of flow} through the
micro-orifice, at which the vortex migration process sets in, 
and $J_c$ the corresponding mass current (see for a derivation below).
%\newpage       
%\begin{center}{\it
 
In Figures \ref{wirbelbewI} and \ref{wirbelbewII}, 
we represent a quantized vortex 
crossing the micro-orifice designated with quantities 
$S_w$, $l_w$ in Figure \ref{zavosz}. 
%, in the Figures \ref{wirbelbewI} and \ref{wirbelbewII}. 
%} \end{center} 
The lines emanating from 
the vortex center (black dot) in Figures \ref{wirbelbewI} and 
\ref{wirbelbewII} are lines of constant phase, standing 
perpendicular on the orifice wall (which is a streamline). For ease 
of representation we have chosen the branch cut of the phase to be 
exactly parallel to the direction of motion of the vortex. 
The whole phase slip process is then maximally symmetric. 
The shaded areas represent the walls of the orifice  
$S_w,l_w$ on the left hand side of Figure \ref{zavosz}. The points A, B are 
chosen to lie 
sufficiently far away from the orifice, as indicated by the dots. 
Initially, at $t=t_1$ (first drawing), 
when the vortex starts on the left side of the
orifice, the phase difference between A and B, $\theta_A-\theta_B$, 
is zero.  
When the vortex reaches the line joining A and B, the phase
difference is $\pi$ (second picture in Figure \ref{wirbelbewI}). 
Finally, as shown above, 
the vortex has crossed all the streamlines through the orifice, flowing, as 
indicated in the first Figure, in the vertical direction, 
and disappears to the right. 
The drawings above give a representation of the effect of 
migration of the order 
parameter zero, the vortex, across the matter flow through the orifice, 
by means of the phase of the order parameter. The physical result of this 
migration is invariantly given by the energy change, expressed  
in terms of the critical mass current $J_c$ 
for triggering the generation of the vortex, $\Delta E = \kappa J_c$, 
and does not depend on a representation 
in terms of the phase (or, for that matter, on the choice of the 
branch cut), {\it i.e.} is gauge invariant; 
it is also invariant under changes of the location of the points A and B 
(as long as they are situated far away from the micro-orifice).
Multiplying the thermodynamic Josephson relation with the critical number 
current through the orifice, results in a time rate of energy decrease 
of the external flow driven by the oscillating
membrane in Figure \ref{zavosz}, 
$\dot E = (\kappa J_c/2 \pi) \dot \theta $,  
from which the energy change for one phase slip, necessarily 
causing $\theta$ to change by exactly $2\pi$, 
follows by time integration.\footnote{The resulting expression 
for the energy change neglects a small 
correction due to the diminishment of the external current during the 
phase slip \cite{huggins}.} 
%In terms of the electrodynamical analogy of the preceding section
%(cf. eq. (\ref{Fdef})), the analogous energy change is  
%given by the electrostatic expression $\Delta E = |q {\bm E}_c| S_w$, where 
%${\bm E}_c= \rho_0 {\bm v}_c \times {\bm X}'$ %($v_c>0$) 
%is the critical electric field, perpendicular to 
%$\bm v_c $ and the orifice wall, driving away 
%the vortex from the wall and separating it from its image therein.  

\begin{center}
\begin{figure}[hbt]
\hspace{-2em}
\rotate[r]{\includegraphics[width=0.22\textwidth]{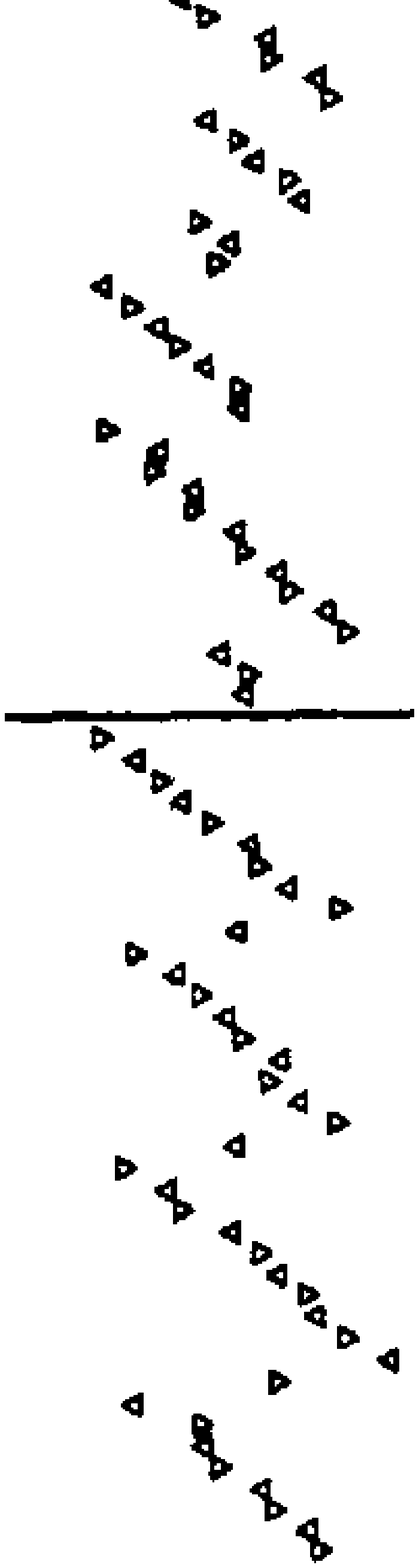}}%\vspace*{-1.5em}
\caption{\label{elps} Elementary phase slip processes in two different
runs \cite{AV1}. 
%On the left-hand side there is a zero bias of phase
%difference across the orifice, on the right-hand side this bias is
%$(\theta_A-\theta_B)_{\rm bias}=\pi$. 
The resonator amplitude of the diaphragm, which is
ramped up in time by applying a %constant 
voltage bias on the coated diaphragm, 
is recorded every half-cycle (triangles
pointing up and down, respectively). A phase slip occurs if there is
a breakdown of the amplitude to the next half-cycle.} 
\end{figure}\end{center}
\begin{center}
\begin{figure}[hbt]\hspace*{-5em}
\rotate[l]{\includegraphics[width=0.8\textwidth]{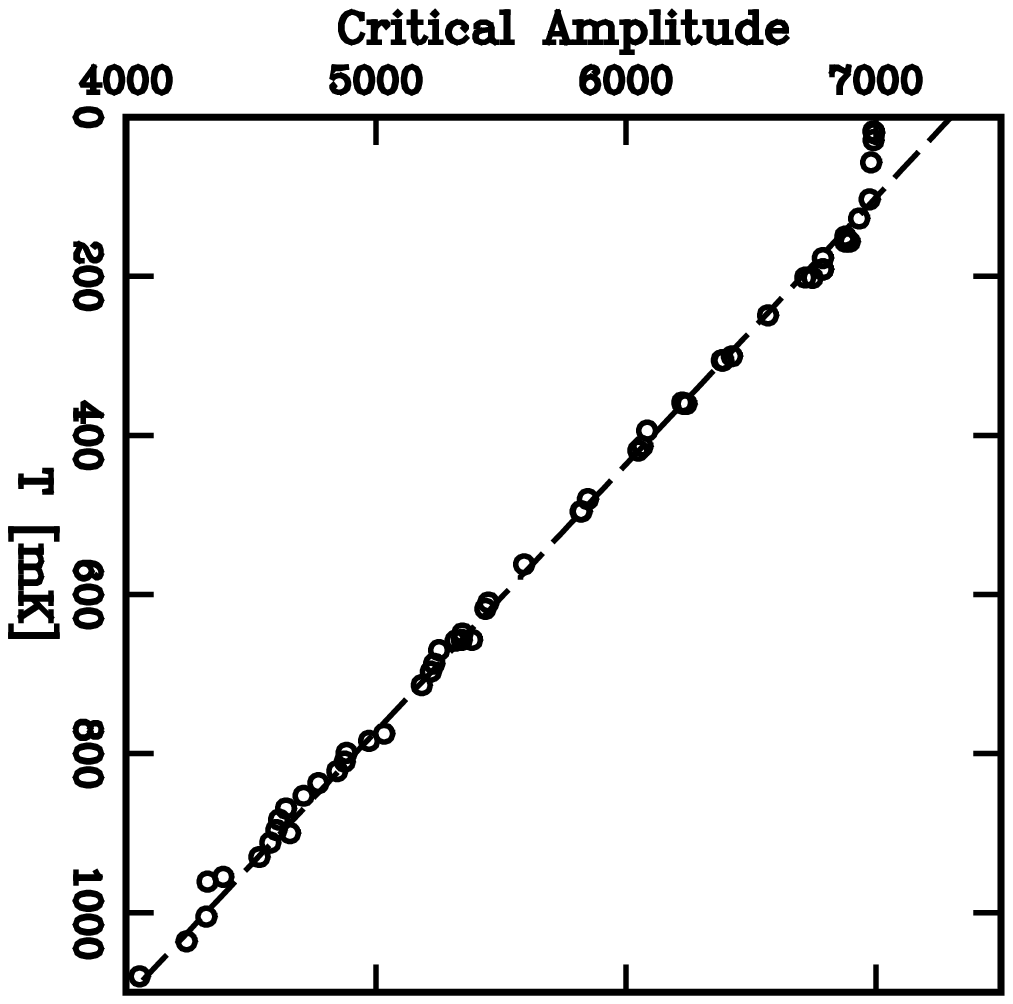}}\vspace*{-2em}
%\rotate[l]{\includegraphics[width=0.5\textwidth]{Minn-3.ps}}\vspace*{-1.5em}
\caption{\label{adc} The critical velocity through the orifice 
(in terms of instrument units of the critical amplitude of the diaphragm) 
\cite{AIV}. The 
line is a fit to a (thermal) `half-ring model', discussed in the text.
The temperature region below 200 mK is magnified in Fig. \ref{magadc}.}
\end{figure}\end{center}
\subsection{Principal findings}
\begin{center}
\begin{figure}[hbt]
\hspace*{-4.5em}
\rotate[l]{\includegraphics[width=0.6\textwidth]{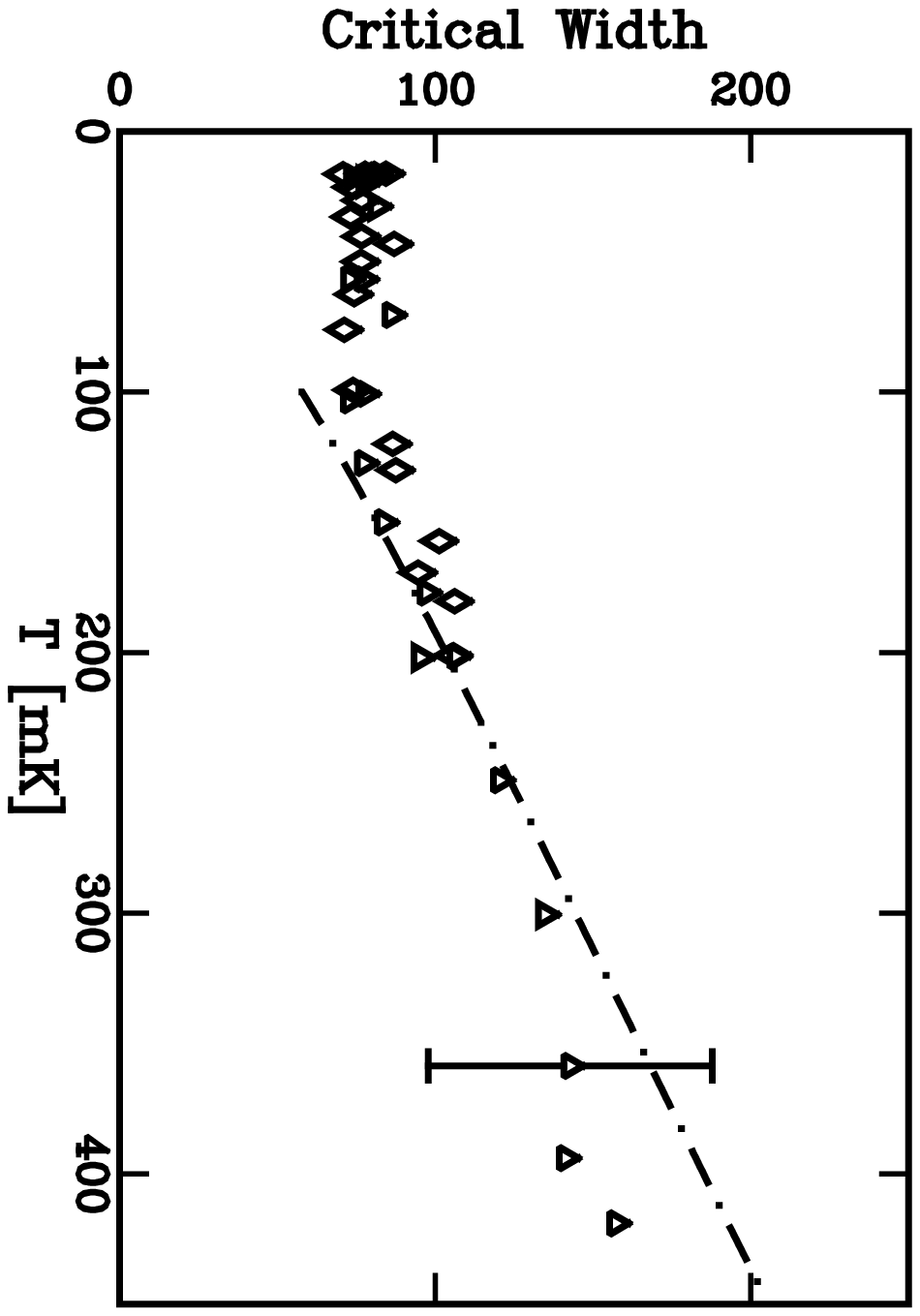}}\vspace*{-1.5em}
\caption{\label{dadc} The critical velocity 
statistical distribution.}
%, which is shown in Figs. \ref{adc} and in \ref{magadc}, 
%magnified below 200 .}
\vspace*{1em}
\end{figure}\end{center}

In the experiments, the critical velocity $v_c=v_c(T)$ as a function
of temperature $T$ is measured. 
The critical amplitude of the diaphragm, corresponding to $v_c$, 
is that for which there is a diminished 
amplitude in the next half-cycle of oscillation.  
The Figure \ref{elps} shows a typical
experimental run of measured resonator amplitudes.   
An important feature of $v_c$ is that it has a statistical
distribution, which has also been recorded. 
The results, obtained from the statistical analysis of the 
series of phase slips like those in Fig.\,\ref{elps} 
are shown in Figs.\,\ref{adc}, \ref{dadc}, \ref{magadc} 
(received from E. Varoquaux and reproduced here with kind permission).   
The salient results are 
that the mean critical velocity first rises linearly with temperature and then
saturates at $T_0\simeq 150$ mK. Correspondingly, the statistical
width decreases linearly and saturates at approximately the same
temperature. 
\subsection{Interpretation}  
According to the phase slip picture we have developed above, a possible 
interpretation of the experimental data is as follows. 
A vortex half-ring, standing with its axis antiparallel to the flow 
\cite{sonin}, is generated at the orifice wall and expands under the 
influence of the diverging flow field through the orifice. During this
process, it crosses all the streamlines through the orifice,
completing the phase slip, and is 
finally transported away from the orifice by the flow.
The fact that there is a certain critical amplitude of the diaphragm
for this procedure to happen, can be associated with the fact that 
there is a potential barrier opposing the process. Furthermore, the
fact that the critical amplitude ({\it viz.} the critical velocity $v_c$),
has a statistical distribution, supports the idea that the existence
of a phase
slip critical velocity has the statistical origin of barrier 
crossing events. Additional support is provided by the linearity of
the amplitude respectively its distribution with temperature, a
signature of thermal activation over barriers \cite{eckern}.  
It is also measured that the (average) flow velocity through the orifice 
as a function of
temperature, needed to trigger the phase slips, as well as its 
statistical distribution, saturates at a temperature of $\simeq$ 150 mK. 
This can be ascribed to a non-thermal process of surmounting the
existing barrier: The possible explanation of the observed behaviour
is quantum tunnelling of half-ring vortices at boundaries.  

Within a phenomenological approach \cite{AIV,ihasAV}, 
a model of half-rings with axis antiparallel 
to the applied flow, and standing perpendicular to the walls of the orifice, 
 has been developed. 
%fitted to the available data. 
The Hamiltonian is essentially that in 
(\ref{EYdotX}), {\em save for} the kinetic term. It turns out that, to make 
this model conform with the available measurements, it is necessary to
postulate, 
a) that the coherence length increases to $\xi\simeq 9\cdots 10$\,\AA\,  at
boundaries and b) that the vortex half-ring be described by a collective
co-ordinate on $O(\xi)$-scales.
This approach must consequently be understood as the parameter-fitting of a 
simplified model to the available data.  
From a more fundamental point of view, it does not describe crucial,
indeed salient features of the actual problem: 
\begin{itemize}
\item[{\it i.}] The model does not incorporate Galilean invariance
violation. We have seen that this is a necessary requisite for any 
(hydrodynamic) formalism making, in particular, use of the notion 
of velocity,
to describe the quantum tunnelling of vortices at temperatures close 
to absolute zero.    
\item[{\it ii.}] It cannot be reasonably expected from a description
of the entity vortex on $\xi$-scales to make sense for the tunnelling
exponent beyond crude order of magnitude estimates, like in the case 
of the prefactor (where these estimates are sufficient). For the
semiclassical tunnelling exponent, all nontrivial 
dependence on coherence length
(respectively microscopic)
physics, in whichever form, should be excluded, such 
a dependence only being allowed in the form of an ultraviolet cutoff.     
\item[{\it iii.}] It is certainly not permissible to neglect the
dynamic influence of the kinetic energy of the vortex in the
tunnelling exponent, if we approach scales 
of order the coherence length. 
\end{itemize}
Let us now further analyse the experimental outcome. 
First of all, we rely on the hydrodynamic 
relation (\ref{defR0}) to deduce the scale of 
materialisation of the nascent vortex half-ring. Scaling the velocity
with 10 m/sec, the order of magnitude of the (local) flow velocity, 
we get 
\begin{equation}\label{R0scaled}
R_0 = \frac{1.59 \,{\rm nm}}{u[10 {\rm m/sec}]}
\, \ln\left(\frac{9.89}{\xi[\sigma_{\rm LJ}]\,u[10 {\rm m/sec}]}
\right) \,, 
\end{equation}
where use was made of the Roberts-Grant result \cite{robgrant} for 
$C=1.615$, valid within Gross-Pitaevski\v{\i} theory 
\cite{pita}-\cite{gross4}. In the experiments, the
measured velocity through the hole is 5-10 m/sec. This can only be
measured as an average over the cross-section of the orifice, locally 
the velocity can of course be higher. Nevertheless, it can be concluded that 
the radius $R_0$ should be of the order of nanometers. This small 
mesoscopic scale makes it difficult to decide if a hydrodynamic formalism is 
applicable in a rigorous sense (one should also
bear in mind that the formula above is strictly valid only in the
low-velocity limit). In particular, the value of the coherence length
is not exactly known under the circumstances considered. The neglect 
of any large density variations ($\delta\rho/\rho_0= {O}(1)$)
in the formalism makes it necessary at least to assume that
$R_0\gg\xi$, so that the knowledge of $\xi$ is crucial indeed.%\hspace*{4em}
\begin{center}
\begin{figure}[hbt]
%\begin{center}
\hspace*{-1.5em}
\rotate[l]{\includegraphics[width=0.75\textwidth]{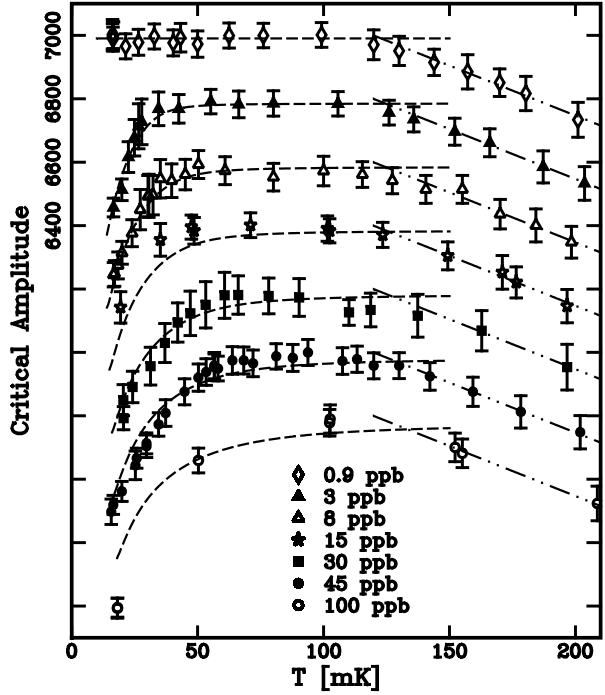}}
\vspace*{-1em}
%\end{center}
\caption{\label{magadc} 
Magnified portion of the temperature region below $\simeq$
200 mK \cite{AV3He,AIV}. 
The figure shows the influence of minute impurity concentrations
of $^3\!$He, given in ppb,
on the critical velocity of vortex phase slip for very low
temperatures (for clarity, the curves are shifted downward 
with respect to one another by 200 instrument units).
The dashed curves are a fit to the data, according to a  
phenomenological theory  
that $^3\!$He atoms, bound to the vortex core at the tunnelling site, 
reduce the critical velocity \cite{AV3He}. The dash-dotted curves are 
the high-temperature linear fits shown in Fig. \ref{adc}.
For the purest sample (0.9 ppb $^3\!$He), the critical velocity is 
temperature independent within experimental accuracy down to 
the lowest temperature $\simeq 15 $\,\,mK.}    
\end{figure}
\end{center}

Next we consider the value of the cross-over temperature $T_0\simeq
150$ mK. (Ref. \cite{jcdavis} reports a value $T_0\simeq 200 $ mK.)
The crossover temperature between thermal and quantum 
behaviour gives in general a measure of the equality of 
quantum-mechanical and thermal energies of the `particle', trying 
to surmount the barrier with the aid of these energies \cite{weiss}.  
For zero damping (dissipation) the crossover temperature is given by 
$\hbar\omega_b= 2\pi k_B T_0 $, where $\omega_b$ is the frequency of
oscillation at the {\em top} of the barrier, connected to the trivial 
solution of the Euclidean equations of motion of the `particle' sitting at the 
top (the bottom of the inverted potential) \cite{weiss,affleck,gorokhovquant}. 
This leads to
\begin{equation}\label{omegab}
\omega_b=8.2\cdot 10^{10}{\rm Hz}\cdot T_0 [{\rm 100 mK}]\,.
\end{equation}
%\newpage
If we compare the value of $\omega_b$ in (\ref{omegab})
with the frequencies (\ref{omega0}),
(\ref{omegas}), we see that it is smaller by a relatively 
large factor, up to one order of magnitude, provided 
we assume the coherence length to have its bulk value, which is about
$\sigma_{\rm LJ}$.  
However, a direct comparison of the experimental value and these
estimates cannot give more than an order of magnitude agreement.
This has several reasons. 
First of all, 
these estimates can give only a very approximate idea about the
true dynamical behaviour of a quantum many-body vortex near the boundary.  
It is conceivable, for example,  that the effective `spring constant'
of the vortex against deformations is lowered compared to the
semiclassical estimate in (\ref{omega0}), because of the many-body 
quantum uncertainty of its position.
Then, the prefactor 
is in general a function of driving velocity $u$ and temperature $T$
\cite{gorokhovquant}, and at the measured crossover temperature 
in terms of the critical velocity 
% plateau to temperature dependent behaviour)  
not necessarily equal to its value at zero temperature.
Finally, as already mentioned,
we do not know the (effective) value of the coherence
length at the boundary. It is conceivable that the value of $\xi$ is
enlarged as compared to the bulk, because of boundary conditions, 
{\it i.e.} the depleted superfluid density \cite{sobyanin}.
What one can thus definitely claim to have observed 
from Fig.\,\ref{magadc},  
is that there exists a crossover temperature
from a temperature dependent to a temperature independent r\'egime, 
whose energy equivalent $k_B T_0$ is in {\em order of magnitude} agreement
with the estimates for the quantum oscillator energies $\hbar\omega_0$ and 
$\hbar\omega_s$. 
To conclude this section, we 
give an idea about the number of particles involved in the
tunnelling procedure, therein following the statement of equation
(\ref{N}) and the scaled tunnelling probability in
eq. (\ref{P(E)scaled}). Assuming from the above discussion 
that the prefactor can vary in its order of magnitude  
between $A\approx 10^{10}\cdots 10^{12}$ Hz, its logarithm  ln$\,A
\simeq  23\cdots 28$. Corresponding to these conceivable values of the
prefactor, the total number of particles in the tunnelling volume 
should then be somewhere in the range $N^{(3)}=4\cdots 6 $, say, for tunnelling
events to be observable within a reasonable span of experimental time.
Again, like the value for the materialisation radius $R_0$ 
in (\ref{R0scaled}), this number indicates a rather small 
%mesoscopic 
scale of tunnelling.
\subsection{Concluding evaluation of the experiment}\label{concludeeval}
We have seen that the available data on critical velocities of phase slips 
can be interpreted to be in phenomenological  accordance with the picture of 
the quantum tunnelling of vortices at boundaries below some
temperature $T_0$. 
But the very fact of tunnelling at boundaries 
certainly needs further proof, so that the
predictions of tunnelling theory
can be compared to that of classical instability
mechanisms, which can be temperature independent as well. 
Theoretical investigations in this classical direction are found
under Refs. \cite{kuznetsov}--\cite{srivastava} (vortex nucleation 
as a process of classical flow instability in experiments using
$^3\!$He-B was discussed in Ref. \cite{parts}). 
Provided that the hydrodynamic, large scale 
picture we developed here is applicable with
sufficient predictive power for the actual materialisation scales of
the vortex, one such proof could consist in the comparison of critical
velocities for chemically identical orifices of equal global sizes, 
having different (microscopic respectively mesoscopic) surface structures. 
If the result of such measurements is negative, {\it i.e.} there is no
reproducible difference in critical velocities, there 
% exist the following possibilities. Either the tunnelling mechanism is not
% describable by semiclassical, hydrodynamic means or there is 
is no quantum process taking place 
describable by hydrodynamic means
of semiclassical tunnelling at irregular boundaries.  
%The experimental information
%at present is too sparse to give a final and conclusive answer.

%\newpage
%\thispagestyle{empty}
%\thispagestyle{plain}
 
\section{Aspects of vortex tunnelling in Fermi superfluids}
\label{chapFermi}
The analysis so far has been concerned with dense superfluids which are 
uncharged, and in which the fundamental constituents, {\it i.e.}
the particles carrying the superfluid current, bear no internal 
degrees of freedom like spin. % or angular momentum. 
The spinless elementary bosons, 
which are $^4\!$He atoms, form the only hitherto known example of such a 
superfluid. There exists, however, a large number of superfluids
which are constituted by elementary fermionic particles. By far the
most of these superfluids are charged: The charge carriers in 
superconductors represent a charged superfluid. 
Besides He II, the only other charge neutral dense superfluid known 
in laboratories on Earth is its isotope $^3\!$He. In the following, we 
%expound and 
give an overview of some general features of these
Fermi superfluids, with particular emphasis on the vortex dynamical
equations. 
The  main intention of this
section is to set the complications arising in the treatment of vortex
motion in Fermi superfluids
along the comparatively elementary hydrodynamic problem of unpaired 
bosons in He II.    
\subsection{Introduction}
For all Fermi superfluids, there has to exist a mechanism binding the 
fermions into Cooper pairs \cite{BCS}, which constitute the bosons 
of the superfluid condensate. In distinction from the elementary $^4\!$He
bosons, these effective particles, arising from paired fermions,
have in general the internal degrees of freedom %like 
spin and angular momentum. According to the symmetry of the order parameter, 
these superfluids are classified to be $s$-, $p$- or $d$-wave superfluids. 
In the case of isotropic superfluids, the 
value of the internal angular momentum $L=0,1,2$ of the Cooper pairs 
corresponds to $s$-, $p$- or $d$-wave, respectively. 
A relatively weak 
effective interaction binding the fermions together leads 
%in general 
to coherence lengths $\xi \gg k^{-1}_F$, 
where $k_F$ signifies the Fermi momentum.
The corresponding Cooper-pairs are then objects bound together 
over distances  by far exceeding the microscopic scales 
relevant for many-body quantum mechanics.
One  consequence is that,
{\it e.g.}, the Ginzburg-Landau or mean-field 
levels of description, not sufficient to describe 
scales of order $\xi$ in He II, are indeed useful for such paired
Fermi superfluids on these scales \cite{degennes}. 
%\footnote{With the 
The notable exception are high-$T_c$ superconductors, 
where $k_F\xi \gtrsim O(1)$. 
More properly, this is to be written as 
$\xi\gtrsim d_L$, where $d_L$ is the lattice spacing, 
because the Fermi surface is in general 
not a single continous surface  %no well-defined quantity 
for the high-$T_c$ materials.
%}

The fact that the Cooper pairs have internal angular momentum
and/or spin leads to a richly structured order parameter, which
can support symmetries much more complex than the global and local
U(1) symmetries associated with particle conservation and the 
electromagnetic field. 
In particular, as one of the most important 
consequences of these enriched symmetries, 
%for the dynamics of vortices to be discussed in the following, 
there can occur nodes of the energy gap in momentum space for 
quasiparticle excitations above the superfluid ground state
\cite{grishabook}. 
%, whereas in isotropic $s$-wave superfluids with spin singlet
% pairing, or more general for isotropic superfluids with 
% total angular momentum equal to zero, the gap has no nodes.

\subsection{Vortex motion in Fermi superfluids}\label{secVMFermi}  
Around a vortex line, there exists a potential well for quasiparticles, 
{\it i.e.} the pair potential 
is position dependent, $\Delta = \Delta ({\bm x})$
(we designate $\Delta_\infty$ to be the bulk value of the gap,
infinitely far away from the vortex line). 
This leads to the existence 
of bound quasi-particle states in the vortex core, of extension $O(\xi)$.  
These bound states are obtained by solving the Bogoliubov-deGennes mean-field
equations for the wave functions in particle-hole space, in which 
$\Delta ({\bm x})$ plays the role of the potential 
\cite{degennes,caroli}. They lead to a profound alteration of the
low-energy dynamical behaviour of vortices because, among these bound states,
there exist the so-called zero modes crossing the zero of energy 
as a function of a component of
the quasiparticle 
wave vector (the other core bound states excitation branches have 
energies at least of $O(\Delta_\infty)$, {\it i.e.} of order 
the bulk energy gap).    
These zero modes lead to a exchange of quasiparticle momentum 
between the superfluid vortex vacuum,  moving with velocity 
$\partial{\bm X}/\partial t$,
and the quasiparticle heat bath, moving with velocity ${\bm v}_n$, if 
the relative velocity  $\partial {\bm X}/\partial t- {\bm v}_n$ is 
non-vanishing:  The quasiparticles are driven by the effective
(electric-like) field, stemming from this relative velocity,  
%of vortex line and quasiparticle bath 
from the occupied negative energy levels to those
of positive energy (energies are counted from the Fermi energy $E_F$),
thereby transferring momentum to the quasiparticle bath \cite{stone2}.
The motion of the quasiparticles on the zero mode branch 
is a process for which the notion of {\em spectral flow} has been
coined. A contribution to spectral flow is stemming only from chiral 
particles (cf. \cite{grishabook}, chapter 6).  
%and the creation of chiral charge, {\it i.e.} chiral part 
Due to the momentum exchange between superfluid vacuum and
quasiparticle bath, there results an additional transverse force 
on the vortex, to be added to the usual contributions which occur 
in superfluids without zero mode bound states \cite{grisha3,kopningrisha}.
 
The equation of motion for the vortex 
in the time domain is in general regions 
of the parameter space non-local. 
In frequency space, the equation of motion is local, because the 
convolutions of Green's functions with the vortex co-ordinates 
in the time domain become products after Fourier transformation.  
The equation of motion may then be written 
in a form which satisfies Galilean
invariance, which implies that only relative 
velocities of line and heat bath (the lattice) as well as 
 line and superflow are to occur in the equation of motion. 
We thus assume isotropy and 
the existence of only one charge carrier, for simplicity of
representation. 
%In the general case, Galilean invariance is broken in a superconductor.}
The force balance equation between Magnus and dissipative as well as
transverse forces on the vortex is then expressed by 
\cite{otterlo,kopningrisha,koljamutual} 
%(using $\hat{\bm \Gamma}_s \equiv \vec{\Gamma}_s/|\vec{\Gamma}_s|$):
\begin{eqnarray}
{\bm F}_{M} & = & 
\frac12 h\rho_s\left({\bm v}_s (\omega)
-\dot {\bm X} (\omega)\right)
\times {\bm X}' \nonumber\\
& = &   
D(\omega) \left(\dot {\bm X} (\omega) -{\bm v}_n(\omega)\right)
+ D'(\omega ){\bm X}'\times\left(\dot {\bm X} (\omega)
-{\bm v}_n(\omega)\right)\,.
\label{eqmotion1}
\end{eqnarray}
It is assumed that the vortices are singly quantized, with 
a {constant} circulation vector, the factor of 1/2 in the Magnus force 
taking account of the paired nature of the superfluid. 
%${\bm \Gamma}_s = \kappa {\bm X}'$. 
For the point vortex we will deal with, 
${\bm X}'=\partial {\bm X}/\partial \sigma = \pm {\bm e}_z$, the upper/lower
sign valid for a positive/negative circulation vortex. 
The first line contains the conventional Magnus force (where
$\rho_s=\rho_s(T)$ is the superfluid density), %, $\rho_0=\rho(T=0)$), 
and the second line the dissipative and reactive mutual friction forces from 
momentum exchange between vortex line and quasiparticle bath.
We further assume in what follows for the sake of simplicity 
that the normal component is clamped, {\it i.e.} ${\bm v}_n=0$ (this,
of course, destroys the property of Galilean invariance afforded by 
(\ref{eqmotion1})). 
Then, a convenient writing for the above equation of motion is 
\begin{equation}
\frac{h \rho_s}2 {\bm v}_s (\omega)\times {\bm X}' %n_{ij} v_s^j (\omega) 
%{\bm J}_s (\omega)\times {\bm \Gamma}_s
%{\bm X}' 
 = D(\omega) \dot {\bm X} (\omega) + 
\left(D'(\omega )-\frac12 h\rho_s\right)
{\bm X}'\times\dot {\bm X} (\omega)\,, \label{eqmotion2}
\end{equation}
that is, there is a coefficient $D(\omega)$ of the force linear in
the velocity of the vortex, the damping force, and a coefficient 
$D'(\omega )-h\rho_s/2 $ % = D'(\omega )-m\kappa\rho_s$ 
of a force perpendicular to the vortex 
velocity, which represents 
the Magnus force part proportional to vortex velocity (the Hall term).   
The driving term %${\bm J}_s \times{\bm \Gamma}_s= 
$(1/2) h\rho_s {\bm v}_s\times {\bm X}'$ 
(the superfluid current part of the Magnus force)
is proportional to the circulation vector and the superflow 
current density  ${\bm j}_s = \rho_s {\bm v}_s$.
In the case of superconductors, the driving force is 
equal to a Lorentz force exerted by the flux line on the electrons, 
and is given as ${\bm j}_s\times{\bm \Gamma}_s
\equiv {\bm J}_{\rm el}\times {\bm \Phi}_0$, 
with $|{\bm \Phi}_0| = h/ 2e $ equal to the flux 
quantum. The electrical superfluid current of electrons  
${\bm J}_{\rm el}=-e\rho_s {\bm v}_s$.
%, with ${\bm v}_s =(\hbar/m)(\nabla \theta + (2e/\hbarc){\bm A})$.
% (where the quantity $\rho_s$ is the bulk superfluid number density of
% electrons).  
%For normal fluid velocities different from zero, the replacement 
%$\dot {\bm X} \Rightarrow \dot {\bm X} - {\bm v}_n$ yields the 
%equations of motion.
%\subsection{Mutual friction in relaxation time approximation}

The mutual friction coefficients from equations (\ref{eqmotion1}) respectively
(\ref{eqmotion2}) are determined by the momentum (and energy) exchange 
of the quasiparticle bath with the superfluid vacuum containing 
the topological defect structure vortex.  
In the quasiclassical limit, their value is governed 
%in the quasiclassical limit\footnote{The quasiclassical 
%limit is defined to be the one in which $k_F\xi \gg 1$ holds,  
%and quasiparticles at the Fermi level 
%move along straight lines to a good approximation.}
by the quasiparticle kinetic equation \cite{stone2,makhlin,BGK}.
\subsubsection{The $s$-wave case}
We now consider the simple case of a singly quantized 
vortex in a two-di\-mensio\-nal isotropic superfluid, for which 
the quasiparticle momentum has components ${\bm k}=(k_r,k_\phi)$,  
and the zero mode branch 
%for each of the two spin directions takes the form 
\begin{equation}\label{zeromode}
E(k_\phi) = - \omega_0 k_\phi \,.
\end{equation}
The level spacing $\omega_0$ is of order $\hbar/(m\xi^2)\sim
\Delta^2_\infty/E_F$, where $m$ is the (effective) fermion mass. 
The energy levels correspond to those of electrons on anomalous Landau
levels, which are linear in momentum \cite{stone2}. 
 
On the approximation level of a single relaxation-time in the quasiparticle 
kinetic equation, the coefficients are then given by  
\cite{otterlo,stone2,koljamutual}\footnote{We remark that in the
literature the setting $\hbar\equiv 1$ is frequently taken, which 
means that Planck's quantum of action $h\equiv 2\pi$.}
\begin{eqnarray}
D(\omega) = \frac{i hC_0\omega_0}4
\left(\frac 1{\omega-\omega_0 + i\tau^{-1}}
+
%\left(\begin{array}{c} + \\ - \end{array}\right)
\frac 1{\omega+\omega_0 + i\tau^{-1}} \right)\nonumber\\  
= \frac{(hC_0/2) ( \tau^{-1} -i\omega )/\omega_0}
{1+\left((\omega_0\tau)^{-1}- i\omega/\omega_0\right)^2}\,,
\label{Domega}
\end{eqnarray}
for the longitudinal coefficient in equation (\ref{eqmotion2}) and, 
for the transverse coefficient,
\begin{eqnarray}
D'(\omega ) - \frac12 h\rho_s = \frac{hC_0\omega_0}4
\left(\frac{1}{\omega-\omega_0 + i\tau^{-1}}
-
%\left(\begin{array}{c} + \\ - \end{array}\right)
\frac{1}{\omega+\omega_0 + i\tau^{-1}}\right)\nonumber\\  
= \frac{- hC_0/2}
{1+\left((\omega_0\tau)^{-1}- i\omega/\omega_0\right)^2}
\label{D'omega}\,.
\end{eqnarray}
In these equations, 
the parameter density $C_0= k_F^2/2\pi$ (in three dimensions: $C_0=
k_F^3/3\pi^2$) is the normal state density.\footnote{ 
The parameter $C_0$ is in general a measure of the density at the
location of gap nodes; it is zero if no gap nodes are present. 
In our case, the gap nodes are at the position of the vortex line, 
other possible occurences of nodes are in the bulk of $p$- or $d$-wave 
superfluids.}  
The parameter $\tau$ is a
constant relaxation time in the collision term of the kinetic
equation. The relations above hold 
provided that the conditions $\omega_0\ll (\hbar\beta)^{-1}=k_B T/\hbar$,
$\tau\gg \hbar\beta$ and $\hbar \omega\ll k_B T_c$, $T\ll T_c$ are met. 
The first condition means the thermal population of the levels 
represented by (\ref{zeromode}), such that the sums over  
the bound state levels in the general expressions for $D,D'$
\cite{koljamutual} can be converted into integrals. 
This condition is not very stringent,
and allows for a use of the formulas above for actually quite low 
temperatures, because $\omega_0\sim T_c (\Delta_\infty(T)/\Delta_\infty(T=0))
(\Delta_\infty (T) /E_F)\ll T_c$. The second condition implies that the 
broadening of levels by scattering is much less than the temperature.
Finally, the condition $\hbar \omega\ll k_B T_c$ restricts the energy 
equivalent of the vortex motion frequency, $\hbar\omega$,  
to be much less than the bulk gap $\Delta_\infty\sim k_B T_c$. 
% amplitude (the vortex velocity will then be
%much smaller than the pair-breaking velocity).   
%The term of dimension mass $M_c\equiv hC_0/2\omega_0$.
% and the 
% upper/lower signs in the equation determining 
% $D'(\omega )-h\rho_s$ are for a vortex with positive/negative unit 
% circulation. The dissipative mutual friction coefficient $D(\omega )$
% is independent of circulation's direction.    

It is important to point out that the validity of 
(\ref{Domega}) and (\ref{D'omega})
relies on the transverse and longitudinal coefficients in the low energy 
limit being determined by 
the core level spacing $\omega_0$ and the scattering frequency 
$\tau^{-1}$ alone.  
The relations for the Hall and longitudinal coefficients 
%, under such an assumption, 
are then given irrespective of an electric 
charge of the particles carrying the superfluid current.

Two limits of the equations of motion are particularly well known.
The first is provided by $\omega_0\tau\gg 1$, $\omega \ll \omega_0$
and corresponds to massive vortex motion under influence of the 
superfluid Magnus force ($ \dot X^ i(\omega) =  -i \omega X^i (\omega)$): 
\begin{equation}\label{taubig}
M_c \ddot{\bm X} = \frac h 2 \rho_s \, {\bm X}'\times 
\left(\dot {\bm X}-{\bm v}_s\right)\,. 
\end{equation}
At the low temperatures considered, we neglected the small 
contribution of $(h/ 2) (C_0-\rho_s) {\bm X}'\times \partial {\bm X}
/\partial t \simeq (h/ 2)\rho_n {\bm X}'\times \partial {\bm X}/\partial t$, 
where $\rho_n$ is the normal density in the superfluid state. 
%, which equals approximately 
This last expression represents the Iordanski\v\i\, 
force \cite{sonin2} (remember that we fixed ${\bm v}_n=0$), which is 
also present in Bose superfluids and not related to spectral flow 
\cite{grisha3}.
%(if we reinstate ${\bm v}_n$, {\it i.e.}
%put $\dot {\bm X}\rightarrow \dot {\bm X}-{\bm v}_n$).
 
The Magnus force dominates in the above  limit of `slow' vortex motion 
with $\omega\ll\omega_0$
the massive term
(the Hall term on the right-hand side is larger by a
factor of $\omega_0/\omega$ than the inertial term). 
%This domination is a necessary consequence of the definition of 
%the dynamic vortex mass
%$M_c$ as the coefficient of $\partial^2\!{\bm X}/\partial t^2 
%=-i\omega\partial{\bm X}/\partial t$.   
Still, the vortex core dynamical 
mass $M_c = h C_0/(2\omega_0) \sim m C_0\pi \xi^2$ 
is much larger than the mass %(\ref{M0})
arising from the compressible superflow outside the vortex core. 
%(with the logarithm  infrared cutoff $R$ for a neutral superfluid):  
%\begin{equation}
%\frac{M_0}{M_c}\sim \frac{m\rho_s\kappa^2}{4\pi c_s^2}\frac{\ln (R/\xi)}
%{mC_0\pi\xi^2}\sim \left(\frac a\xi\right)^2 \ln (R/\xi) \ll 1  
%\end{equation}
Their ratio is of order $M_0/M_c \sim (d/\xi)^2$, where  
the quantity $d$ signifies the interparticle spacing
$d\ll\xi$, so that $M_0$ may be neglected in the equation of motion 
(\ref{taubig}).  
%The smallness of $M_h$ as compared to $M_c$
%constitutes the reason for neglecting $M_h$ in the equations of
%motion.

Dissipative motion prevails in the limit $\omega_0\tau\ll 1$,
$\omega\tau\ll 1$:
\begin{equation}\label{tausmall}
\frac h2  \rho_s {\bm v}_s \times {\bm X}'
= % M_h \ddot {X^i} +
\frac h2 C_0(\omega_0\tau) \dot {\bm X} 
+ \frac h2 
C_0(\omega_0 \tau)^2 \dot {\bm X}\times {\bm X}'\,.
\end{equation}
The vortex motion is overdamped,
%$M_0\partial^2!{\bm X}/\partial t^2$ appearing on the right-hand side),  
% if $\tau \kappa_0/a^2 \gg \omega/\omega_0$. 
% which will be fulfilled 
with friction coefficient given by the expression  
$\eta= (1/2) h C_0\omega_0\tau $.\footnote{ 
We do not show the hydrodynamic and 
transverse mass terms on the right-hand side of 
(\ref{tausmall}), which are small compared to the friction term. 
The transverse mass \cite{kolyavinokur}, relating vortex velocity 
and momentum in different directions, can be defined dynamically
from the equations of motion (\ref{eqmotion2}),
 like $M_c$.} 
%The transverse inertial term
%is small compared to the friction 
%term, just as the inertial term in (\ref{taubig})
%is small compared to the Hall term.} 
%The transverse mass, relating vortex velocity and 
%momentum in different directions, is defined from the coefficient
%$D(\omega)$ in the same way as the longitudinal mass $M_c$ is defined 
%from $D'(\omega)-h\rho_s/2$.} 
It is observed that time inversion 
invariance is spoilt by the fact that the first term on the right-hand
side of the equation above does not have the factor ${\bm X}'$, which
changes sign under time inversion. 
% = (h/2)C_0(\omega_0\tau) \sim \hbar |\kappa_0| C_0\tau /\xi^2$. 
The dissipation is in the given  
limit of ohmic nature with a longitudinal 
conductivity independent of the driving frequency $\omega$.   
The Hall
part of the Magnus force is suppressed by $\omega_0\tau$ relative to 
the friction term, which is the result of the spectral flow phenomenon
discussed above. Spectral flow is made possible because the minigap 
$\omega_0$ between zero mode levels is broadened by a large quasiparticle
collision frequency $\tau^{-1}\gg \omega_0$. On the other hand, 
in the case of (\ref{taubig}), spectral flow is impeded 
by the presence of the minigap 
(save for tunnelling events between the levels \cite{grisha3}),  
and the Magnus force obtains. 
It is to be mentioned that in a superconductor the energy levels 
above the gap are also 
quantized into Landau levels with the interlevel distance
$\hbar\omega_c$, where $\omega_c$ is the usual 
magnetic cyclotron frequency. For sufficiently small magnetic field,
however, the spectrum is quasi-continous and no change of the above
results applies \cite{kopningrisha}. 
\subsubsection{The $d$-wave case}
For $p$- and $d$-wave superconductors, the gap has nodes not only 
at the location of the vortex itself, but also in the bulk superfluid.
The $d$-wave case distinguishes itself by the fact 
that these nodes occur on {\em lines} in momentum
space, whereas in the $p$-wave case they only occur at {\em points} 
\cite{grishabook}.  
%Hence quasiparticles are easily excited even far away from the
%vortex core, which 
The lines of gap nodes in $d$-wave superconductors 
lead to a profound alteration of the effective vortex 
dynamical equations \cite{fluxflow,kolyaresonant,makhlin}. 

To explain the essential features of these changes, we observe, first
of all, that the minigap $\omega_0=\omega_0(\alpha)$
is a function of the angle $\alpha$, which indicates the position 
of the gap nodes in momentum space, 
where the gap modulus $\Delta = \Delta_0\sin (2\alpha)$.  
The average minigap $\Omega_0\equiv\, 
<\!\omega_0(\alpha)\!>$ is of
the same order $\Delta_\infty^2/E_F$ as the constant $\omega_0$
in the $s$-wave case, $\Omega_0=O(\omega_0)$. 
There exists, however, 
an additional energy scale, the true quantum-mechanical 
interlevel distance in the vortex core $E_0$, 
which is defined according to a Bohr-Sommerfeld type of
quantization prescription for the canonically conjugate 
{\em quantum} variables $k_\phi(\alpha)$ and $\alpha$. 
In a $\alpha$-dependent version of (\ref{zeromode}),
$k_\phi(\alpha)= -E/\omega_0(\alpha)$, the Bohr-Sommerfeld
quantization reads $\oint k_\phi (\alpha) d\alpha = h(m+\gamma)$ (with $\gamma
=1/2$), so that the true interlevel distance is given by  
$E_0^{-1}=h^{-1}\int_0^{2\pi}\!d\alpha \,\omega^{-1}_0(\alpha)$ 
(the integral is rendered finite by the existence of a magnetic field 
\cite{kolyaresonant}). 
There are several different regimes of vortex motion 
corresponding to the ratio of the relaxation rate $\tau^{-1}$ 
and the frequency of vortex motion $\omega$, not only to the (average) minigap
$\Omega_0$, as in the $s$-wave case discussed above, but also to 
$E_0=\hbar \sqrt{\Omega_0 \omega_c} \ll \hbar \Omega_0$.\footnote{There
is a further complication, which 
we do not take into account here because of the low temperatures $T\ll T_c$
considered. If  $T\lesssim T_c$, %delocalized modes come into play, and
the relevant energy scale approaches the cyclotron level spacing
$\hbar\omega_c$. Furthermore, the results have in general to be written
in a form taking into account both particle- and holelike Fermi
surface parts, see \cite{kolyaresonant}.}

There is a parameter region which yields a comparatively 
simple and unique result. This is the case of $\Omega_0\tau\ll 1$, 
which implies $E_0\tau\ll \hbar$. Assuming we have
in addition $\omega\tau\ll 1$, we are led back to an equation of
motion of the type (\ref{tausmall}). On the other hand, if 
$\Omega_0\tau\gg 1$ (and $\omega\tau\ll 1$), there is a `universal
region' \cite{fluxflow,kolyaresonant}, which is realized if 
$E_0\tau\ll \hbar$: 
The dissipative and Hall coefficients are 
%The coefficients $D,D'-h\rho_s/2$ are 
independent of the relaxation time for very low temperatures and 
small magnetic fields %$H\ll H_{c2}$ 
and have the same order of magnitude. 
% There is finite dissipation even if the vortex motion frequency equals zero. 
Finally, if $\Omega_0\tau\gg 1$
and  $E_0\tau\gg \hbar$, the equation of motion is of the type 
(\ref{taubig}), with dominating Magnus force. This is a regime which
will presumably not be realizable with respect to 
practially achieved sample purities,
whereas the universal regime should be observable. 
It appears useful at this point to insert a short treatise on the
terminology in the literature. 
Superconductors 
are classified as being `superclean', `clean', `moderately clean' and
`dirty', according to the value of the ratio $l/\xi$,
where $l=v_F\tau$ is the quasiparticle mean free path.
The superclean limit $\omega_0\tau\gg 1$ (or $\Omega_0\tau\gg 1$)
corresponds to $l\gg \xi (E_F/\Delta_\infty)$,
representing a much stricter condition than its clean counterpart
$l\gg \xi$. The moderately clean and dirty limits correspond to 
$l\gtrsim \xi$ and $l\ll \xi$, respectively. 
We will agree to call a superconductor `moderately clean' 
if it has $\omega_0\tau\ll 1$ (but still simultaneously  
$l\gtrsim \xi$). Because 
of the above discussed $d$-wave pecularities, there is yet another notion 
of `extremely clean', which ought to be introduced. This corresponds
to the extreme limit  $\Omega_0\tau\gg 1$ and $E_0\tau\gg \hbar$, or 
$l\ggg \xi (E_F/\Delta_\infty)$.

The discussion of $s$-wave vortex motion to follow is thus valid  
for the $d$-wave case in a straightforward sense only if we are in  
the moderately clean region $\Omega_0\tau\ll 1$, $E_0\tau\ll 1$,
and have additionally $\omega\tau\ll 1$, 
with the local form of the  %predominantly dissipative 
equation of motion in (\ref{tausmall}). 
%It should also be remarked 
%that the semiclassical approach (the Andreev limit of the 
%Bogoliubov-deGennes equations \cite{kosztin}), on which the validity of 
%the present treatment relies, can become questionable 
%for high-$T_c$ superconductors.  In these, $d$-wave symmetry, or some 
%variety of this pairing symmetry with small deviations from pure
%$d$-wave, is believed to be present \cite{tsuei}.

%because there the condition for the 
%applicability of the semiclassical treatment, $k_F^{-1}\gg \xi$, 
%is in general not fulfilled.  
\subsubsection{Nonlocal motion in the time domain}
In order to obtain the real time motion of the vortex, 
we rewrite the general equation of motion in its convoluted form.
To this end, we use that $ \dot X^ i(\omega) =  -i \omega X^i (\omega)$   
and define $K^D (\omega) \equiv  i\omega D(\omega)$, 
$K^H(\omega) \equiv i \omega (D'(\omega)-h\rho_s/2)$. Then, by     
multiplying (\ref{eqmotion2}) with 
$(1/2\pi)\int d\omega \exp(-i\omega t) $, we have  
\begin{equation}\label{eqmotiont}
\frac {h\rho_s}2  {\bm v}_s \times {\bm X}'
= - \int dt' K^D(t-t') {\bm X} (t') 
-{\bm X}'\times\int dt' K^H (t-t') {\bm X}(t') \,,
\end{equation} 
where the kernels are, by use of 
(\ref{Domega}) and (\ref{D'omega}), the Fourier transforms
%\renewcommand{\arraystretch}{2}
%\begin{eqnarray}
$$
K^D(t-t') = %&=& 
\frac{1}{2\pi}
\int^{+\infty}_{-\infty} 
%\int 
d\omega\, \exp[ -i\omega (t-t') ]\,
\frac{i\omega \,(hC_0/2) ( \tau^{-1} -i\omega )/\omega_0}
{1+\left((\omega_0\tau)^{-1}- i\omega/\omega_0\right)^2}\,,
%\label{JDt}
%\nonumber\\
$$
$$
K^H(t-t') =  %&=& 
\frac{1}{2\pi}
\int^{+\infty}_{-\infty} 
%\int 
d\omega\, \exp[-i\omega (t-t') ]\,
\frac{-i\omega\, hC_0/2}
{1+\left((\omega_0\tau)^{-1}- i\omega/\omega_0\right)^2}\,.\,\,\,\,\,
\nonumber %\label{JHt}
$$
%\end{eqnarray}
%\renewcommand{\arraystretch}{1.0}
The integrals 
can be  evaluated {\it via} complex contour integration. The poles 
$z_1=\omega_0-i\tau^{-1}$, $z_2= -\bar z_1$ 
are lying both in the lower half plane, so that 
the kernels are causal:
\begin{eqnarray}
\left(\begin{array}{c} K^D(t-t')\\ K^H(t-t') \end{array}\right)
&=&
\left(\begin{array}{c} -1 \\ i \end{array}\right)
\frac{h C_0 \omega_0}4 \,\frac{1}{2\pi}
\oint dz  \, e^{ -iz(t-t') }
\left(\frac z {z-z_1}\pm \frac z {z+\bar z_1} \right)
\nonumber\\
& = & \left(\begin{array}{c} i \\ 1  \end{array}\right)
\frac{hC_0\omega_0}4\,\theta (t-t')\,
\left[ {\rm Res}(z_1) + {\rm Res} (-\bar z_1) \right]\,.\nonumber%\\
\end{eqnarray}
Calculating ${\rm Res}(z_1) + {\rm Res} (-\bar z_1)$ , 
they are given by ($\Delta t \equiv t-t'$):
\begin{eqnarray}
K^D(\Delta t)&=&
\theta (\Delta t)\, \frac{hC_0}2\,\omega_0^2\, 
\exp\left[-\frac{\Delta t} \tau\right]
\left(\sin(\omega_0\Delta t) + \frac 1{\omega_0\tau}\cos (\omega_0\Delta
t)\right)\,,\nonumber\\
K^H(\Delta t) & = & 
\theta (\Delta t)\, \frac{hC_0}2\,\omega_0^2\, 
\exp\left[-\frac{\Delta t}\tau\right]
\left(\cos(\omega_0\Delta t) - \frac 1{\omega_0\tau}
\sin(\omega_0\Delta t)\right)\,.\nonumber%\\\label{kernelsJt}
\end{eqnarray}
The equations of motion (\ref{taubig}) and (\ref{tausmall}), which are
local in time, result from the non-local equations
(\ref{eqmotiont}) only in case 
that the frequency of vortex motion is sufficiently 
smaller than the bigger of the two frequencies
$\omega_0,\omega_\tau\equiv \tau^{-1}$,
{\it i.e.} such that $\omega\ll (\omega_0^2 +\omega_\tau^2)^{1/2}$
and either $\omega_0\tau \gg 1$, like in (\ref{taubig}), 
or $\omega_0\tau\ll 1$, like in (\ref{tausmall}), holds.

The action corresponding to (\ref{eqmotiont}) 
is obtained by
integrating with respect to the vortex position ${\bm X}(t)$.   
We arrive at 
\begin{eqnarray}
S[{\bm X}(t)] = \frac{h\rho_s}2 \int dt \,\psi [{\bm X}(t)]
+%\frac12 
\int dt \int^t dt' K^D(t-t') {\bm X} (t)
\cdot {\bm X} (t') \nonumber\\
-%\frac12 
{\bm X}'\cdot \int dt \int^t dt' K^H (t-t')\, 
%{\rm sgn}(t-t')\, %\left(
{\bm X} (t) \times  {\bm X} (t') % \right)
\,.%\nonumber\\
\label{actiongeneral}
\end{eqnarray}
%written here in a symmetrized form.
The gradient of the 
stream function $\psi$ 
describes the superflow perpendicular to the vortex,  
$\nabla \psi = {\bm v}_s\times {\bm X}'$.

Making use of the 
phase angle $\Phi=\arctan [(\omega_0\tau)^{-1}]$ and standard 
addition theorems, we can cast the action into the suggestive form 
\begin{eqnarray}
S[{\bm X}(t)] 
=  \frac{h\rho_s} 2 \int dt \,\psi [{\bm X}(t)]%\,\nonumber\\
+ \frac{hC_0}2\,\omega_0 
\sqrt{\omega_0^2 +\omega_\tau^{2}}
%\omega_0^2 \sqrt{1+(\omega_0\tau)^{-2}}
\int dt \int^t dt'\exp[-\omega_\tau\Delta t ]\nonumber\\
\,\,\,\times \left\{ 
{\bm X} (t) \cdot {\bm X} (t')
\sin (\omega_0\Delta t + \Phi)
%\arctan (\omega_0\tau)^{-1}
%\right.\nonumber\\
- %\left. 
{\bm X}' \cdot ({\bm X} (t) \times {\bm X} (t'))\,  
%{\rm sgn}\,(t-t')
\cos (\omega_0 \Delta t + \Phi
%\arctan (\omega_0\tau)^{-1}
)\right\}\,.\nonumber%\\\!\!\!
\end{eqnarray}
%\begin{equation}
%{ }
%\end{equation}
The factors multiplying the dot- and cross-products of ${\bm X}(t)$ 
and ${\bm X}(t')$ in this nonlocal Lagrangian  
are for any value of $\omega_0\tau$ and thus $\Phi$ 
just $3\pi/2$ out of phase.
In case that $\omega_0\tau\ll 1$, $\Phi \simeq\pi /2$, 
the first term with the dot-product dominates, whereas if 
$\omega_0\tau\gg 1$, $\Phi \simeq 0$, the second one involving the 
cross-product does. 
\subsection{Euclidean vortex motion}
%\subsection{The Euclidean action nonlocal in time}
For a description of tunnelling motion, 
we have to use the Euclidean action in the interval
$[-\hbar\beta/2,\hbar\beta/2)$.
% of periodic imaginary time motion. 
Performing the Wick rotation in (\ref{actiongeneral})
through the replacement 
$t\rightarrow -it_e$, % (in frequency space $\omega\rightarrow i\omega$), 
gives the Euclidean action $S_e[{\bm X}(t_e)] 
\equiv -i S[{\bm X}(t\rightarrow -it_e)]$: 
\begin{eqnarray}
S_e[{\bm X}(t_e)] &=& -(h\rho_s/2)
\int^{\hbar\beta/2}_{-\hbar\beta/2}  dt_e \,\psi [{\bm X}(t_e)]\nonumber\\
& + & %\frac12
\int^{\hbar\beta/2}_{-\hbar\beta/2} dt_e 
\int^{t_e}_{-\hbar\beta/2}dt_e' \left[-K^D(t_e-t_e') 
%K^D(|t_e-t_e'|) 
{\bm X} (t_e)\cdot {\bm X} (t_e')\right.\nonumber\\
&+& % \frac12 {\bm X}'\cdot 
%\int^{\hbar\beta/2}_{-\hbar\beta/2} dt_e
%\int^{\hbar\beta/2}_{-\hbar\beta/2} dt_e'
\left. K^H (t_e-t_e')\,
%K^H (|t_e-t_e'|)\, 
%{\rm sgn}(t_e-t_e')\, %\left(
{\bm X}'\cdot\left\{ {\bm X} (t_e) \times  {\bm X}
(t_e')\right\}\right]\,\label{Euclideanaction}
% \right)\,,
\end{eqnarray}
where, under the condition that the real frequency 
$\omega\ll\omega_0,\omega_\tau$, 
%\sqrt {\omega^2_0 +\omega^2_\tau}$, 
the kernels are approximately given as 
\begin{eqnarray}
\left(\begin{array}{c} K^D(t_e-t_e')\\ K^H(t_e-t_e') \end{array}\right)&=&
%\left(\begin{array}{c} i \\ -i \end{array}\right)
\frac{hC_0\omega_0}{2\pi}\int^{\infty}_{0} 
d\omega\, e^{ -\omega |t_e-t_e'| }\frac{\omega}{\omega_0^2+\omega^2_\tau}
\left(\begin{array}{c}\omega_\tau \\ -\omega_0 \end{array}\right)
%\left(\frac{\omega}{-\omega_0 + i\tau^{-1}}\pm
%\left(\begin{array}{c} + \\ - \end{array}\right)
%\frac{\omega}{\omega_0 + i\tau^{-1}}  \right)
\nonumber\\
&= & \frac{hC_0}{2\pi}\,\frac{1}{(t_e-t_e')^2}\,\,
\frac{\omega_0}{\omega_0^2+\omega_\tau^2}
\left(\begin{array}{c}\omega_\tau  \\ -\omega_0 
\end{array}\right)\,,\label{kernelslowomega}
\end{eqnarray}
and include nonlocality in lowest order.
% of the frequency.
% of imaginary time motion. 
The dissipative kernel is of the Caldeira-Leggett-form 
(\cite{CLPRL}--\cite{shin})   
for ohmic dissipation, 
$K^D=(\eta/\pi)(\Delta t_e)^{-2}$, with the friction coefficient 
$\eta = (1/2) h C_0 
(\omega_0\omega_\tau)/(\omega_0^2+\omega_\tau^2)$.   
%\simeq  (1/2) h C_0\omega_0\tau $ (if $\omega_0\tau\ll 1$). 
The dissipation due to bound states
is at a maximum if $\omega_0\tau= O(1)$ and vanishes
in the limits $\omega_0\tau\rightarrow \infty$  and 
$\omega_0\tau\rightarrow 0$. 
The nonlocality of the Hall term in the action is of the same 
importance as that of the friction term if $\omega_0\simeq \omega_\tau$.
%It should be mentioned that, in this order of frequency,
%a mass term like that in (\ref{taubig}), does not appear. 
%If we were to include, {\it e.g.}, a mass term, 
%it is useful to remind ourselves 
%that terms of different order in frequency play different roles 
%in the equations of motion and should be treated properly 
%in the Wick rotation process. 	
%In general, a term in the mutual
%friction coefficient (\ref{Domega}), 
%which is of even order in $i\omega$, gives rise
%to dissipation, whereas a term of odd order is non-dissipative.
\vspace*{-0.85em}
\subsection{Different regions of parameter space} 
%\subsection{Tunnelling in different regions of parameter space} 
It is rather obvious that the motion of a vortex for arbitrary competing 
contributions in the action (\ref{Euclideanaction}) 
can be quite complicated.
% (also cf. the basic real time equation of a 3d vortex of arbitrary shape
% in an unpaired, dissipation-free superfluid in 
% (\ref{originaleqmotion})). 
In principle, the following contributions 
in the action are conceivable. 
In addition to the terms appearing in (\ref{Euclideanaction}), 
there can be contributions arising from the self-interaction, 
like in (\ref{SEQ}), that is, the hydrodynamic mass 
and elasticity terms (the self-energy is usually absorbed into the
potential). The hydrodynamic mass was argued to be in general
completely negligible as compared to the dynamic core mass in paired Fermi 
superfluids. The elasticity arises from the generalization 
of a 2d or rectilinear vortex to one of arbitrary shape and the
additional self-energy this creates. 
Hence, in order to
actually describe an imaginary time motion and thus evaluate the tunnelling 
process probability, only certain classes of metastability problems have been
investigated.

\begin{table}[t]\begin{center}
\begin{tabular}
{|p{1cm}|p{1cm}|p{1.5cm}|p{1.5cm}|p{2cm}|p{3.5cm}|}\hline
Ref. & Mass & Elasticity & Magnus & Dissipation & Potential \\\hline\hline
\cite{gorokhovquant} & No & No & \checkmark & No &
$U_0(\sqrt{x^2+y^2})-Fx$\\\hline
\cite{chudnovsky} & No & \checkmark & \checkmark & No & 
$V_g(q)[\epsilon \ll 1]$ \\\hline
\cite{morais} & No & \checkmark & No & \checkmark (ohmic) &
$V_g(q)[\epsilon \ll 1]$\\\hline
\cite{feigelman} & No & No & \checkmark & \checkmark(ohmic) & 
$V_g(q)[\epsilon = 1]$\\\hline
\cite{shin} & No & No & \checkmark & \checkmark (ohmic) & 
$V_g(q)[{\rm any}\, \epsilon]$\\\hline
\end{tabular}\vspace*{1.5em}
\caption{\label{qtoverview} Comparison of different approaches to
quantum tunnelling with respect to dominant contributions employed in the
calculation of the Euclidean tunnelling action (shown 
by an entry with a checkmark, \checkmark).  
%and the form of the potential. 
The Magnus column indicates if
a linear coupling of the vortex velocity to the superfluid background, 
generating the vortex velocity part of the Magnus force, has been
used. In the potential $V_g(q)$, $\epsilon \ll 1$ means that the case 
of a near critical potential has been treated. In Ref. \cite{morais}, 
this was done by an analytical method, in \cite{chudnovsky} by a numerical
one. In the case of entry \cite{feigelman}, both {\it limitis} 
of dominant Magnus and dissipation contributions have been considered, but not 
the intermediate case of equal strengths. This intermediate case has been 
examined numerically in \cite{shin}, far off 
and near criticality.}\end{center}
\end{table}

The (stream function) potential is, for reasons of
(analytical) solvability, frequently represented as a 
quadratic plus cubic potential for a one-dimensional generalized
co-ordinate $q$. This potential is conventionally 
parameterized with two or three quantities, a height $h$ and width 
$w$ of the potential, and possibly with the additional parameter 
of closeness to criticality $\epsilon$.\footnote{The potential 
(\ref{Vepsilon}) corresponds to
the so-called `tilted washboard potential near criticality' if 
$\epsilon \ll 1$, see, {\it e.g.}, \cite{ivlev}.} 
The potential thus has the general representation  
\begin{equation}\label{Vepsilon}
-\psi = V_g(q) = 3 V_0 \left[\epsilon \left(\frac q{q_0}\right)^2
- \frac 23  \left(\frac q{q_0}\right)^3\right]\,,
\end{equation} 
where $\epsilon =\sqrt{1-v_s/v_{cb}}$ measures the closeness to a 
critical external velocity $v_{cb}$, for which the barrier vanishes. 
A measure of the typical curvature radius of the potential is $q_0$. 
The zeros of the potential are at $q=0$ and $q=(3/2)\epsilon q_0$, 
so that the width of the potential may be defined to be $w= (3/2)\epsilon
q_0$. The maximum is at $q_{\rm max}=\epsilon q_0$ and its height equals 
$h=V_0\epsilon^3$. 
The different cases and approximations investigated in a 
selection of recently published 
papers on quantum tunnelling are brought together 
for comparison in Table \ref{qtoverview}. 
% (for a more 
%complete overview, see the corresponding bibliography section). 
For the detailed results and methods used, 
the reader is referred to the cited works. 
%We confine ourselves here to same salient features of the results.  

%The general neglect of a `mass' term in any of the calculations
%({\it i.e.}, a term of the form $(1/2)M\dot q^2$ in the effective
%vortex action),  
%is a result of the treatment of the problem 
%within a central collective co-ordinate approach. This 
%can be understood as follows. 
%Consider first the Magnus case ($\omega_0\tau\gg 1$) with equation of 
%motion of the type (\ref{taubig}).
In relation to what we have found in the preceding section, 
% which preceded this one,
we can make the following observations. 
The core level spacing is 
of order $\omega_0\sim \hbar/(m\xi^2)$, and the Magnus force 
dominates over the mass term in (\ref{taubig})
if the vortex motion frequency $\omega\ll\omega_0$.
% (which is simultaneously the condition for the definability of the mass). 
The collective co-ordinate approach implies that  the 
curvature radius of the potential $q_0\gg \xi$, because the motion 
of a massless vortex in the potential is of typical frequency 
$\omega\sim \hbar/(mq_0^2)$, the `cyclotron' frequency 
associated with $q_0$. The condition for the collective
co-ordinate approach, $q_0\gg \xi$, is thus equivalent to the
dominance of the Magnus contribution over the core 
mass term in the superclean limit $\omega_0\tau\gg 1$. 
%This is true even for the relatively 
%large value of the core mass $M_c=hC_0/(2\omega_0)$, which is,
%however, in any case suppressed by the factor $\omega/\omega_0$. 
The fundamental hydrodynamic analysis of the last
section, treating the Magnus force as dominant, 
therefore remains valid for tunnelling 
in the case of a superclean $s$-wave fermionic superfluid.
The case of a $d$-wave superfluid is, as already argued, more
intricate. The dominance of the Magnus force only obtains in the 
extremely clean limit, whereas in a superclean limit, which has 
$E_0\tau\ll \hbar$, the dissipative and Hall force components are of 
comparable magnitude. This necessitates a complete treatment of the tunnelling 
phenomenon in two dimensions for this `universal' \cite{fluxflow}
parameter regime, even for a point vortex, 
because, in the plane, one has to solve two coupled differential
equations of motion. We will further discuss this case and its possible 
occurence in high-$T_c$ superconductors below.
In the extremely
clean limit respectively for large enough cyclotron level spacing, and 
if the temperature goes to zero, {\it i.e.} is
less than any of the energy scales associated with the average and
true minigap $\Omega_0=<\!\omega_0\!>$ and $E_0$, 
the superfluid Magnus force is the
only remaining nondissipative force on the vortex,  
%in the extremely clean limit respectively for large enough 
%cyclotron level spacing, 
and the analysis of tunnelling in the last section is valid.
% for any superfluid in which 
% the remaining dissipation is negligible.   
 
%In the case of large quasiparticle collision frequency, with 
%the moderately clean condition $\omega_0\tau\ll 1$, 
%the dominance of friction over the hydrodynamic mass 
%and Hall terms 
%is expressed by $\eta/(M_0\omega)\gg 1$, equivalent 
%to the condition $\omega_{0a}\tau \gg \omega/\omega_0$, with $\omega_{0a}= 
%\kappa/a^2$, the (large) 
%`cyclotron' frequency associated with the interparticle
%spacing. 
%%The frequency $\omega$ has to be much less than $\omega_0$,  
%%if one is using solely a collective co-ordinate, as already argued above.
%This condition is certainly fulfilled, because it ultimately
%expresses the validity limit of a hydrodynamic, collective co-ordinate 
%treatment.
%\subsection{Possible observation of tunnelling
%in high-$T_c$ superconductors}
\subsection{Quantum tunnelling in high-$T_c$ superconductors}
\label{sechighTctunnel} 
In conventional superconductors, the dissipative component  
in the vortex equations of motion usually is very significant, so that quantum 
tunnelling is largely 
suppressed; the temperature region above zero, in which temperature 
independent quantum tunnelling is of importance, is
exceedingly small \cite{RHF}. 
There is, however, the intensely investigated class of 
high-$T_c$ superconductors, which can very well be in a clean or even 
superclean limit. In this latter case the Hall angle can %be large and 
approach the value $\pi /2$
(the vortex then moving more with the local superflow, rather than
perpendicular to it) \cite{harris}. In addition, and even more
important, in contrast to conventional superconductors these materials
can exhibit quite small coherence lengths in the order of 
the lattice spacing,
%inverse Fermi momentum, and a ratio $E_F/k_B T_c\sim O(1\cdots 10)$. 
and a ratio $T^*/T_c\sim O(1\cdots 10)$ (cf. the energy barrier 
considerations relevant for vortex tunnelling 
in the introductory considerations. 
%on page \pageref{energyconsider}). 
%$E(R_0)/k_B T = (T^*/T)(R_0/d)\ln (R_0/R_c)$ 
%approaching unity. %According to the criteria explained on        
These facts lead to the possibility 
that in some of these superconductors, quantum
tunnelling of flux lines might be observable at low temperatures.  
The ratio of the crossover 
temperature $T_0$ from thermally activated to quantum behaviour for the flux
line depinning to the critical temperature $T_c$ was 
measured in very different materials (\cite{zhang}--\cite{monier}).
It is found, depending on the material, that
$(T_0/T_c)_{\rm high-T_c}\sim 0.03\cdots 0.09$, 
where $T_0$ is defined in these measurements to be 
the temperature at which flux motion deviates from the one 
expected for purely thermal activation.\footnote{In Ref.  
\cite{stein}, however, quite large ratios up to 
$(T_0/T_c)_{\rm high-T_c}\sim 0.22$ have been reported.} 
% and $T_c$ is the critical temperature. 
We compare this with He II, where $T_0\sim 150$ mK and $T_c\sim 2.2$ K, 
so that $(T_0/T_c)_{\rm He II}\sim 0.07$.
Considering the fact that the ratio of the critical temperatures for 
these high-$T_c$ superconductors and helium II can be  
up to a factor of 50, the values of $T_0/T_c$ are 
comparatively close, differing at most by a factor of two.
This suggests a common physical origin of the deviation from 
thermal activation behaviour in helium II and high-$T_c$ superconductors, 
be it quantum tunnelling or some other, classical, flow instability mechanism.
This is only natural from the point of view that both systems represent, 
on a fundamental level, strongly coupled %interacting 
superfluids.
%, in which quantum fluctuations play a major role.
%, in which atomic scales.
%In both systems 
%It should also be remarked 
%This, however, also implies %that a quasiclassical approach 
%(the Andreev limit of the 
%mean field Bogoliubov-deGennes equations \cite{degennes}),
%that a quasiclassical approach,\footnote{The 
%quasiclassical limit is defined to be the one in which $k_F\xi \gg 1$ holds,  
%and quasiparticles at the Fermi level 
%move along straight lines to a good approximation.}
%on which the validity of the formulas we presented relies, 
%can become questionable for high-$T_c$ superconductors.  

%Provided we assume such 
%a quasiclassical treatment %based on a mean field Hamiltonian 
%to be a viable description,and b
There are high-$T_c$ superconductors presumably  
belonging to the class of $d$-wave superconductors \cite{wollman,tsuei}, 
or some variety of this pairing symmetry with small deviations from pure
$d$-wave.
%, the aforementioned complications in the dynamical behaviour
%of vortices arise. 
The considerations of section \ref{secVMFermi} 
for the $d$-wave case then apply, 
provided we assume a quasiclassical, low energy treatment of 
vortex motion in linear response %,  presupposed there, 
to be valid, at least qualitatively, in these superconductors   
%This is no trivial assumption, in particular 
and on the scales of tunnelling.
 
The $d$-wave superconductor, for 
practically achieved maximal sample purities and sufficiently small %external
magnetic fields, will not be in an `extremely clean'
region (Magnus force dominating), but rather in the superclean
`universal' region (for the exact conditions on $H$ and $T$, see
\cite{kolyaresonant}). This implies that there is a magnetic field region 
in which, as already explained, the vortex tunnelling motion 
is not dominated either by the dissipative or the Magnus force 
(the Hall term), even for very low temperatures, but is governed 
by both forces. 
There are indications that in clean high-$T_c$ materials
%, the values of $\omega_0\tau$ are indeed of $O(1)$   \cite{dalen,hoekstra}, 
an intermediate regime between purely dissipative and Hall 
tunnelling may indeed be realized \cite{dalen,hoekstra}. 
The intermediate case thus clearly needs further investigation,
because the (measurable) temperature dependence of the Euclidean action can 
be different from that expected for purely dissipative or Hall motion,  
and the transition from quantum to thermally activated depinning 
of the vortices be of first or second order 
\cite{gorokhov,gorokhovd+n,monier}. 
A first step in this direction has been made in \cite{shin}, where the
problem of quantum tunnelling was investigated with the 
%the frequency independent 
static (low frequency) versions of the formulas for the $s$-wave
case, (\ref{Domega}) and (\ref{D'omega}).
%,for low frequencies, when they become frequency independent. 
It is found that the tunnelling rate displays a minimum for 
$\omega_0\tau\sim O(1)$. This is in accordance with the fact 
that dissipation due to spectral flow is at a maximum for values 
of the parameter $\omega_0\tau$ which are of order unity. Whereas the 
nonlocality of the ohmic dissipation was considered there {\it \`a la} 
Caldeira-Leggett, that of the Hall term, however, was not taken into account. 
It is apparent from (\ref{kernelslowomega}) that this is 
not justifiable in the intermediate case of interest, 
in which $\omega_0\sim \omega_\tau=\tau^{-1}$. Within the formalism 
we presented, %in section \label{secVMFermi}
the Hall term can be treated 
locally in the action only in the limits %for $\omega_0\tau$.  
$\omega_0\tau\rightarrow\infty$ and $\omega_0\tau\rightarrow 0$. 
%and $\omega_0\tau\rightarrow\infty$.    
Apart from this objection, the vortex dynamical behaviour in $d$-wave 
superconductors is in general more complicated than in conventional
$s$-wave superconductors, as we have already pointed out. 
For example, under certain conditions there can be  %conceivable that there 
resonances in the vortex response, if vortex tunnelling frequencies are near 
$(2k+1)E_0/\hbar$, with $k$ an integer number \cite{kolyaresonant}. 
%In clean high-$T_c$ materials, the values of $\omega_0\tau$ can indeed be 
%of $O(1)$ \cite{dalen,hoekstra}, so that an intermediate regime between pure
%Hall and dissipative tunnelling may be realized.  
A complete treatment of the tunnelling 
problem for a $d$-wave system in the intermediate range
% which is of direct experimental relevance, 
then necessitates an incorporation of 
nonlocality in the longitudinal and transverse vortex response, as well 
as possible resonances with collective modes induced by the moving vortex.
%\newpage
%\thispagestyle{empty}
%\newpage
%\thispagestyle{empty}
%\thispagestyle{plain}
\vspace*{-0.75em}
\section{Concluding remarks}
%\addcontentsline{toc}{chapter}{\protect Concluding remarks}
%\markboth{\protect Concluding remarks}{\protect 
%Concluding remarks}

The present work has treated the consequences and limitations of a 
%long wavelength, 
large scale description of the motion and tunnelling 
generation of quantized vortices % as quantum objects 
in dense superfluids. 
%condensed matter systems.  

We may summarize the salient assumptions, pertaining 
to this treatment, in a compact way as follows. 
The vortex object cannot be described in its {\em genesis}, 
because we do not know precisely in which way 
a vortex should be represented on %length (
curvature scales of order the coherence length.  
The intrinsic nucleation process, in all its quantum many-body subtleties, 
happens on these scales. Applying the formalism used in this treatise, 
we thus have to assert that the vortex somehow comes into a topologically 
ensured existence. We can, then, assign a collective co-ordinate to the 
singular center of topological stability. Furthermore, if we wish to describe
the vortex as a string (in three dimensions), or point object 
(in two dimensions), which is tunnelling through a potential barrier, 
we have to adopt the point of view that we are allowed to 
quantize vortex position and momentum, in a canonical manner. These two 
assumptions and their validity lie at the heart of our treatment of
the problem of vortex tunnelling %generation 
in a superfluid at the absolute zero of temperature.
There is, then, no theory 
of vortex {\em nucleation} in a dense, real life condensed matter system,
because we drastically reduce the number of (quantum) 
degrees of freedom actually relevant for 
nucleation. 
There is, though, a consistent theory of vortex quantum tunnelling in the 
large scale domain, which we represented here in its formal requirements 
and geometric implications. 

%Once given these prerequisites, we can develop the theory of vortex motion 
%by using the standard hydrodynamical conservation laws of number and 
%vorticity conservation, as well as the force acting on any object moving
%in an ideal fluid, around which a circulation exists, the Magnus force. 
%In the superfluid, the moving vortex is an elementary object because
%of topological restrictions, which require that vorticity can only exist 
%quantized. We have argued that then, in the long wavelength limit which 
%we assume, the 
%vortex moves like a quantized charged string, the flow field acting upon 
%it like electric and magnetic fields locally on the string.  

The necessity of employing the long wavelength limit affects the 
contributions of different origin in the tunnelling exponent. The 
volume contribution, associated with the incompressible superfluid, always 
dominates over the area contribution. This latter contribution 
is  associated with the vortex effective mass, 
and is thus depending on the detailed dynamical 
behaviour of the vortex on the tunnelling path.  The dominant 
volume contribution, in contrast, depends on the shape of the vortex path in 
configuration space only. 

At absolute zero, the Galilean invariance of 
the bulk superfluid is required to be broken for tunnelling to be 
energetically possible. This breaking of invariance can be attained by 
considering vortex motion in the presence of obstacles. 
The geometrical implications imprinted by a specific 
obstacle chosen play an import part in our analysis 
and give a central result. Namely, if the vortex moves near 
the boundary, trapped by the pinning potential generated by the obstacle, 
only those paths are allowed in which the vortex center remains at least 
within a distance of order the coherence length from the obstacle surface. 
If the flow obstacle, then, 
has curvature perpendicular to the flow passing 
at infinite distance over the obstacle
much larger than parallel to this flow, we have shown that 
the (constant) energy of the tunnelling vortex
cannot be less than a given minimal energy. This minimal energy is needed
by the vortex to be completely describable by the collective co-ordinate    
with which we have equipped it, because else it would come 
within a distance less
than the coherence length to the obstacle boundary.   
This result is generalizable to the case of 
relativistic vortices in spacetime, where the necessary 
breaking of (local) Lorentz invariance for timelike currents will 
lead to the same kind of prediction. 

The theory we have developed can claim to make exact predictions  
on observable tunnelling probabilities in the semiclassical limit, as long as
the collective co-ordinate approach makes (geometrical) sense {\em and} the 
Magnus force contribution is dominating the tunnelling action. We have seen, 
by considering equation (\ref{R0scaled}), that the tunnelling
scales in He II, using the available data, will be of order nanometers. 
Correspondingly, the number of particles 
participating in the tunnelling event, {\it i.e.} those 
contained in the volume determining the tunnelling action, 
 is comparatively small. It is of 
order $N^{(3)}= 4\cdots 6$, given physically realistic estimates of the 
prefactor and the tunnelling rates to be expected. 
Hence, the applicability of our theory, for the actual physical conditions
encountered in He II, is restricted, 
in the sense that it can only give lower bounds for tunnelling rates, 
valid on large enough scales.   
It is, first of all, not entirely obvious that the dominance of the Magnus 
force still holds on the %mesoscopic 
scales relevant for tunnelling. 
Second, we have no really direct means to compare the tunnelling rates 
observed by varying experimental conditions with the predictions of the 
theory. This, of course, stems from the very nature of 
the theory as a geometric theory. 
The predictions it actually makes concern primarily 
the variation of tunnelling 
rates with the geometrical parameters of flow obstacles. Thus the only 
conceivable possibility of checking the validity of the theory 
is the observation of a 
variation of tunnelling rates with micro-orifice surface roughness;  
such an experiment, in a reproducible fashion, has not been carried out yet. 
%More direct means of checking a candidate theory for the observed intrinsic 
%vortex generation processes are provided by checking their       
%dependence on temperature and pressure. 
%The dependence on these thermodynamic
%quantities, in the expression for the tunnelling probability 
%in terms of the tunnelling volume is, however, 
%effectively only 
%contained in parameters like density and the coherence length 
%in the ultraviolet cutoff, of order the coherence length, 
%occuring in a logarithm 
%and thus difficult to observe. 
In order to make further progress in relating the experimental findings to 
a suitable theory, it appears from these arguments that one is
required to go beyond the Magnus force dominance in the tunnelling action
and consider the modifications of this dominance in the mesoscopic scale 
domain, in particular by the interaction of the vortex with the elementary
excitation spectrum of the superfluid. This interaction should, on these 
scales, change the relevant effective forces acting on the vortex, leading 
to measurable effects on the tunnelling probability.    
The same considerations essentially apply to the possible observation 
of tunnelling of flux quanta ({\it i.e.}, of the magnetization) 
in high-$T_c$ superconductors, however
with a considerable amount of complications, caused by different 
parameter domains, as mentioned in the third section. 
In addition, in the general case,  
the requirement that proper electromagnetic gauge invariance 
is to be satisfied should play an important role. 

\nopagebreak
Future research, along
 the directions we have been alluding to above, is needed   
to shed further light on the intrinsic process of the genesis of 
quantized vortices in superfluids.  
%\newpage
%\thispagestyle{empty}

%\renewcommand{\bibname}{}
%\renewcommand{\bibname}{Bibliography}
%\vspace*{-4em}
%\thispagestyle{plain}

\end{document}